\documentclass[twocolumn]{aa}

\usepackage{graphicx}
\usepackage{amsmath,amsfonts,amssymb}
\usepackage{txfonts}
\usepackage{color}
\usepackage{natbib}
\usepackage{float}
\usepackage{dblfloatfix}
\usepackage{afterpage}
\usepackage{ifthen}
\usepackage[morefloats=12]{morefloats}
\usepackage{placeins}
\usepackage{multicol}
\bibpunct{(}{)}{;}{a}{}{,}
\usepackage[switch]{lineno}
\definecolor{linkcolor}{rgb}{0.6,0,0}
\definecolor{citecolor}{rgb}{0,0,0.75}
\definecolor{urlcolor}{rgb}{0.12,0.46,0.7}
\usepackage[breaklinks, colorlinks, urlcolor=urlcolor,
    linkcolor=linkcolor,citecolor=citecolor,pdfencoding=auto]{hyperref}
\hypersetup{linktocpage}
\usepackage{bold-extra}

\def\setsymbol#1#2{\expandafter\def\csname #1\endcsname{#2}}
\def\getsymbol#1{\csname #1\endcsname}

\def\Planck{\textit{Planck}}

\newbox\tablebox    \newdimen\tablewidth
\def\leaderfil{\leaders\hbox to 5pt{\hss.\hss}\hfil}
\def\endPlancktablewide{\tablewidth=\textwidth 
    $$\hss\copy\tablebox\hss$$
    \vskip-\lastskip\vskip -2pt}
\def\tablenote#1 #2\par{\begingroup \parindent=0.8em
    \abovedisplayshortskip=0pt\belowdisplayshortskip=0pt
    \noindent
    $$\hss\vbox{\hsize\tablewidth \hangindent=\parindent \hangafter=1 \noindent
    \hbox to \parindent{$^#1$\hss}\strut#2\strut\par}\hss$$
    \endgroup}
\def\doubleline{\vskip 3pt\hrule \vskip 1.5pt \hrule \vskip 5pt}

\def\L2{\ifmmode L_2\else $L_2$\fi}

\def\DeltaT{\ifmmode \Delta T\else $\Delta T$\fi}
\def\deltat{\ifmmode \Delta t\else $\Delta t$\fi}
\def\fknee{\ifmmode f_{\rm knee}\else $f_{\rm knee}$\fi}
\def\Fmax{\ifmmode F_{\rm max}\else $F_{\rm max}$\fi}
\def\solar{\ifmmode{\rm M}_{\mathord\odot}\else${\rm M}_{\mathord\odot}$\fi}
\def\Msolar{\ifmmode{\rm M}_{\mathord\odot}\else${\rm M}_{\mathord\odot}$\fi}
\def\Lsolar{\ifmmode{\rm L}_{\mathord\odot}\else${\rm L}_{\mathord\odot}$\fi}
\def\inv{\ifmmode^{-1}\else$^{-1}$\fi}
\def\mo{\ifmmode^{-1}\else$^{-1}$\fi}
\def\sup#1{\ifmmode ^{\rm #1}\else $^{\rm #1}$\fi}
\def\expo#1{\ifmmode \times 10^{#1}\else $\times 10^{#1}$\fi}
\def\,{\thinspace}
\def\lsim{\mathrel{\raise .4ex\hbox{\rlap{$<$}\lower 1.2ex\hbox{$\sim$}}}}
\def\gsim{\mathrel{\raise .4ex\hbox{\rlap{$>$}\lower 1.2ex\hbox{$\sim$}}}}

\def\simprop{\mathrel{\raise .4ex\hbox{\rlap{$\propto$}\lower 1.2ex\hbox{$\sim$}}}}
\def\deg{\ifmmode^\circ\else$^\circ$\fi}
\def\pdeg{\ifmmode $\setbox0=\hbox{$^{\circ}$}\rlap{\hskip.11\wd0 .}$^{\circ}
          \else \setbox0=\hbox{$^{\circ}$}\rlap{\hskip.11\wd0 .}$^{\circ}$\fi}
\def\arcs{\ifmmode {^{\scriptstyle\prime\prime}}
          \else $^{\scriptstyle\prime\prime}$\fi}
\def\arcm{\ifmmode {^{\scriptstyle\prime}}
          \else $^{\scriptstyle\prime}$\fi}
\newdimen\sa  \newdimen\sb
\def\parcs{\sa=.07em \sb=.03em
     \ifmmode \hbox{\rlap{.}}^{\scriptstyle\prime\kern -\sb\prime}\hbox{\kern -\sa}
     \else \rlap{.}$^{\scriptstyle\prime\kern -\sb\prime}$\kern -\sa\fi}
\def\parcm{\sa=.08em \sb=.03em
     \ifmmode \hbox{\rlap{.}\kern\sa}^{\scriptstyle\prime}\hbox{\kern-\sb}
     \else \rlap{.}\kern\sa$^{\scriptstyle\prime}$\kern-\sb\fi}
\def\ra[#1 #2 #3.#4]{#1\sup{h}#2\sup{m}#3\sup{s}\llap.#4}
\def\dec[#1 #2 #3.#4]{#1\deg#2\arcm#3\arcs\llap.#4}
\def\deco[#1 #2 #3]{#1\deg#2\arcm#3\arcs}
\def\rra[#1 #2]{#1\sup{h}#2\sup{m}}

\def\dots{\relax\ifmmode \ldots\else $\ldots$\fi}
\def\WHzsr{\ifmmode $W\,Hz\mo\,sr\mo$\else W\,Hz\mo\,sr\mo\fi}
\def\mHz{\ifmmode $\,mHz$\else \,mHz\fi}
\def\GHz{\ifmmode $\,GHz$\else \,GHz\fi}
\def\mKs{\ifmmode $\,mK\,s$^{1/2}\else \,mK\,s$^{1/2}$\fi}
\def\muKs{\ifmmode \,\mu$K\,s$^{1/2}\else \,$\mu$K\,s$^{1/2}$\fi}
\def\muKRJs{\ifmmode \,\mu$K$_{\rm RJ}$\,s$^{1/2}\else \,$\mu$K$_{\rm RJ}$\,s$^{1/2}$\fi}
\def\muKHz{\ifmmode \,\mu$K\,Hz$^{-1/2}\else \,$\mu$K\,Hz$^{-1/2}$\fi}
\def\MJysr{\ifmmode \,$MJy\,sr\mo$\else \,MJy\,sr\mo\fi}
\def\MJysrmK{\ifmmode \,$MJy\,sr\mo$\,mK$_{\rm CMB}\mo\else \,MJy\,sr\mo\,mK$_{\rm CMB}\mo$\fi}
\def\microns{\ifmmode \,\mu$m$\else \,$\mu$m\fi}

\def\muK{\ifmmode \,\mu$K$\else \,$\mu$\hbox{K}\fi}
\def\microK{\ifmmode \,\mu$K$\else \,$\mu$\hbox{K}\fi}
\def\muW{\ifmmode \,\mu$W$\else \,$\mu$\hbox{W}\fi}
\def\kms{\ifmmode $\,km\,s$^{-1}\else \,km\,s$^{-1}$\fi}
\def\kmsMpc{\ifmmode $\,\kms\,Mpc\mo$\else \,\kms\,Mpc\mo\fi}
\providecommand{\sorthelp}[1]{}

\def\WMAP{WMAP}

\def\commanderthree{\texttt{Commander3}}

\renewcommand{\d}[0]{\vec{d}}

\newcommand{\n}[0]{\vec{n}}

\definecolor{orange}{RGB}{255,127,0}

\newcommand{\s}[0]{\vec{s}}
\renewcommand{\a}[0]{\vec{a}}
\newcommand{\m}[0]{\vec{m}}

\newcommand{\B}[0]{\tens{B}}

\renewcommand{\L}[0]{\tens{L}}
\newcommand{\g}[0]{\vec{g}}

\newcommand{\N}[0]{\tens{N}}
\newcommand{\Z}[0]{\tens{Z}}

\newcommand{\M}[0]{\tens{M}}

\renewcommand{\r}[0]{\vec{r}}

\renewcommand{\P}[0]{\tens{P}}

\newcommand{\Dbp}[0]{\Delta_{\mathrm{bp}}}

\newcommand{\BP}{\textsc{BeyondPlanck}}

\newcommand{\sroll}[0]{\texttt{SRoll}}
\newcommand{\srollTwo}[0]{\texttt{SRoll2}}

\def\inv{^{-1}}

\begin{document}

\title{\bfseries{\scshape{BeyondPlanck}} X. \Planck\ LFI frequency maps\\with sample-based error propagation}
\newcommand{\oslo}[0]{1}
\newcommand{\helsinkiA}[0]{2}
\newcommand{\helsinkiB}[0]{3}
\newcommand{\milanoA}[0]{4}
\newcommand{\milanoC}[0]{5}
\newcommand{\milanoB}[0]{6}
\newcommand{\triesteB}[0]{7}
\newcommand{\planetek}[0]{8}
\newcommand{\princeton}[0]{9}
\newcommand{\jpl}[0]{10}
\newcommand{\nersc}[0]{11}
\newcommand{\haverford}[0]{12}
\newcommand{\mpa}[0]{13}
\newcommand{\triesteA}[0]{14}
\author{\small
A.~Basyrov\inst{\oslo}\thanks{Corresponding author: A.~Basyrov; \url{
	artem.basyrov@astro.uio.no}}
\and
A.-S.~Suur-Uski\inst{\helsinkiA, \helsinkiB}
\and
L.~P.~L.~Colombo\inst{\milanoA}
\and
J.~R.~Eskilt\inst{\oslo}
\and
S.~Paradiso\inst{\milanoA, \milanoC}
\and
K.~J.~Andersen\inst{\oslo}
\and
\textcolor{black}{R.~Aurlien}\inst{\oslo}
\and
\textcolor{black}{R.~Banerji}\inst{\oslo}
\and
M.~Bersanelli\inst{\milanoA, \milanoB}
\and
S.~Bertocco\inst{\triesteB}
\and
M.~Brilenkov\inst{\oslo}
\and
M.~Carbone\inst{\planetek}
\and
H.~K.~Eriksen\inst{\oslo}
\and
\textcolor{black}{M.~K.~Foss}\inst{\oslo}
\and
C.~Franceschet\inst{\milanoC}
\and
\textcolor{black}{U.~Fuskeland}\inst{\oslo}
\and
S.~Galeotta\inst{\triesteB}
\and
M.~Galloway\inst{\oslo}
\and
S.~Gerakakis\inst{\planetek}
\and
E.~Gjerl{\o}w\inst{\oslo}
\and
\textcolor{black}{B.~Hensley}\inst{\princeton}
\and
\textcolor{black}{D.~Herman}\inst{\oslo}
\and
M.~Iacobellis\inst{\planetek}
\and
M.~Ieronymaki\inst{\planetek}
\and
\textcolor{black}{H.~T.~Ihle}\inst{\oslo}
\and
J.~B.~Jewell\inst{\jpl}
\and
\textcolor{black}{A.~Karakci}\inst{\oslo}
\and
E.~Keih\"{a}nen\inst{\helsinkiA, \helsinkiB}
\and
R.~Keskitalo\inst{\nersc}
\and
G.~Maggio\inst{\triesteB}
\and
D.~Maino\inst{\milanoA, \milanoB, \milanoC}
\and
M.~Maris\inst{\triesteB}
\and
B.~Partridge\inst{\haverford}
\and
M.~Reinecke\inst{\mpa}
\and
T.~L.~Svalheim\inst{\oslo}
\and
D.~Tavagnacco\inst{\triesteB, \triesteA}
\and
H.~Thommesen\inst{\oslo}
\and
D.~J.~Watts\inst{\oslo}
\and
I.~K.~Wehus\inst{\oslo}
\and
A.~Zacchei\inst{\triesteB}
}
\institute{\small
Institute of Theoretical Astrophysics, University of Oslo, Blindern, Oslo, Norway\goodbreak
\and
Department of Physics, Gustaf H\"{a}llstr\"{o}min katu 2, University of Helsinki, Helsinki, Finland\goodbreak
\and
Helsinki Institute of Physics, Gustaf H\"{a}llstr\"{o}min katu 2, University of Helsinki, Helsinki, Finland\goodbreak
\and
Dipartimento di Fisica, Universit\`{a} degli Studi di Milano, Via Celoria, 16, Milano, Italy\goodbreak
\and
INAF-IASF Milano, Via E. Bassini 15, Milano, Italy\goodbreak
\and
INFN, Sezione di Milano, Via Celoria 16, Milano, Italy\goodbreak
\and
INAF - Osservatorio Astronomico di Trieste, Via G.B. Tiepolo 11, Trieste, Italy\goodbreak
\and
Planetek Hellas, Leoforos Kifisias 44, Marousi 151 25, Greece\goodbreak
\and
Department of Astrophysical Sciences, Princeton University, Princeton, NJ 08544,
U.S.A.\goodbreak
\and
Jet Propulsion Laboratory, California Institute of Technology, 4800 Oak Grove Drive, Pasadena, California, U.S.A.\goodbreak
\and
Computational Cosmology Center, Lawrence Berkeley National Laboratory, Berkeley, California, U.S.A.\goodbreak
\and
Haverford College Astronomy Department, 370 Lancaster Avenue,
Haverford, Pennsylvania, U.S.A.\goodbreak
\and
Max-Planck-Institut f\"{u}r Astrophysik, Karl-Schwarzschild-Str. 1, 85741 Garching, Germany\goodbreak
\and
Dipartimento di Fisica, Universit\`{a} degli Studi di Trieste, via A. Valerio 2, Trieste, Italy\goodbreak
}

\authorrunning{Basyrov et al.}
\titlerunning{\BP\ LFI frequency maps}

\abstract{We present \Planck\ LFI frequency sky maps derived within
  the \BP\ framework. This framework draws samples from a global
  posterior distribution that includes instrumental, astrophysical and
  cosmological parameters, and the main product is an entire ensemble
  of frequency sky map samples, each of which corresponds to one
  possible realization of the various modelled instrumental systematic
  corrections, including correlated noise, time-variable gain, far
  sidelobe and bandpass corrections. This ensemble allows for
  computationally convenient end-to-end propagation of low-level
  instrumental uncertainties into higher-level science products,
  including astrophysical component maps, angular power spectra, and
  cosmological parameters. We show that the two dominant sources of
  LFI instrumental systematic uncertainties are correlated noise and
  gain fluctuations, and the products presented here support -- for
  the first time -- full Bayesian error propagation for these effects
  at full angular resolution. We compare our posterior mean maps with
  traditional frequency maps delivered by the \Planck\ collaboration,
  and find generally good agreement. The most important quality
  improvement is due to significantly lower calibration uncertainties
  in the new processing, as we find a fractional absolute calibration
  uncertainty at 70\,GHz of $\Delta g_0/g_0 = 5\cdot10^{-5}$, which is
  nominally 40 times smaller than that reported by
  \Planck\ 2018. However, we also note that the original \Planck\ 2018
  estimate has a non-trivial statistical interpretation, and this
  further illustrates the advantage of the new framework in terms of
  producing self-consistent and well-defined error estimates of all
  involved quantities without the need of ad hoc uncertainty
  contributions. We describe how low-resolution data products,
  including dense pixel-pixel covariance matrices, may be produced
  directly from the posterior samples without the need for
  computationally expensive analytic calculations or simulations. We
  conclude that posterior-based frequency map sampling provides unique
  capabilities in terms of low-level systematics modelling and error
  propagation, and may play an important
  role for future CMB $B$-mode experiments aiming at nanokelvin
  precision.  }


\maketitle

\tableofcontents

\section{Introduction}
\label{sec:introduction}

The current state-of-the-art in all-sky CMB observations is provided
by the \Planck\ satellite \citep{planck2016-l01}, which observed the
microwave sky in nine frequency bands (30--857\,GHz) between 2009 and
2013. This experiment followed in the footsteps of COBE
\citep{1982OptEn..21..769M} and \WMAP\ \citep{2003ApJ...583....1B},
improving frequency coverage, sensitivity and angular resolution,
and ultimately resulted in constraints on the primary cosmic microwave
background (CMB) temperature fluctuations that are limited by cosmic
variance, rather than instrumental noise or systematics.

The \Planck\ consortium produced three major data releases, labelled
2013, 2015 and 2018
\citep{planck2013-p02,planck2014-a03,planck2016-l02}, respectively,
each improving the quality of the basic frequency maps in terms of
overall signal-to-noise ratio and systematics control. Several
initiatives to further improve the quality of the \Planck\ maps have
also been made after the \Planck\ 2018 legacy release, as summarized
by \srollTwo\ \citep{2019A&A...629A..38D}, \Planck\ PR4 (sometimes
also called \texttt{NPIPE}) \citep{npipe}, and \BP\ \citep{bp01}.

The \BP\ project builds
directly on experiences gained during the last years of the
\Planck\ analysis, and focuses in particular on the close relationship between the various steps involved in the data analysis, including low-level time-ordered data processing, mapmaking, high-level component separation, cosmological parameter estimation.
 As described in \citet{bp01}, the main goal of \BP\ is to
implement and deploy a single statistically coherent analysis pipeline
to the \Planck\ Low-Frequency Instrument (LFI;
\citealp{planck2016-l02}), processing raw uncalibrated time-ordered
data into final astrophysical component maps and angular CMB power
spectra within one single code, without the need for intermediate
human intervention.  This is achieved by implementing a Bayesian
end-to-end analysis pipeline, which for convenience is integrated in
terms of a software package called \commanderthree\ \citep{bp03}. The
current paper focuses on the 30, 44, and 70\,GHz LFI frequency maps,
and compares these novel maps with previous versions, both from
\Planck\ 2018 and PR4.

Map-making in a CMB analysis chain compresses time-ordered data (TOD)
into sky maps. These TOD contain contributions from many instrumental
effects, and those strongly affect the statistical properties of the
resulting maps. For LFI, instrumental effects have traditionally
been modelled in terms of three main components, namely white
(uncorrelated) thermal noise, $1/f$ (correlated) noise, and general
systematics \citep{Mennella_2010, Zacchei_2011}. The white noise is by
definition uncorrelated between pixels, and its statistical properties
are therefore straightforward to characterize and propagate to
higher-level science products. In contrast, the correlated $1/f$ noise
component manifests itself as extended stripes in maps, which
significantly correlate pixels along the scanning path of the
main beam. The same also holds true for general classes of systematic
effects, including for instance gain fluctuations or bandpass mismatch
between detectors. The goal of mapmaking is to suppress these effects
as much as possible, while ideally leaving the astrophysical signal
and white noise unchanged.

Traditionally, the actual mapmaking step in CMB analysis pipelines
has focused primarily on the correlated noise component
\citep[e.g.,][and references therein]{tegmark1997,ashdown:2007}, while
general systematics corrections have been applied in a series of
pre-processing steps. One particularly flexible method that has
emerged during this work is the so-called destriping
\citep{1998A&AS..127..555D,1999astro.ph..6360B,1999A&AS..140..383M,2002A&A...387..356M,2004A&A...428..287K,2005MNRAS.360..390K,2009A&A...506.1511K,2010A&A...510A..57K},
in which the $1/f$ noise component is modelled by a sequence of
constant offsets, often called baselines. Furthermore, in generalised
destriping, as for instance implemented in the {\tt Madam} code
\citep{2005MNRAS.360..390K,2010A&A...510A..57K}, prior knowledge of
noise properties can be utilized to better constrain the baseline
amplitudes, at which point the optimal maximum-likelihood solution may
be derived as a limiting case.

Toward the end of the \Planck\ analysis period, however, a major effort was undertaken
to include the treatment of some instrumental systematic effects
(other than $1/f$ noise) directly into the pipeline, and several codes 
that simultaneously account for mapmaking,
systematic error mitigation and component separation were
developed. In particular, these integrated approaches were pioneered
by \sroll\ \citep{planck2016-l03} and \Planck\ PR4 \citep{npipe}, both of
which adopted a template-based linear regression algorithm to solve
the joint problem.

In contrast, the \BP\ framework described in \citet{bp01} and its
companion papers adopts a more general approach, in which a full
joint posterior distribution is sampled using standard Markov Chain
Monte Carlo methods. In this framework, the correlated noise is simply
sampled as one of many components, together with corrections for gain
fluctuations, bandpass mismatch, etc., and for each iteration in the
Markov Chain a new frequency map is derived. This ensemble of
frequency maps then serves as a highly compressed representation of
the full posterior distribution that can be analyzed in a broad range
of higher-level analyses. Indeed, this approach implements for the first time 
a true Bayesian end-to-end error propagation for high-resolution
CMB experiments, from raw uncalibrated time-ordered data to final
cosmological parameter estimates.

\begin{figure*}[t]
  \center
  \includegraphics[width=\linewidth]{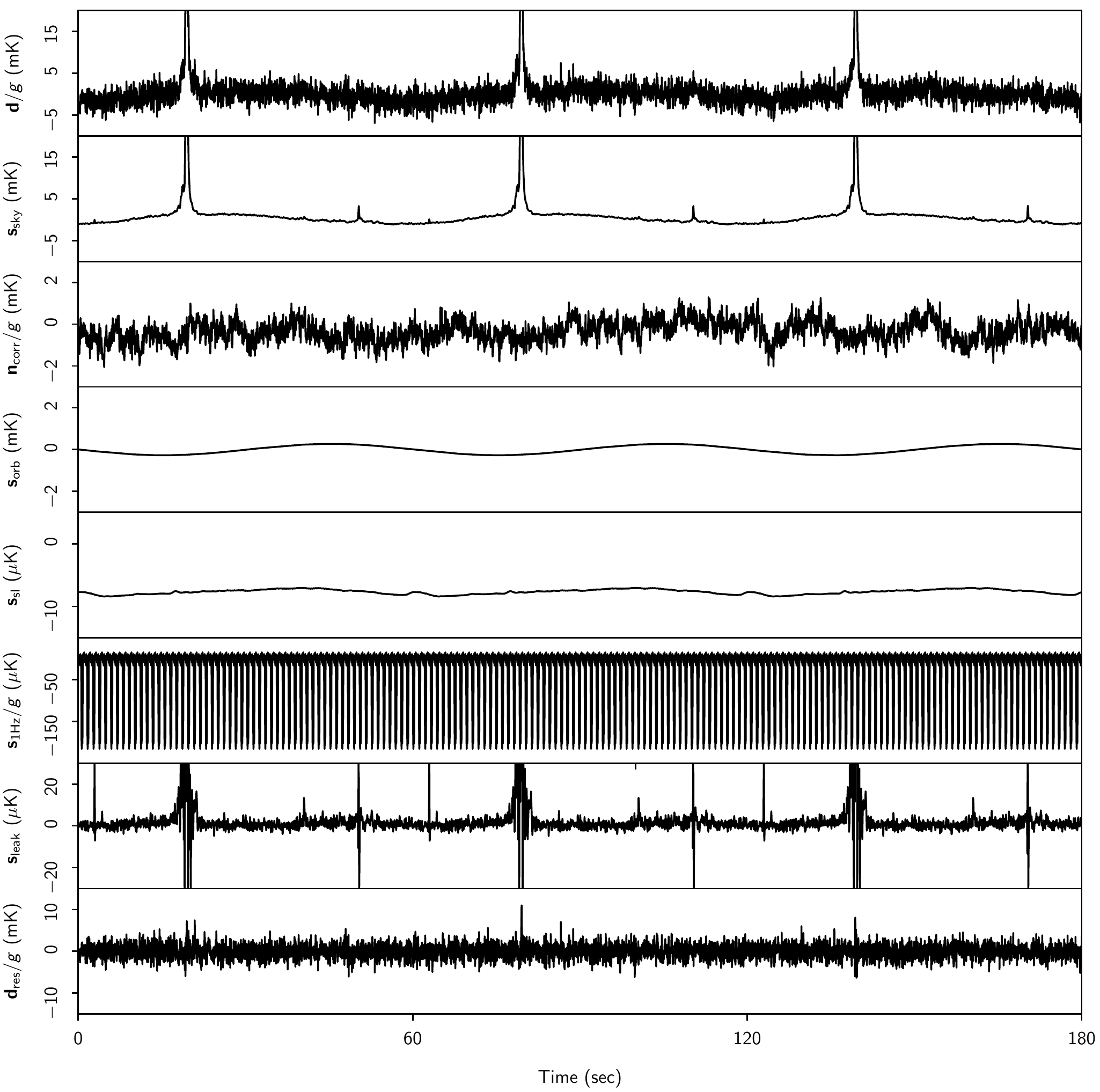}
  \caption{Time-ordered data segment for the 30\,GHz LFI 27M
    radiometer. From top to bottom, the panels show 1) raw calibrated
    TOD, $\d/g$; 2) sky signal, $\s_{\mathrm{sky}}$; 3) calibrated
    correlated noise, $\n_{\mathrm{corr}}/g$; 4) orbital CMB dipole
    signal, $\s_{\mathrm{orb}}$; 5) sidelobe correction, $\s_{\mathrm{sl}}$; 
    6) electronic 1\,Hz spike correction, $\s_{\mathrm{1Hz}}$;
    7) leakage mismatch correction, $\s_{\mathrm{leak}}$;  and 8)
    residual TOD, $\d_{\mathrm{res}} = (\d-\n_{\mathrm{corr}} - \s_{\mathrm{1Hz}})/g -
    \s_{\mathrm{sky}} - \s_{\mathrm{orb}} - \s_{\mathrm{leak}} -
    \s_{\mathrm{sl}}$.}
  \label{fig:todplot}
\end{figure*}

The rest of the paper is organised as follows. In Sect.~\ref{sec:bp}
we give an introduction to and overview of the \BP\ analysis
framework, focusing on the aspects that are most relevant for
mapmaking and systematic error propagation. In
Sect.~\ref{sec:markov_chains} we inspect the Markov chains that result
from the posterior sampling process, and quantify correlations at the
frequency map level. Next, in Sect.~\ref{sec:freqmaps} we present the
resulting \BP\ posterior mean frequency maps, and in
Sect.~\ref{sec:comparison} we compare these with previous
\Planck\ analyses. In Sect.~\ref{sec:error_propagation} we consider
end-to-end error propagation, both in the form of individual posterior
samples and conventional covariance matrices, and in
Sect.~\ref{sec:systematics} we summarize the various systematic error
corrections that are applied to each frequency map. Finally, we
conclude in Sect.~\ref{sec:summary}.


  \section{\BP\ mapmaking and end-to-end error propagation}
  \label{sec:bp}

\subsection{Statistical framework}  
  
The \BP\ framework may be succinctly summarized in terms of a single
parametric model that includes both astrophysical and instrumental
components. The motivation and derivation of the specific \BP\ data
model is described in \citet{bp01}, and takes the following form,
\begin{equation}
  \begin{split}
    d_{j,t} = g_{j,t}&\P_{tp,j}\left[\B^{\mathrm{symm}}_{pp',j}\sum_{c}
      \M_{cj}(\beta_{p'}, \Dbp^{j})a^c_{p'}  + \B^{\mathrm{asymm}}_{j,t}\left(\s^{\mathrm{orb}}_{j}  
      + \s^{\mathrm{fsl}}_{t}\right)\right] + \\
    + &s^{\mathrm{1Hz}}_{j,t} 
    + n^{\mathrm{corr}}_{j,t} + n^{\mathrm{w}}_{j,t},
  \end{split}
  \label{eq:model}
\end{equation}
where $j$ is a radiometer label, $t$ is a time sample, $p$ is a pixel
on the sky, and $c$ is an astrophysical component.  Further, $d_{j,t}$
is the measured data; $g_{j,t}$ is the instrumental gain; $\P_{tp,j}$
is the $N_{\mathrm{TOD}}\times 3N_{\mathrm{pix}}$ pointing matrix;
$\B_{j}$ denotes the beam convolution; $\M_{cj}(\beta_{p}, \Dbp)$ is
an element of a foreground mixing matrix that depends on both a set
of non-linear spectral energy density parameters, $\beta$, and
instrumental bandpass parameters, $\Delta_{\mathrm{bp}}$; $a^c_{p}$ is
the amplitude of a sky signal component; $\s^{\mathrm{orb}}_{j}$ is
the orbital CMB dipole signal; $\s^{\mathrm{fsl}}_{j}$ is a contribution
from far sidelobes; $s^{\mathrm{1Hz}}_{j,t}$ represents an electronic
1\,Hz spike correction; $n^{\mathrm{corr}}_{j,t}$ is correlated
instrumental noise; and $n^{\mathrm{w}}_{j,t}$ is uncorrelated (white)
instrumental noise, which is assumed to be Gaussian distributed with
covariance $N_{\mathrm{wn}}$.

To build intuition regarding the various terms involved in
Eq.~\eqref{eq:model}, Fig.~\ref{fig:todplot} shows an arbitrarily
selected 3-minute segment of the various TOD components for one LFI
radiometer. The top panel shows the raw radiometer measurements, $\d$,
while the second panel shows the sky model contribution described by
the first term on the right-hand side in Eq.~\eqref{eq:model} for
signal parameters corresponding to one randomly selected sample. The
slow oscillations in this function represent the CMB (Solar and
orbital) dipole, while the sharp spikes correspond to the bright
Galactic plane signal. The small fluctuations super-imposed on the
slow CMB dipole signal are dominated by CMB fluctuations, and
measuring these is the single most important goal of any CMB
experiment. The third panel shows the correlated noise,
$\n_{\mathrm{corr}}$, which is stochastic in nature, but exhibits
long-term temporal correlations. The fourth panel shows the orbital
CMB dipole, $\s_{\mathrm{orb}}$, which represents our best-known
calibration source. However, it only has a peak-to-peak amplitude of
about 0.5\,mK, and is as such non-trivial to measure to high
precision in the presence of the substantial correlated
noise and other sky signals seen above. The fifth panel shows the far
sidelobe response, which is generated by convolving the sky model with
the $4\pi$ instrument beam after nulling a small area around the main
beam. The last term in Eq.~\eqref{eq:model} is the contribution from
electronic 1\,Hz spikes, $\s_{\mathrm{1Hz}}$, and is shown in the
sixth panel. The seventh panel shows a correction term for so-called
bandpass and beam leakages, $\s_{\mathrm{leak}}$, which accounts for
differing response functions among the radiometers used to build a
single frequency map. This is not a stochastic term in its own right,
but rather given deterministically by the combination of the sky model
and the instrument bandpasses and beams. The bottom panel shows the
difference between the raw data and all the individual contributions,
which ideally should be dominated by white noise. Intuitively, the
\BP\ pipeline simply aims to estimate the terms shown in the second to
sixth panels in this figure from the top panel.

For notational simplicity, we define $\omega \equiv \{a, \beta, g,
\Dbp, n^{\mathrm{corr}}, \ldots\}$ to be the set of all free
parameters in Eq.~\eqref{eq:model}, and we fit this to $\d$ by
mapping out the posterior distribution, $P(\omega\mid\d)$.  According to
Bayes' theorem, this distribution may be written
\begin{equation}
  P(\omega\,\mid\,\d) = \frac{P(\d\,\mid\,\omega)P(\omega)}{P(\d)}
  \propto \mathcal{L}(\omega)P(\omega),
  \label{eq:joint_posterior_full}
\end{equation}
where $\mathcal{L}(\omega) \equiv P(\d\mid\omega)$ is the
likelihood, and $P(\omega)$ is some set of priors. The likelihood may
be written out explicitly by first defining
$\s^{\mathrm{tot}}(\omega)$ to be the full expression on the
right-hand side of Eq.~\eqref{eq:model}, except the white noise term,
and then noting that $\d-\s^{\mathrm{tot}} = \n^{\mathrm{wn}}$, and
therefore
\begin{equation}
-2\ln\mathcal{L}(\omega) = \left(\d-\s^{\mathrm{tot}}(\omega)\right)^t\N_{\mathrm{wn}}^{-1}\left(\d-\s^{\mathrm{tot}}(\omega)\right).
\end{equation}
The priors are less well defined, and for an overview of both
informative and algorithmic priors used in the current analysis, see
\citet{bp01}.

It is important to note that $P(\omega\mid\d)$ involves billions of free
parameters, all of which are coupled into one highly correlated and
non-Gaussian distribution. As such, mapping out this distribution is
computationally non-trivial. Fortunately, the statistical method of
Gibbs sampling \citep{geman:1984} allows users to draw samples from
complex joint distribution by iteratively sampling from each
conditional distribution. Using this method for global CMB analysis
was first proposed by \citet{jewell2004} and \citet{wandelt2004},
nearly two decades ago. And has finally been applied here for the first time to real data.
As described in \citealp{bp01}, we write the
\BP\ Gibbs chain schematically as
\begin{alignat}{11}
\g &\,\leftarrow P(\g&\,|&\,\d,&\, & &\,\xi_n, &\,\s^{\mathrm{1Hz}},
&\,\Dbp, &\,\a, &\,\beta, &\,C_{\ell}) \label{eq:gain}\\
\n_{\mathrm{corr}} &\,\leftarrow P(\n_{\mathrm{corr}}&\,|&\,\d, &\,\g, &\,&\,\xi_n, &\,\s^{\mathrm{1Hz}},
&\,\Dbp, &\,\a, &\,\beta, &\,C_{\ell}) \label{eq:ncorr}\\
\xi_n &\,\leftarrow P(\xi_n&\,|&\,\d, &\,\g, &\,\n_{\mathrm{corr}}, &\,&\,\s^{\mathrm{1Hz}},
&\,\Dbp, &\,\a, &\,\beta, &\,C_{\ell})\\
\s^{\mathrm{1Hz}} &\,\leftarrow P(\s^{\mathrm{1Hz}}&\,|&\,\d, &\,\g, &\,\n_{\mathrm{corr}}, &\,\xi_n,
&\,&\,\Dbp, &\,\a, &\,\beta, &\,C_{\ell})\\
\Dbp &\,\leftarrow P(\Dbp&\,|&\,\d, &\,\g, &\,\n_{\mathrm{corr}}, &\,\xi_n, &\,\s^{\mathrm{1Hz}},
&\,&\,\a, &\,\beta, &\,C_{\ell})\\
\beta &\,\leftarrow P(\beta&\,|&\,\d, &\,\g, &\,\n_{\mathrm{corr}}, &\,\xi_n, &\,\s^{\mathrm{1Hz}},
&\,\Dbp, & &\,&\,C_{\ell})\\
\a &\,\leftarrow P(\a&\,|&\,\d, &\,\g, &\,\n_{\mathrm{corr}}, &\,\xi_n, &\,\s^{\mathrm{1Hz}},
&\,\Dbp, &\,&\,\beta, &\,C_{\ell})\\
C_{\ell} &\,\leftarrow P(C_{\ell}&\,|&\,\d, &\,\g, &\,\n_{\mathrm{corr}}, &\,\xi_n, &\,\s^{\mathrm{1Hz}},
	&\,\Dbp, &\,\a, &\,\beta\phantom{,}&\,\phantom{C_{\ell}}),\label{eq:cl}
\end{alignat}
where symbol $\leftarrow$ has the meaning that we set variable on the
left-hand side equal to the sample from the distribution on the
right-hand side. 

\begin{figure*}[t]
  \center
  \includegraphics[width=0.49\linewidth]{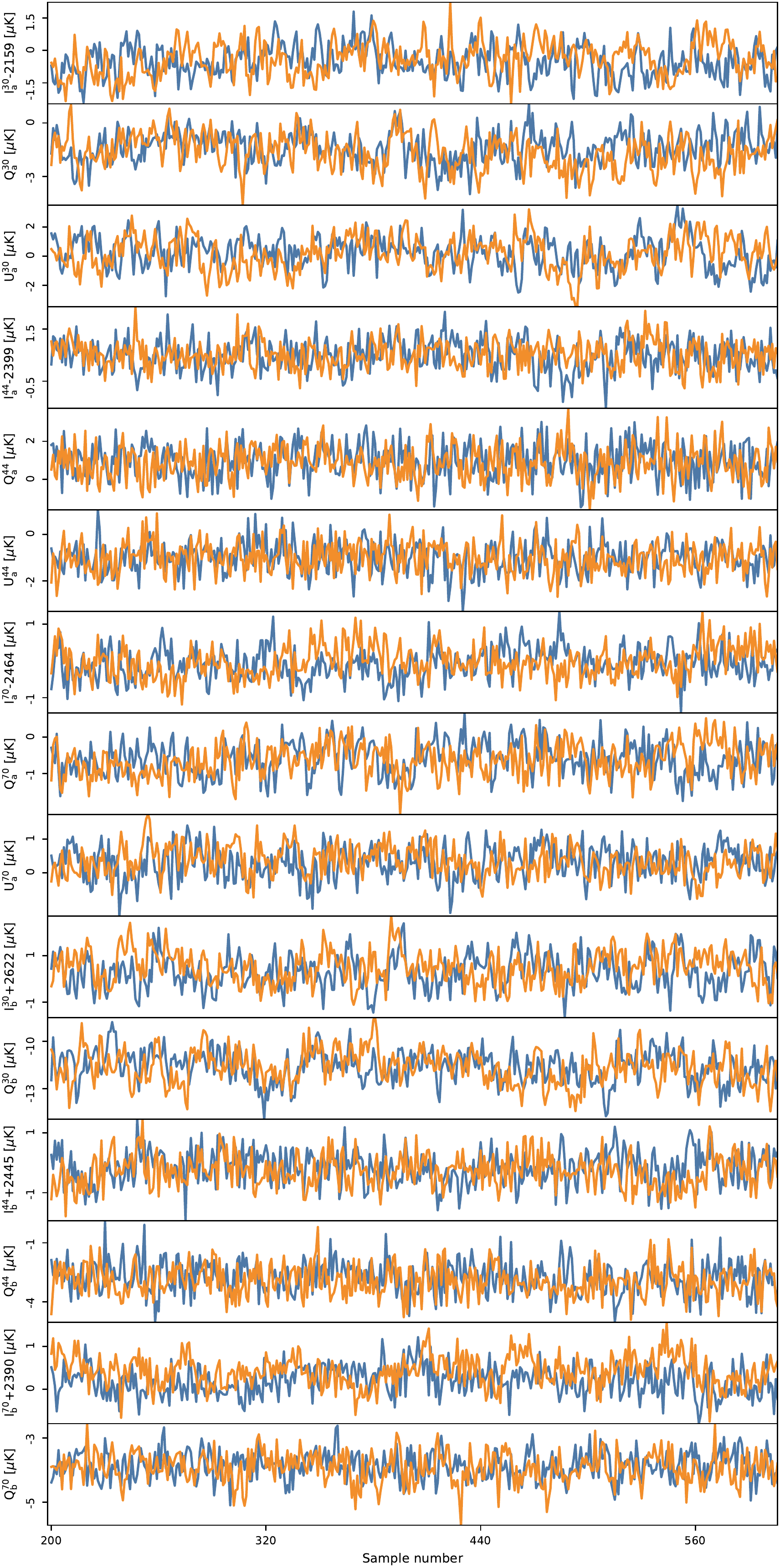}
  \includegraphics[width=0.49\linewidth]{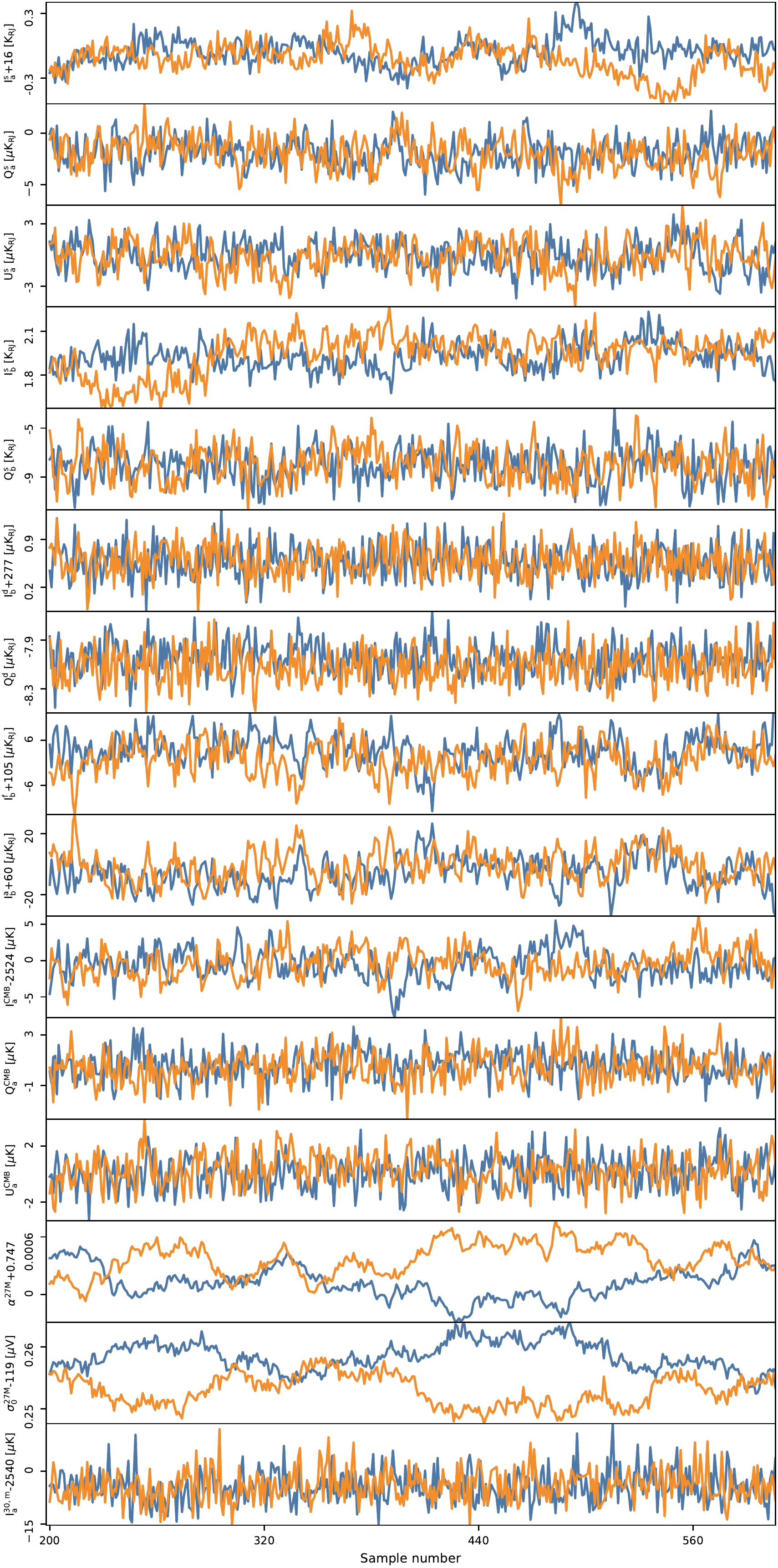}
  \caption{Trace plots of a set of selected frequency map parameters
    and related quantities; see main text for full definitions. The
    different colors indicate two independent Gibbs chains, and the
    `340' and `1960' subscripts indicate HEALPix pixel numbers at
    resolution $N_{\mathrm{side}}=16$ in ring ordering.}\label{fig:traceplot}
\end{figure*}

\begin{figure*}
  \center	
  \includegraphics[width=1\linewidth]{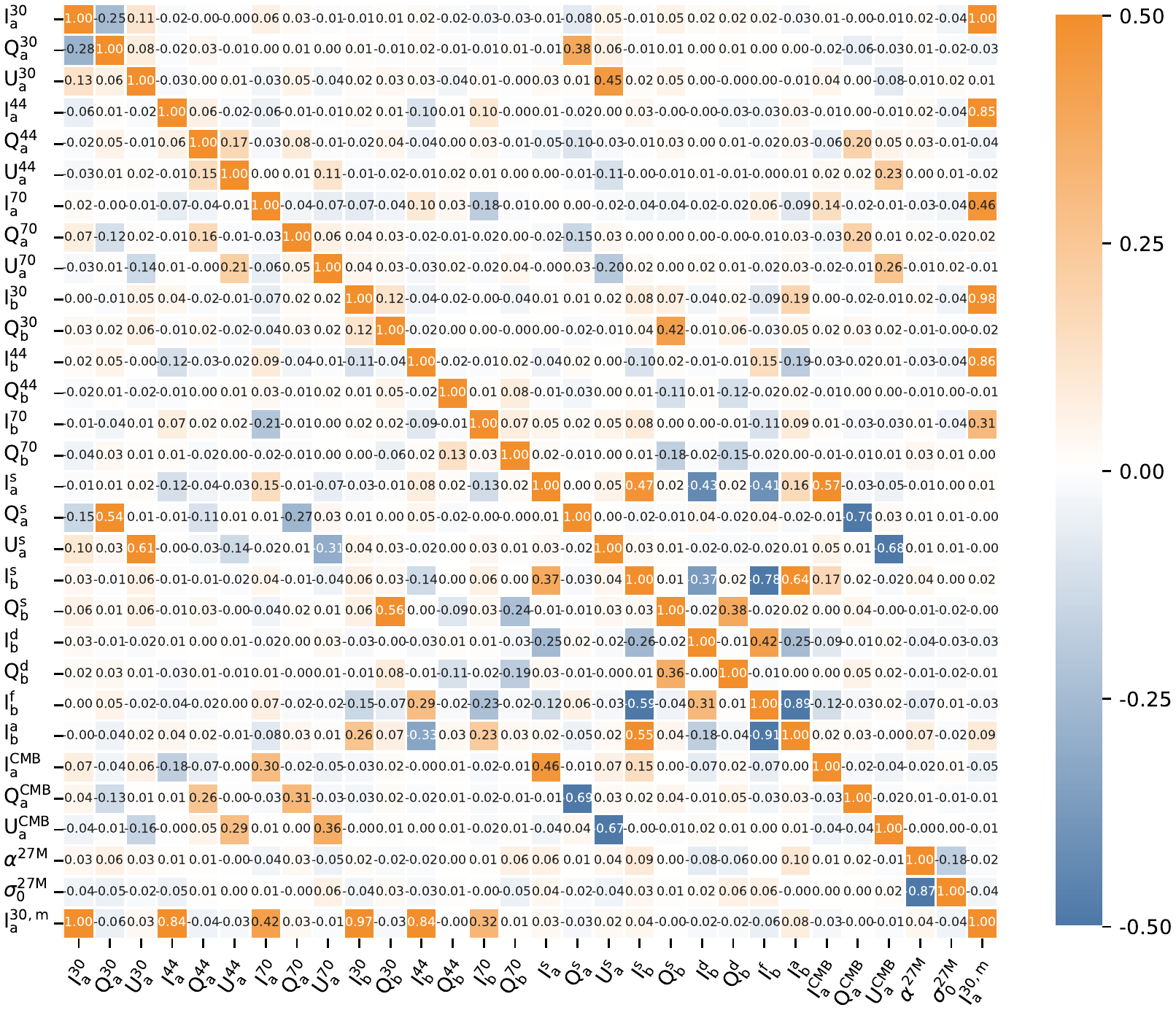}
  \caption{Correlation coefficients between the same parameters as
    shown in Fig.~\ref{fig:traceplot}. The subscripts `a' and `b' relate to the 
    `340' and `1960' HEALPix pixel numbers. The lower triangle shows
    raw correlations, while the upper triangle shows correlations
    after high-pass filtering with a running mean with a 10-sample
    window. For further explanation of and motivation for this
    filtering, see \citet{bp13}.  
    }
  \label{fig:param_corr_local}
\end{figure*}

\subsection{Mapmaking}

In the \BP\ framework, frequency maps are not independent stochastic
variables in their own right, but they serve rather as a deterministic
compression of the full dataset from TOD into sky pixels, conditioning
on other actual stochastic parameters, such as $\g$ and
$\n_{\mathrm{corr}}$. To derive these compressed frequency maps, we
first compute the residual calibrated TOD for each detector,
\begin{equation}
  r_{j,t} = \frac{d_{j,t}- n^{\mathrm{corr}}_{j,t} - s^{\mathrm{1Hz}}_{j,t}}{g_{t,j}} - \left(s^{\mathrm{orb}}_{j,t}  
  + s^{\mathrm{fsl}}_{j,t} + \delta s^{\mathrm{leak}}_{j, t}\right),
  \label{eq:todresidual}
\end{equation}
where
\begin{equation}
\delta s^{\mathrm{leak}}_{j, t} \equiv \P^{j}_{tp}\B_{pp'}^j\left(s^{\mathrm{sky}}_{jp'} - \left<s^{\mathrm{sky}}_{jp'}\right>\right)
\label{eq:leak}
\end{equation}
is the difference between the sky signal as seen by detector $j$ and
the mean over all detectors; this term suppresses so-called bandpass
and beam leakage effects in the final map \citep{bp09}. To the extent
that the data model is accurate, $r_{j,t}$ contains only stationary
sky signal and white noise, given the current estimates of other
parameters in the data model. A proper frequency map may therefore be
obtained simply by binning $r$ into a map, i.e., by solving the
following equation for each pixel,
\begin{equation}
\left(\sum_{j \in \nu} \P_j^t (\N^{\mathrm{w}}_{j})^{-1} \P_j\right) \m_{\nu} =
\sum_j \P_j^t (\N^{\mathrm{w}}_j)^{-1}\d_j.
\label{eq:mapmaking}
\end{equation}

It is important to note that the residual in
Eq.~\eqref{eq:todresidual} depends explicitly on $\omega$, and
different combinations of $g$, $n_{\mathrm{corr}}$, etc., will result
in a different sky map. As such, a frequency map in the \BP\ framework
represents just one possible realization, and in each iteration in the
Markov chain a new frequency map is derived. The final result is an
entire ensemble of frequency maps, $\m_{\nu}^i$, each with different
combinations of systematic effects. The mean of these samples may be
compared to traditional maximum-likelihood estimates, as indeed is
done in Sect.~\ref{sec:comparison}, but it is important
to emphasize that the main product from the current analysis is the
entire ensemble of sky maps, not the posterior mean. The reason is
that only by considering the full set of individual samples is it
possible to fully propagate uncertainties into high-level results.

It is also important to note that this ensemble by itself does
\emph{not} propagate white noise uncertainties, as the residual in
Eq.~\eqref{eq:todresidual} only subtracts instrumental systematic
effects. This is a purely implementational choice, and we could in
principle also have added a white noise fluctuation to each
realization. However, because the white noise is uncorrelated both
in time and between pixels, its $3\times 3$ Stokes $\{T,Q,U\}$ covariance
is very conveniently described in terms of the coupling matrix in
Eq.~\eqref{eq:mapmaking}, and reads
\begin{align}
\N^{\mathrm{w}}_{p} &= \left(\sum_{j \in \nu} \P_j^t
(\N^{\mathrm{w}}_{j})^{-1} \P_j\right)^{-1}\\
  &= \left[\begin{array}{ccc}
      \sum \frac{1}{\sigma_{0,j}^2} & \sum \frac{\cos
        2\psi_{j,t}}{\sigma_{0,j}^2} & \sum \frac{\sin
        2\psi_{j,t}}{\sigma_{0,j}^2} \\
            \sum \frac{\cos 2\psi_{j,t}}{\sigma_{0,j}^2} & \sum \frac{\cos^2
        2\psi_{j,t}}{\sigma_{0,j}^2} & \sum \frac{\cos 2\psi_{j,t} \sin
              2\psi_{j,t}}{\sigma_{0,j}^2} \\
                  \sum \frac{\sin 2\psi_{j,t}}{\sigma_{0,j}^2} & \sum
                  \frac{\sin 2\psi_{j,t} \cos
        2\psi_{j,t}}{\sigma_{0,j}^2} & \sum \frac{\sin^2
        2\psi_{j,t}}{\sigma_{0,j}^2}
    \end{array}\right]^{-1},
  \label{eq:n_wn}
\end{align}
where $\sigma_{0,j}$ and $\psi_{j,t}$ denote the time-domain white
noise rms per sample and the polarization angle for detector $j$ and
sample $t$, and the sums run over all samples. This may be evaluated
and stored compactly pixel-by-pixel, and it allows for
computational efficient and analytical white noise-only error
propagation in higher level analyses. In addition, we note that most
external and already existing CMB analysis tools already expect a mean
map and white noise covariances as inputs, and this uncertainty
description is particularly convenient for those.\footnote{We note
  that the current \commanderthree\ implementation only outputs the
  diagonal of the $3\times3$ noise covariance; the off-diagonal
  elements will be added in a future update.}


\section{Markov chains and correlations}
\label{sec:markov_chains}

The algorithm outlined in Eqs.~\eqref{eq:gain}--\eqref{eq:cl} is a
Markov chain, and must as such be initialized. To reduce the overall
burn-in time and minimize the risk of being trapped in a local
nonphysical likelihood minimum, we perform this operation in
stages. Specifically, we first initialize the instrumental parameters
on the best-fit values reported by the official \Planck\ LFI DPC
analysis \citep{planck2016-l02}, and we fix the sky model at the
\Planck\ 2015 Astrophysical Sky Model \citep{planck2014-a12}. (The
\Planck\ 2018 release did not include separate low-frequency
components -- synchrotron, free-free, and AME -- which are essential
for the current analysis.) Then each frequency channel is optimized
separately, without feedback to the sky model. Once a reasonable fit
is obtained for each channel, the model is loosened up, and a new
analysis is started that permits full feedback between all
parameters. In practice, multiple runs are required to resolve various
issues, whether they are bug fixes or model adjustments, and in each
case we restart the run at the most advanced previous chain, to avoid
repeating the burn-in process, which can take many weeks. The final
results presented in the current data release correspond to the tenth
such analysis iteration, and the parameter files and products
\citep{bp03} are labelled by ``BP10''. We strongly recommend
future analyses aiming to add additional datasets to the current model
to follow a similar analysis path, fixing the sky model on the current
\BP\ chains, and in the beginning only optimize instrumental
parameters per new frequency channel. This is likely to vastly reduce
both the debug and burn-in time, as it prevents the chains from
exploring obviously nonphysical regions of the parameter space.

The final BP10 results consist of four independent Markov chains, each
with a total of 1000 Gibbs samples. We conservatively remove 200
samples from each chain as burn-in \citep{bp12}, but note that we have
not seen statistically compelling evidence for significant chain
non-stationarity after the first few tens of samples. A total of 3200
samples is retained for full analysis.

Figure~\ref{fig:traceplot} shows a small subset of these samples for
various parameters that are relevant for the LFI frequency maps. The
two colors indicate two different Markov chains. From top to bottom,
in the left column the plotted parameters are Stokes $I$, $Q$, and $U$
parameters for each of the 30, 44, and 70\,GHz frequency maps for two
low-resolutions $N_{\mathrm{side}}=16$ pixels, namely pixels 340
(which is located in a low-foreground region of the Northern
high-latitude sky) and 1960 (which is located in a high-foreground
region south of the Galactic center). Similarly, the right column show
synchrotron, free-free, thermal dust, and CMB amplitudes, as well the
bandpass correction and the correlated noise parameters for one
pointing period (PID) for the 30\,GHz 27M radiometer. Overall, we see
that the frequency map pixel values have a relatively short
correlation length, and the chains mix well and appear statistically
stationary. In general, the correlation lengths appear much longer for
the instrumental parameters and some of the astrophysical parameters,
in particular the synchrotron intensity amplitude. This makes
intuitively sense, as we see from the definition of the TOD residual
in Eq.~\eqref{eq:todresidual} that the frequency maps depend only
weakly on the sky model through the far sidelobe and leakage
corrections, both of which are small in terms of absolute amplitudes
(see Fig.~\ref{fig:todplot}), and at second order through the gain,
which uses the sky model for calibration purposes.

\begin{table*}[t]       
\begingroup                                                                            
\newdimen\tblskip \tblskip=5pt
\caption{Frequency map comparison between \Planck\ 2018 and \BP\ products. Entries marked by $-$ indicate that the value is not available, while entries marked by $\cdots$ indicate that the \BP\ value is identical to the \Planck\ 2018 value. \label{tab:overview}}
\nointerlineskip                                                                                                                                                                                     
\vskip -2mm
\footnotesize                                                                                                                                      
\setbox\tablebox=\vbox{ %
\newdimen\digitwidth                                                                                                                          
\setbox0=\hbox{\rm 0}
\digitwidth=\wd0
\catcode`*=\active
\def*{\kern\digitwidth}
\newdimen\signwidth
\setbox0=\hbox{+}
\signwidth=\wd0
\catcode`!=\active
\def!{\kern\signwidth}
\newdimen\decimalwidth
\setbox0=\hbox{.}
\decimalwidth=\wd0
\catcode`@=\active
\def@{\kern\signwidth}
\halign{ \hbox to 2.5in{#\leaderfil}\tabskip=1.0em&
  \hfil$#$\hfil\tabskip=3em&
  \hfil$#$\hfil\tabskip=1em&
  \hfil$#$\hfil\tabskip=3em&
  \hfil$#$\hfil\tabskip=1em&
  \hfil$#$\hfil\tabskip=3em&
  \hfil$#$\hfil\tabskip=1em&  
  \hfil$#$\hfil\tabskip=0em\cr
\noalign{\doubleline}
\omit&\omit&\multispan2\hfil {\sc 30\,GHz}\hfil&\multispan2\hfil {\sc 44\,GHz}\hfil&\multispan2\hfil {\sc 70\,GHz}\hfil\cr
\noalign{\vskip -3pt}
\omit&\omit&\multispan2\hrulefill&\multispan2\hrulefill&\multispan2\hrulefill\cr
\noalign{\vskip 3pt}
\omit\hfil {\sc Quantity}\hfil&\omit\hfil {\sc Unit}\hfil& \hfil \Planck\ \hfil&\hfil {\sc BP}\hfil&\hfil \Planck\ \hfil&\hfil {\sc BP}\hfil&\hfil \Planck\ \hfil&\hfil {\sc BP}\hfil\cr
\noalign{\vskip 3pt\hrule\vskip 5pt}
\hglue 1em HEALPix pixel resolution, $N_{\mathrm{side}}$&  \textrm{None} &   1024& 512& 1024& 512& 1024& \cdots\cr
\hglue 1em Nominal center frequency&  \textrm{GHz} &   28.4& 28.6& 44.1& \cdots& 70.4& \cdots\cr
\hglue 1em TOD-level array sensitivity& \mu\mathrm{K}_{\mathrm{CMB}}\mathrm{s}^{1/2}& 147.9& 145.2& 174.0& 172.7& 151.9& 153.0\cr
\hglue 1em $1/f$ knee frequency, $f_{\mathrm{knee}}$ & \mathrm{mHz}& 113.9& 129& 53.0& 46.0& 19.6& 20.4\cr
\hglue 1em $1/f$ slope, $\alpha$ & \mathrm{None}& -0.92& -0.84& -0.88& -0.95& -1.20& -1.13\cr
\hglue 1em Intermediate frequency noise excess, $A_p/\sigma_0^2$ &
\mathrm{Fraction}& - & 0.53& - & 0.28& - & - \cr
\hglue 1em Total data volume projected into sky maps & \% & 90.99 & 99.64 &  91.30  & 99.76 & 90.97 & 99.88 \cr
\hglue 1em Absolute calibration uncertainty, $\Delta g_0/g_0$ & (10^{-5})& 170& 7& 120& 5& 200& 5\cr
\noalign{\vskip 5pt\hrule\vskip 3pt}
}}
\endPlancktablewide                                                                                                                                            
\endgroup
\end{table*}

This observation is important when comparing frequency maps generated
through Bayesian end-to-end processing with those generated through
traditional frequentist methods; although the Bayesian maps are
formally statistically coupled through the common sky model, these
couplings are indeed weak, and their correlations are fully quantified
through the sample ensemble. It is also important to note that
precisely the same type of couplings exist in the traditional maps, as
they also depend on a sky model to estimate the gain and perform
sidelobe and bandpass corrections in precisely the same manner as the
current algorithm. The only fundamental difference is that the
traditional methods usually only adopt \emph{one fixed sky model} for
these operations \citep[e.g.,][]{planck2016-l02,planck2016-l03}, and
therefore do not propagate its uncertainties, while the Bayesian
approach considers an entire ensemble of different sky models, and
thereby fully propagates astrophysical uncertainties.

In Fig.~\ref{fig:param_corr_local} we plot the Pearson's correlation
coefficient between any two pairs of parameters shown in
Fig.~\ref{fig:traceplot}. The upper triangle shows full correlations
computed directly from the raw Markov chains, while the bottom
triangle shows the same after high-pass filtering each chain with a
boxcar filter with a 5-sample window; the latter is useful to
highlight correlations within the local white noise distribution
(i.e., fast variations in Fig.~\ref{fig:traceplot}), while the former
accounts also for strong long-distance degeneracies (i.e., slow drifts
in Fig.~\ref{fig:traceplot}). Monopoles have been removed from the
intensity frequency maps, except in the last row of Fig.~\ref{fig:param_corr_local}, 
which shows correlations that also including monopole variations.

Going through these in order of strength, we start with the correlated
noise parameters. As described by \citet{bp06}, the current processing
assumes a standard $1/f$ noise power spectral density,
$P_{\n}(f)=\sigma_0^2 (1+(f/f_\mathrm{kneee})^\alpha)$, for the
70\,GHz channel, which is extended with a log-normal term for the 30
and 44\,GHz channels to account for a previously undetected power excess between
0.1--1\,Hz. This additional term introduces a strong degeneracy with
$\alpha$ and $f_{\mathrm{knee}}$ at the 0.8--0.9 level after high-pass
filtering. However, the actual frequency map only depends on the sum
of these components, and internal noise PSD degeneracies are
completely irrelevant for the final sky maps. This is seen by the fact
that the correlation coefficient between any noise parameter and the
sky map pixels is at most 0.06.

The second strongest correlations are seen in the bottom row, which
shows the 30\,GHz intensity in pixel $a$ correlated with all other
quantities. Here we see that pixels $a$ and $b$ are 97\,\% correlated
internally in the 30\,GHz channel map, and pixel $a$ is also 84\,\%
(42\,\%) correlated with the same pixel in the 44\,GHz (70\,GHz)
channel. These strong correlations are due to the fact that the
\BP\ maps have monopoles determined directly from the astrophysical
foreground model, and all frequency maps therefore respond coherently
to changes in, say, the free-free offset. After removing this strong
common mode the correlations among the frequency maps fall to well
below 10\,\%, as seen in the top section of the plot.

Thirdly, we see that there are strong correlations among the sky model
parameters (CMB, synchrotron, free-free, and thermal dust emission),
typically at the 0.3--0.6 level, and some of these also extend to the
actual frequency maps. Starting with the internal sky model
degeneracies, strong correlations between synchrotron, free-free, and
AME are fully expected due to their similar SEDs and the limited
dataset used in the current analysis. Qualitatively similar results
were reported by \citet{planck2014-a12} and \citet{planck2014-a31}
even when including the intermediate HFI frequencies, and excluding
the HFI data obviously does nothing to reduce these correlations. The
large correlations with respect to the frequency maps are perhaps more
surprising, given the statements above that the frequency maps only
depend very weakly on the sky model parameters. However, it is
important to note that the converse does not hold true; the
sky model depends \emph{strongly} on the frequency maps, and these
dependencies are encoded in Fig.~\ref{fig:param_corr_local}. Two
concrete example are CMB polarization, which is most tightly
correlated with the 70\,GHz channel, and the synchrotron polarization
amplitude, which is most tightly correlated with the 30\,GHz channel.

The fourth most notable correlation is seen among the three Stokes
parameters for a single pixel, and in particular between $Q$ and $U$
for which the intrinsic sky signal level is small, and these effects
are therefore relatively more important. Even though white noise does
not contribute to these Markov chains, the correlated noise and
systematic effects also behave qualitatively similar as white noise
with respect to the \Planck\ scanning strategy.


\section{\BP\ frequency maps}
\label{sec:freqmaps}

\subsection{General characteristics}
\label{sec:general_char}

We now turn our attention to the co-added frequency maps, and compare
the \BP\ maps with the corresponding \Planck\ 2018 maps in terms of
various key quantities in Table~\ref{tab:overview}. Unless noted
otherwise, \Planck\ values are reproduced directly from
\citet{planck2016-l02}. Entries marked with three dots indicate that
the \BP\ values are identical to the \Planck\ values, while dashed 
entries for \Planck\ indicate that those values are undefined.

\begin{figure*}[p]
  \center
  \includegraphics[width=12cm]{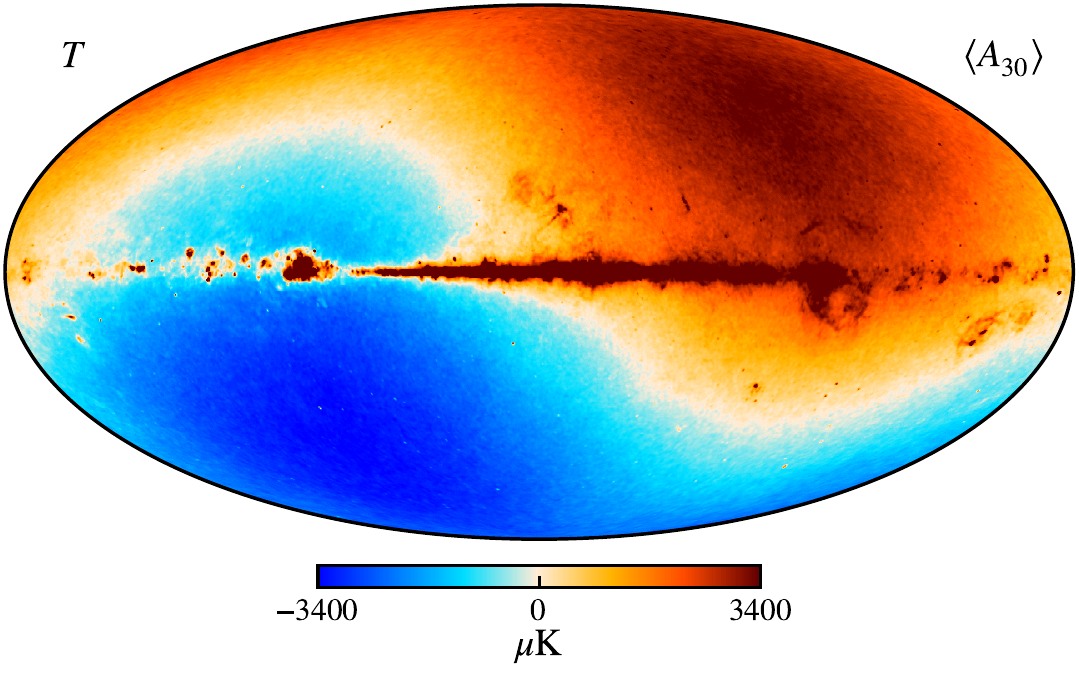}\\
  \includegraphics[width=12cm]{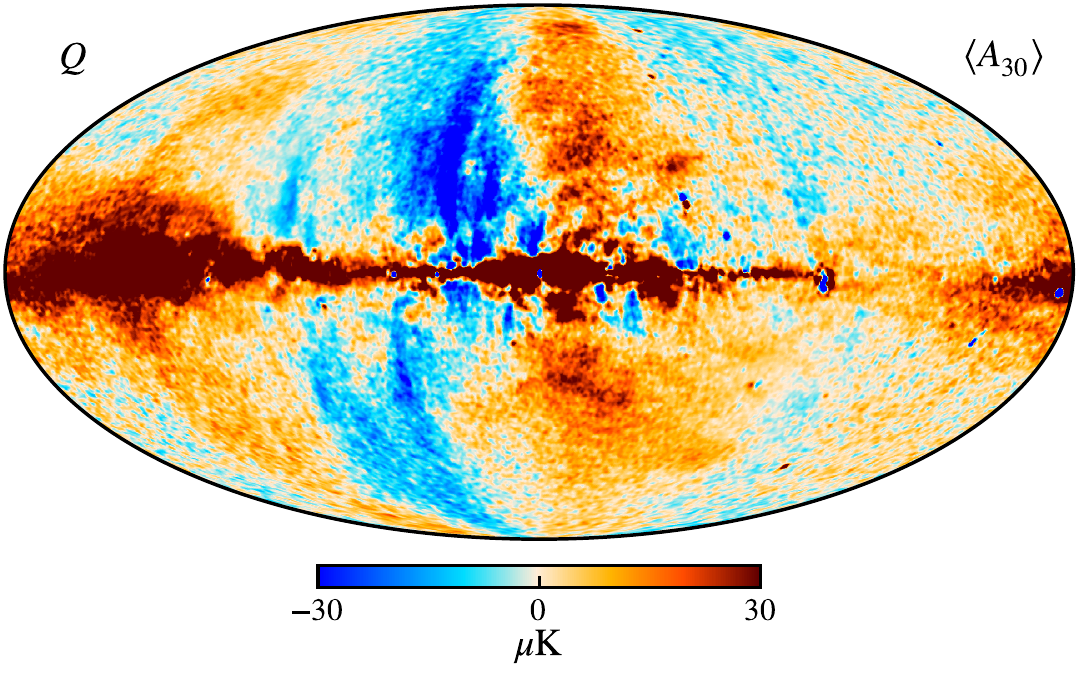}\\
  \includegraphics[width=12cm]{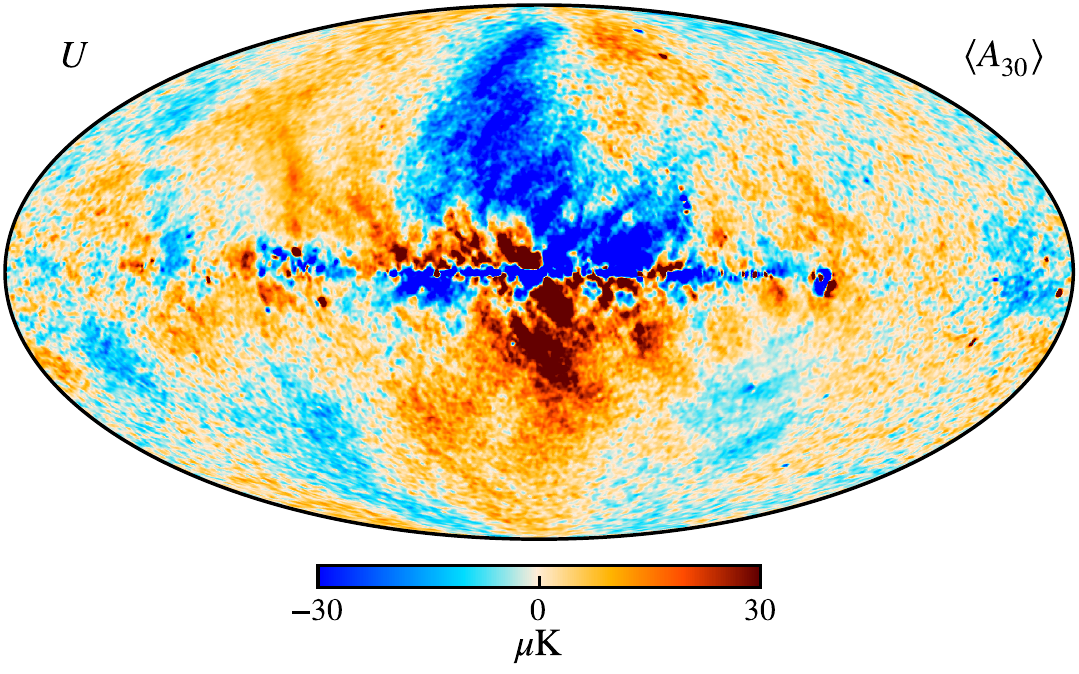}
  \caption{Posterior mean maps for the 30\,GHz frequency channel. From top to bottom: Stokes $I$, $Q$ and $U$ parameters. Maps have resolution $N_{\mathrm{side}} = 512$, are presented in Galactic coordinates, and polarisation maps have been smoothed to an effective angular resolution of $1^\circ$ FWHM.}\label{fig:freq_maps30}
\end{figure*}

\begin{figure*}[p]
  \center
  \includegraphics[width=12cm]{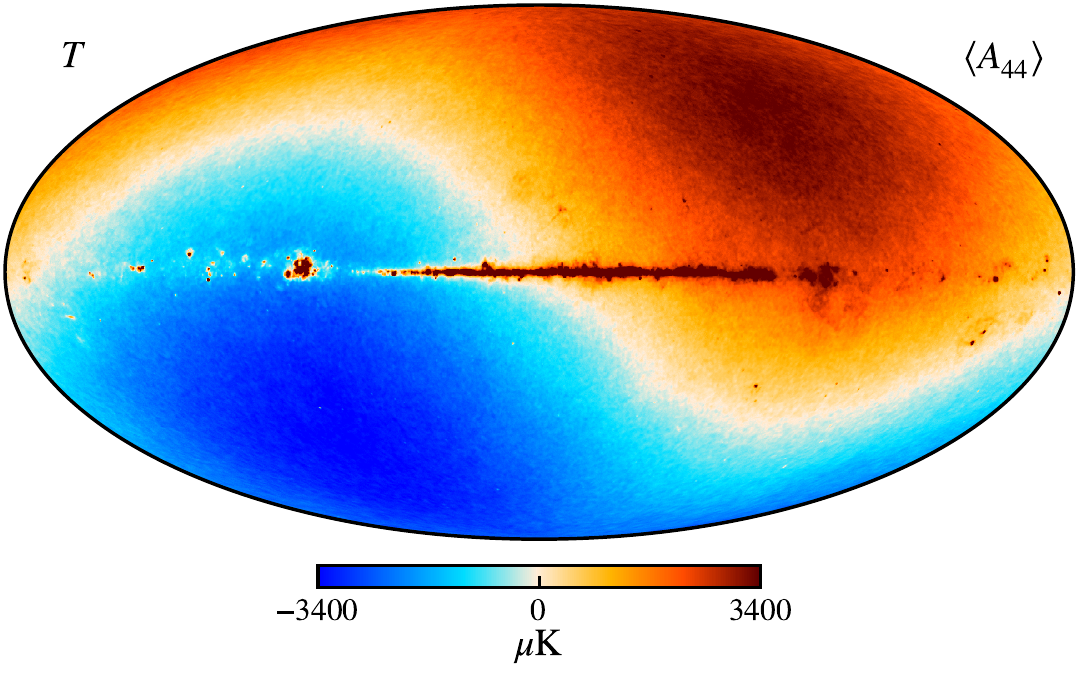}\\
  \includegraphics[width=12cm]{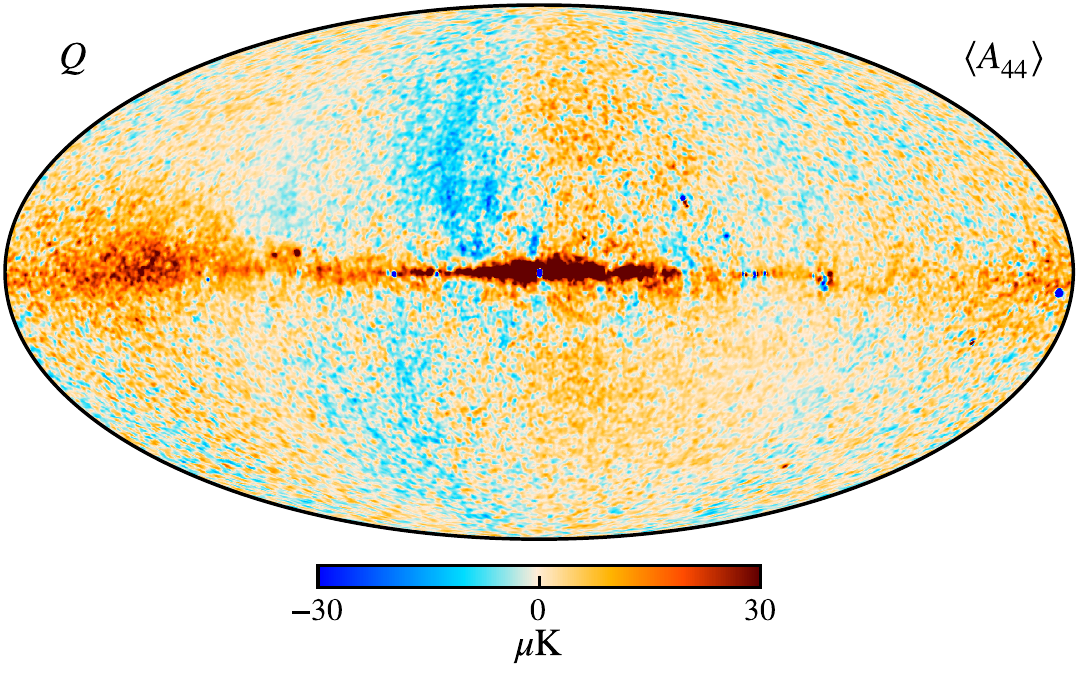}\\
  \includegraphics[width=12cm]{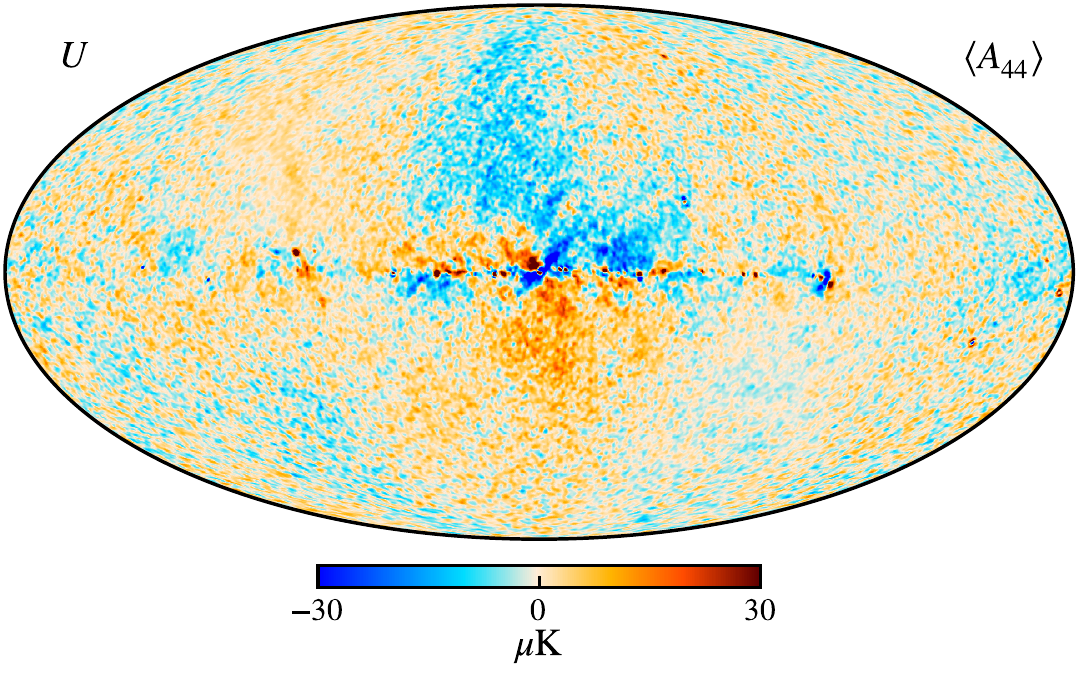}
  \caption{Posterior mean maps for the 44\,GHz frequency channel. From top to bottom: Stokes $I$, $Q$ and $U$ parameters. Maps have resolution $N_{\mathrm{side}} = 512$, are presented in Galactic coordinates, and the polarisation maps have been smoothed to an effective angular resolution of $1^\circ$ FWHM.}\label{fig:freq_maps44}
\end{figure*}

\begin{figure*}[p]
  \center
  \includegraphics[width=12cm]{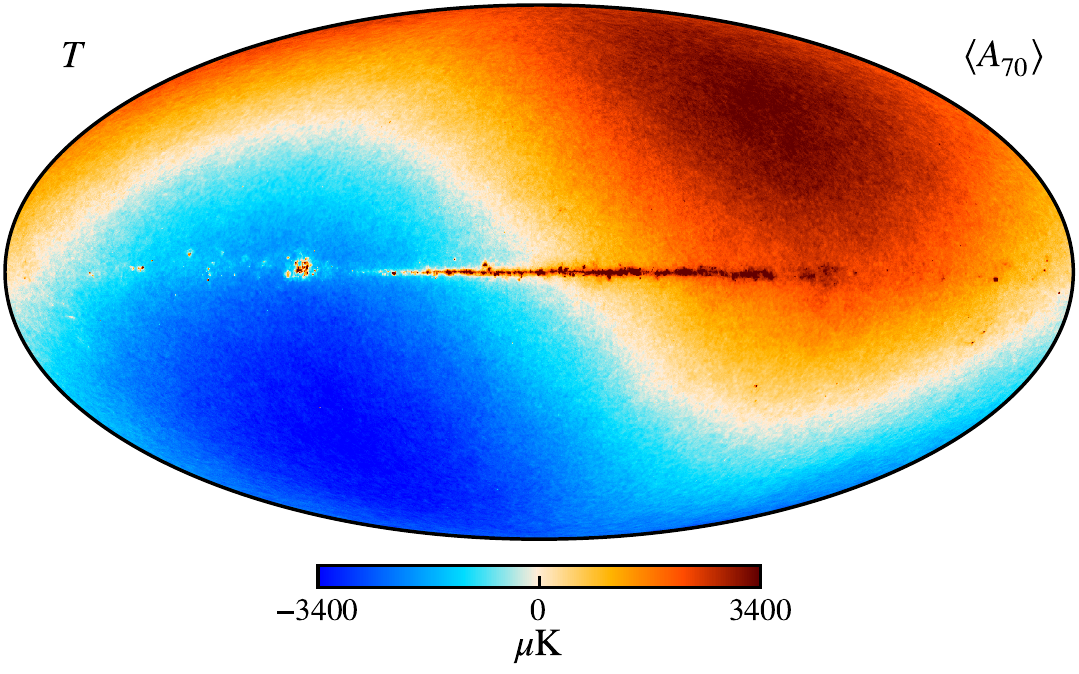}\\
  \includegraphics[width=12cm]{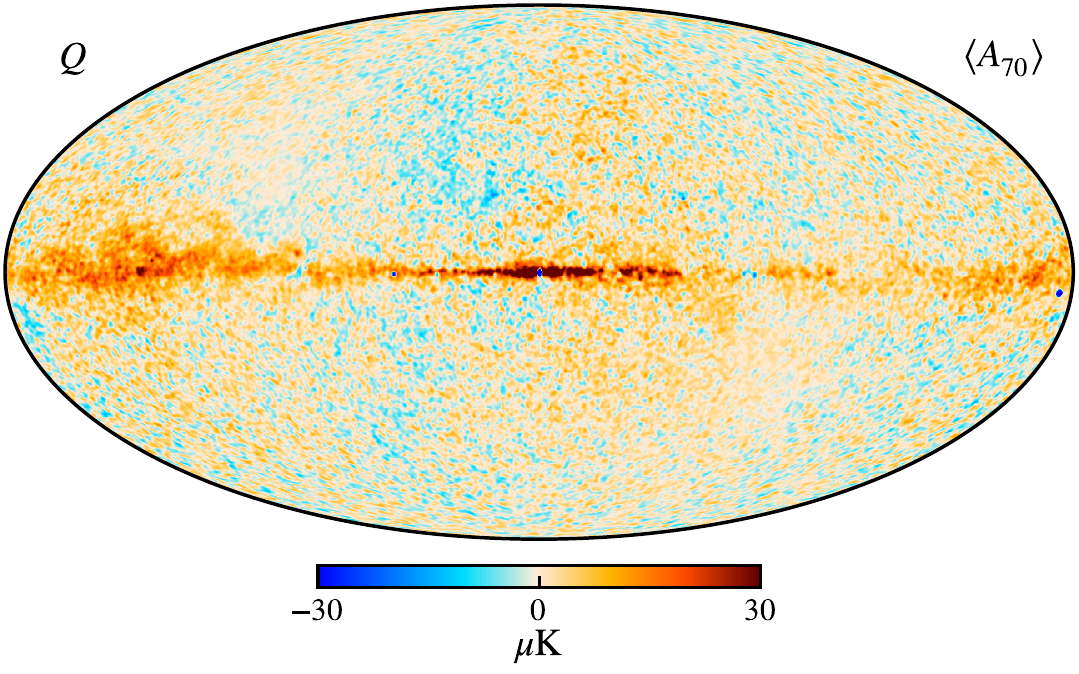}\\
  \includegraphics[width=12cm]{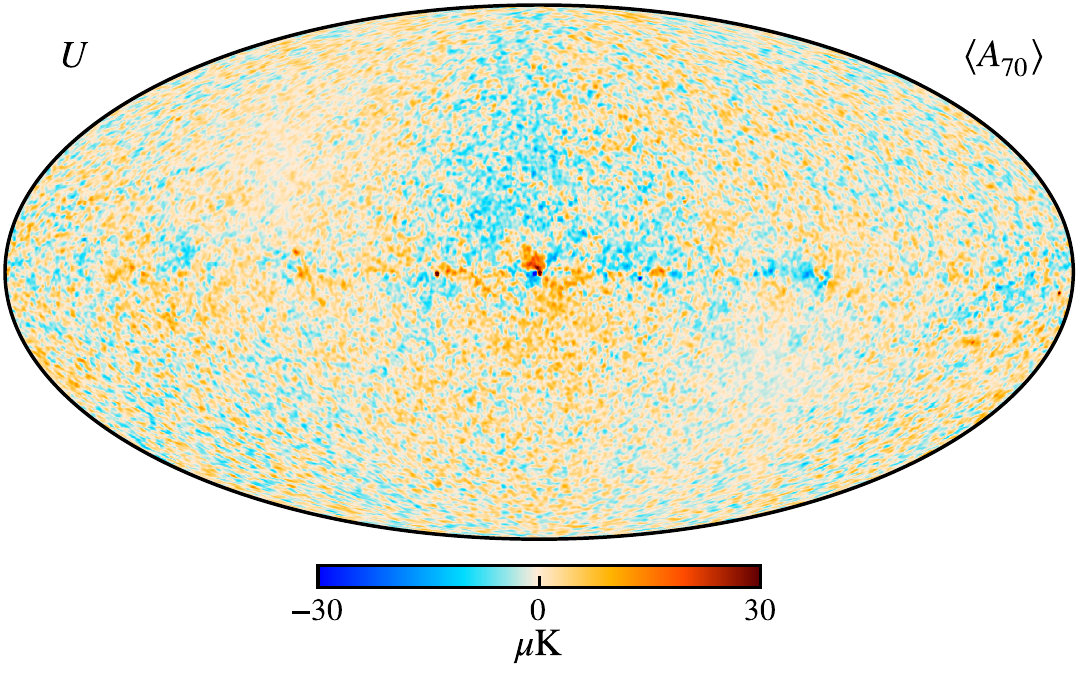}
  \caption{Posterior mean maps for the 70\,GHz frequency channel. From top to bottom: Stokes $I$, $Q$ and $U$ parameters. Maps have a HEALPix resolution of $N_{\mathrm{side}} = 1024$, are presented in Galactic coordinates, and the polarisation maps have been smoothed with a FWHM = $1^\circ$ Gaussian beam.}\label{fig:freq_maps70}
\end{figure*}

\begin{figure*}[p]
  \center
  \center
  \includegraphics[width=0.33\linewidth]{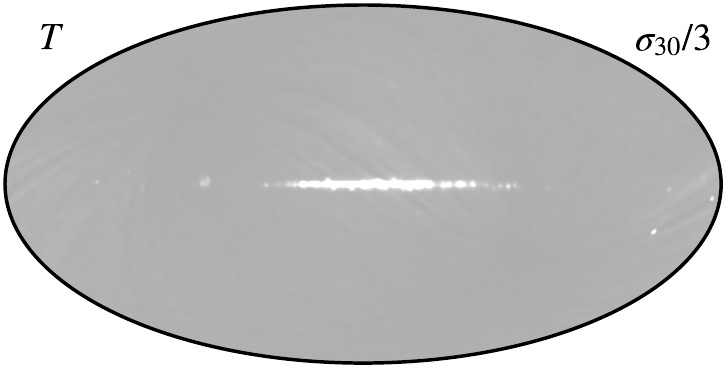}
  \includegraphics[width=0.33\linewidth]{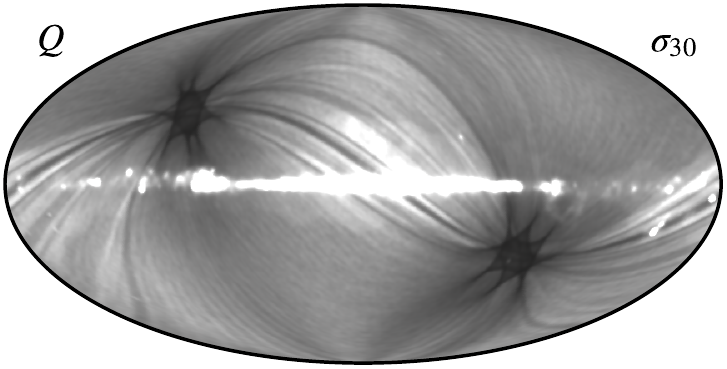}
  \includegraphics[width=0.33\linewidth]{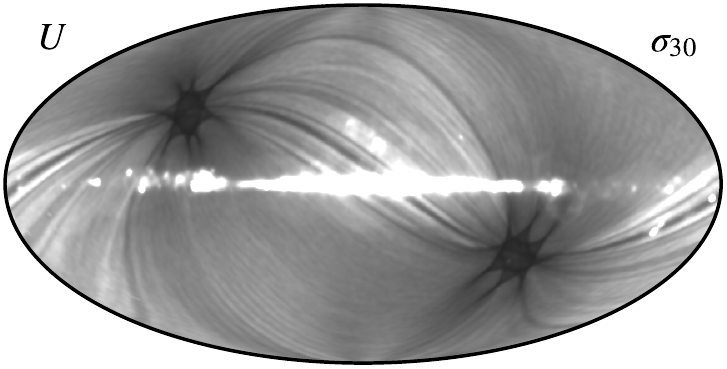}\\
  \includegraphics[width=0.33\linewidth]{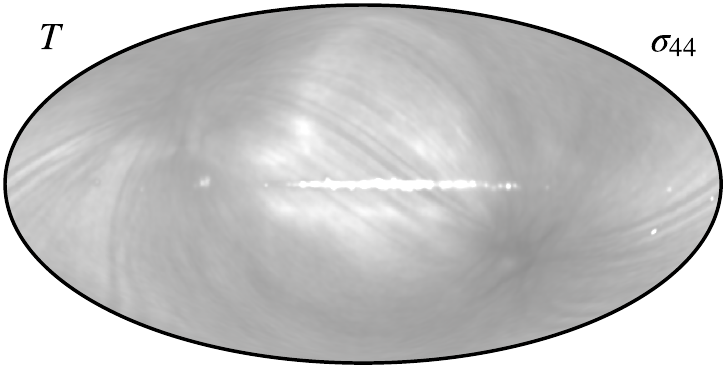}
  \includegraphics[width=0.33\linewidth]{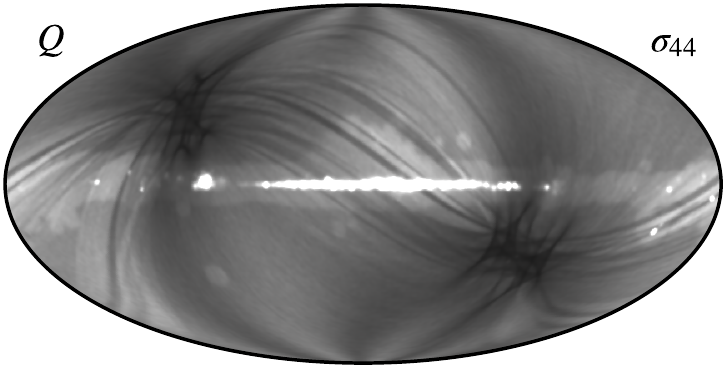}
  \includegraphics[width=0.33\linewidth]{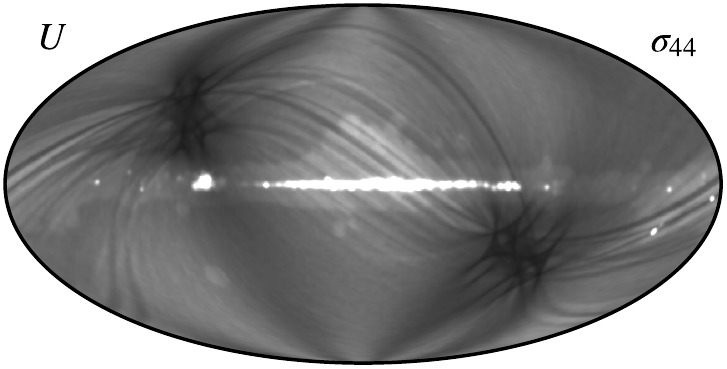}\\
  \includegraphics[width=0.33\linewidth]{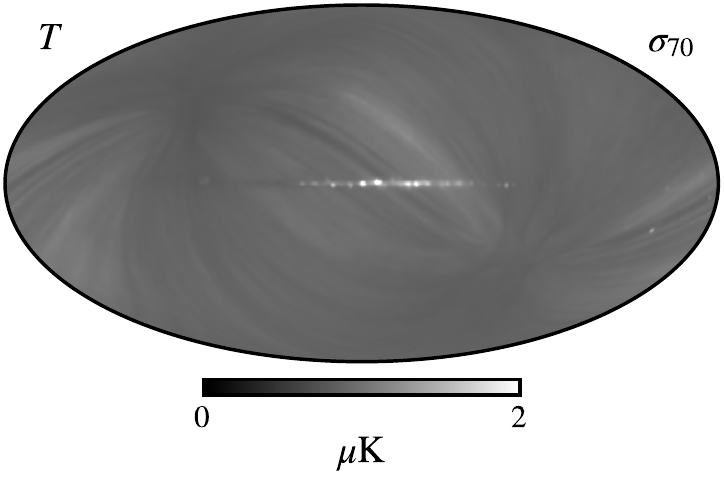}
  \includegraphics[width=0.33\linewidth]{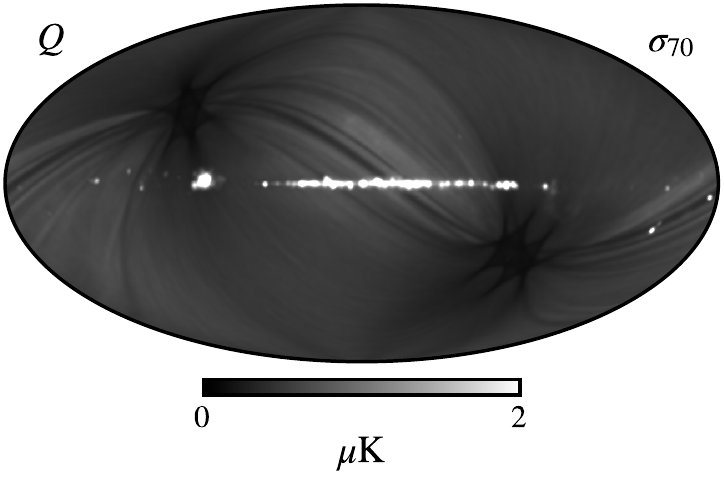}
  \includegraphics[width=0.33\linewidth]{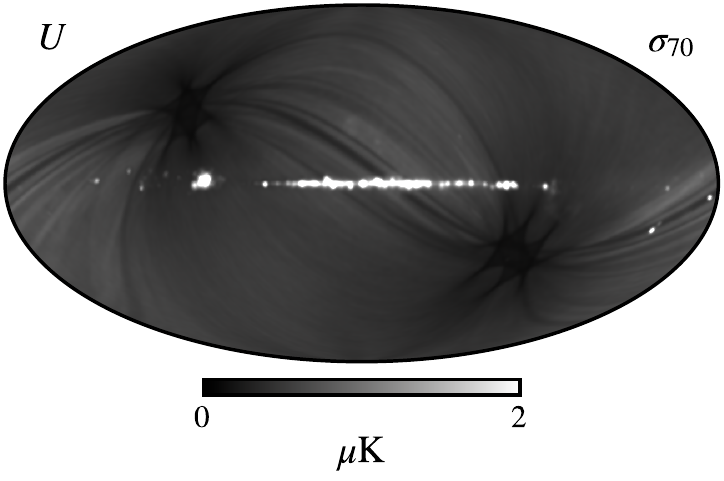}
  \caption{Posterior standard deviation maps for each LFI frequency. Rows show, from
    top to bottom, the 30, 44 and 70\,GHz frequency channels, while
    columns show, from left to right, the temperature and Stokes $Q$
    and $U$ parameters. The 30 \,GHz standard deviation is divided by a factor of 3. 
    Note that these maps do not include
    uncertainty from instrumental white noise, but only variations
    from the TOD-oriented parameters included in the data model in
    Eq.~\eqref{eq:model}. }
  \label{fig:freq_stddev}
  \vspace*{1cm}

  \includegraphics[width=0.33\linewidth]{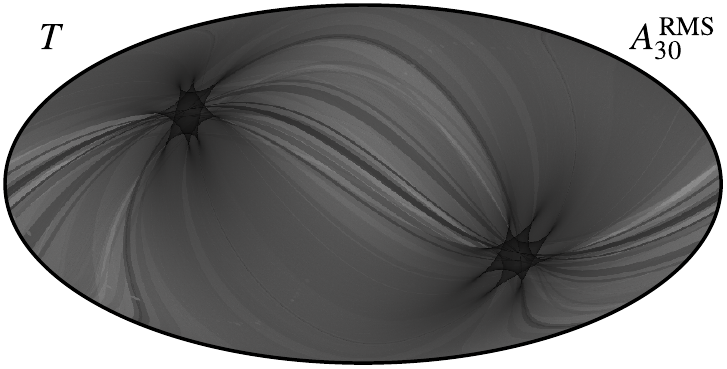}
  \includegraphics[width=0.33\linewidth]{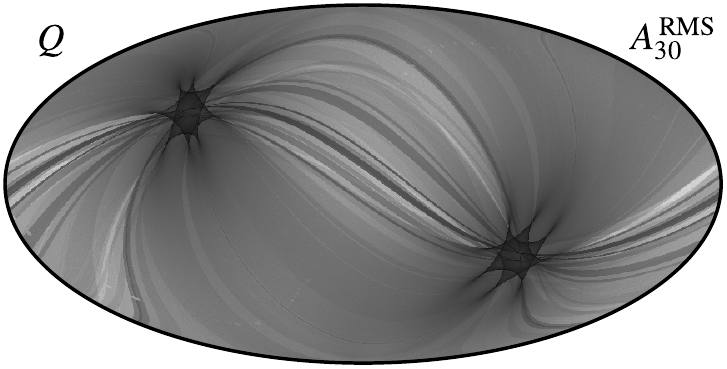}
  \includegraphics[width=0.33\linewidth]{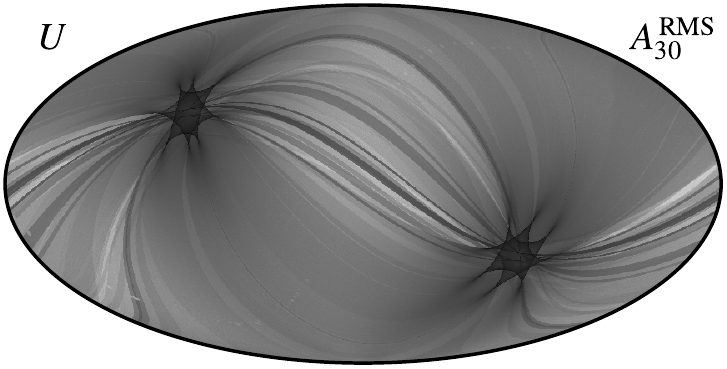}\\
  \includegraphics[width=0.33\linewidth]{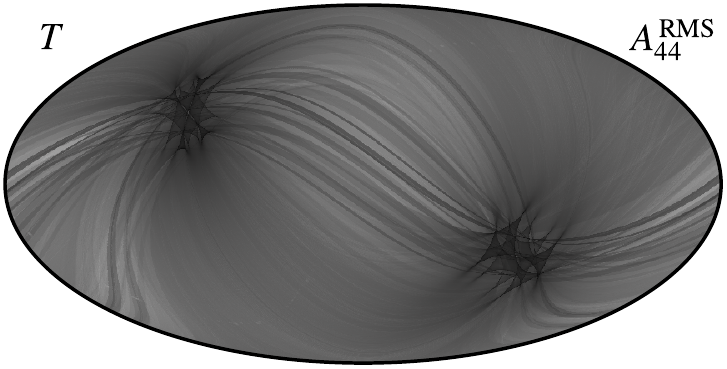}
  \includegraphics[width=0.33\linewidth]{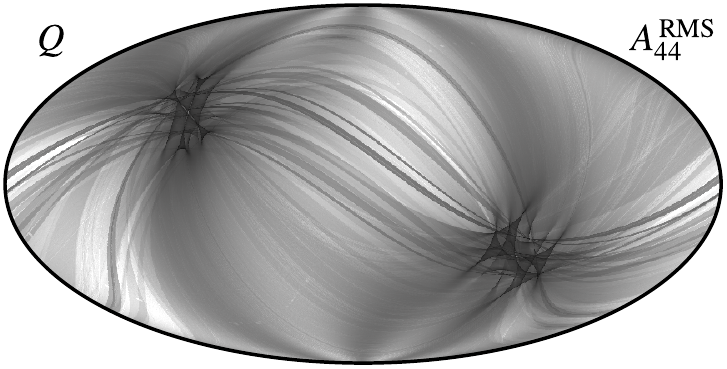}
  \includegraphics[width=0.33\linewidth]{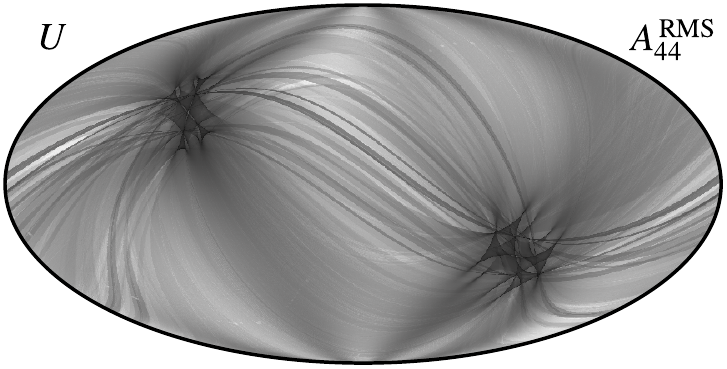}\\
  \includegraphics[width=0.33\linewidth]{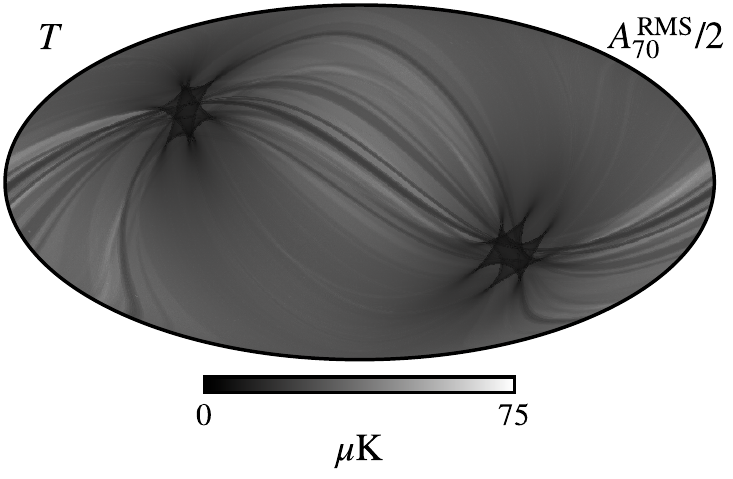}
  \includegraphics[width=0.33\linewidth]{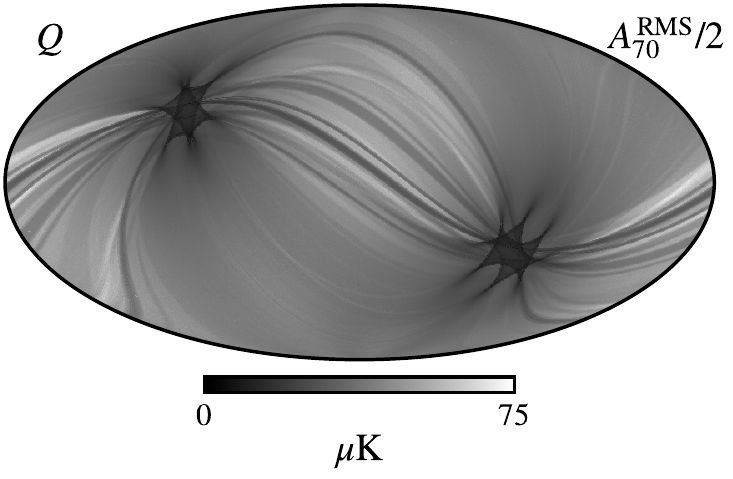}
  \includegraphics[width=0.33\linewidth]{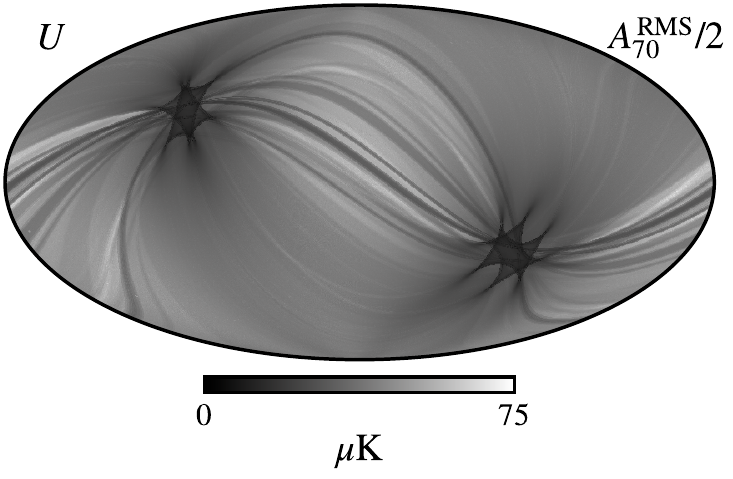}
  \caption{White noise standard deviation maps for a single
    arbitrarily selected sample. Rows show, from top to bottom, the
    30, 44 and 70\,GHz frequency channels, while columns show, from
    left to right, the temperature and Stokes $Q$ and $U$
    parameters. Note that the 70\,GHz maps are scaled by a factor of 2,
    to account for the fact that this map is pixelized at
    $N_{\mathrm{side}}=1024$, while the two lower frequencies are
    pixelized at $N_{\mathrm{side}}=512$.}
  \label{fig:freq_rmswn}    
\end{figure*}


\begin{figure*}[t]
  \center
  \includegraphics[width=0.33\linewidth]{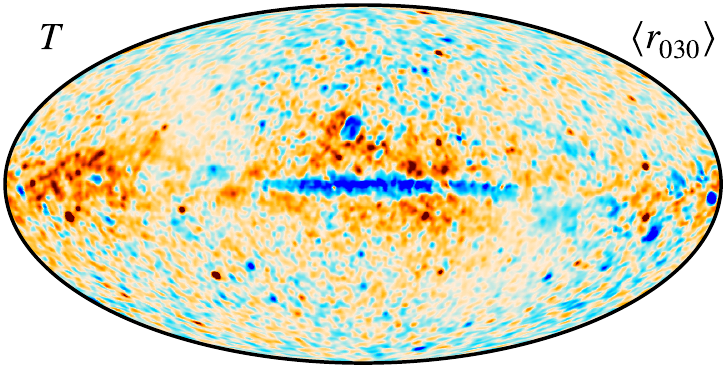}
  \includegraphics[width=0.33\linewidth]{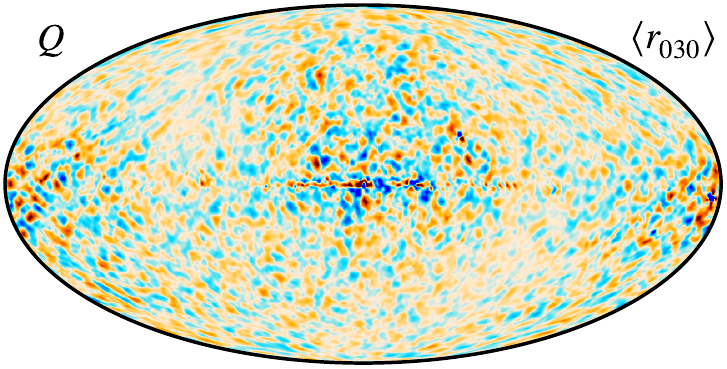}
  \includegraphics[width=0.33\linewidth]{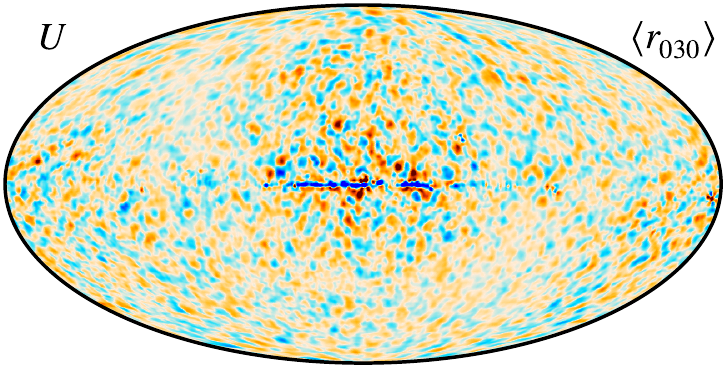}\\
  \includegraphics[width=0.33\linewidth]{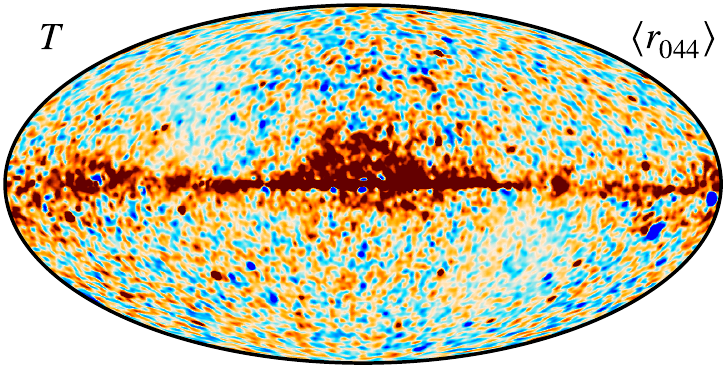}
  \includegraphics[width=0.33\linewidth]{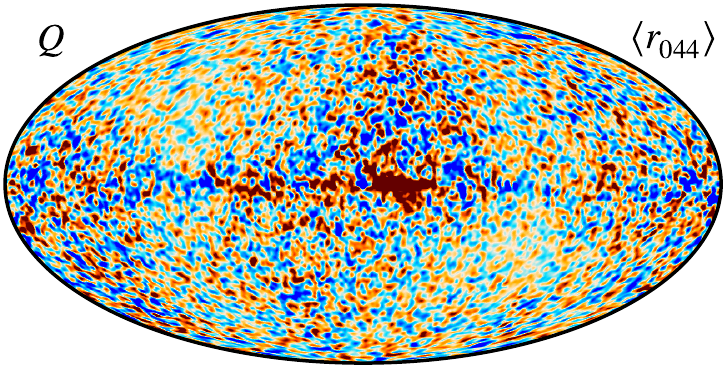}
  \includegraphics[width=0.33\linewidth]{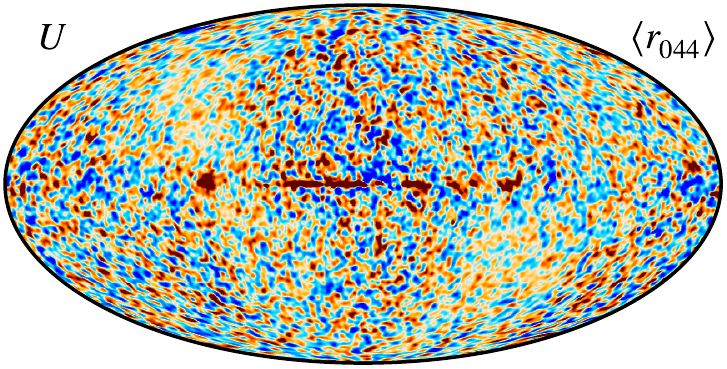}\\
  \includegraphics[width=0.33\linewidth]{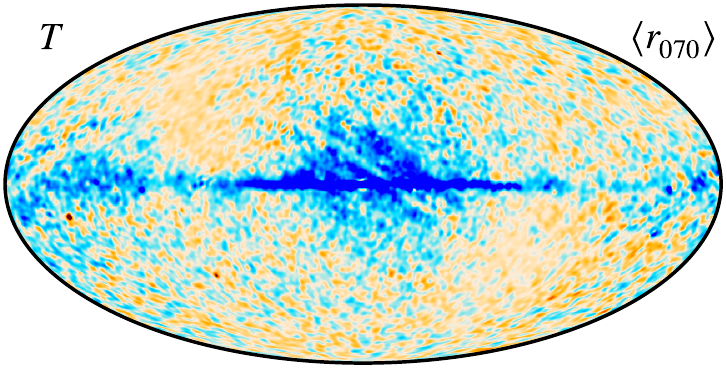}
  \includegraphics[width=0.33\linewidth]{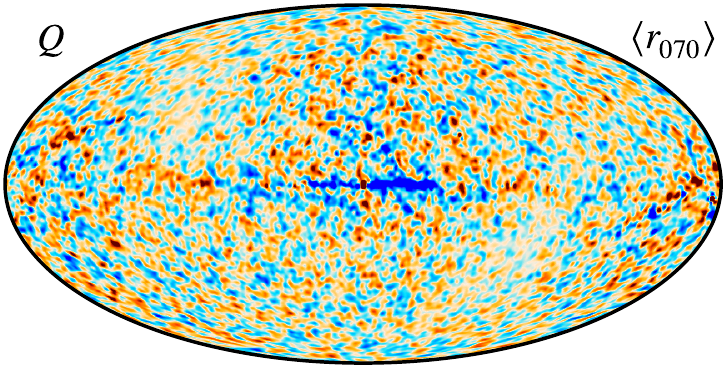}
  \includegraphics[width=0.33\linewidth]{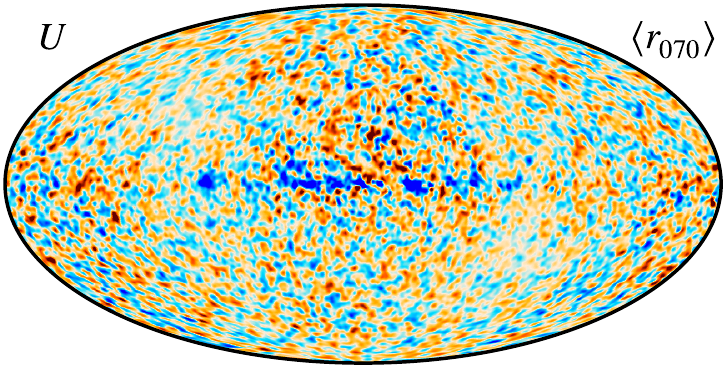}\\
  \includegraphics[width=0.33\linewidth]{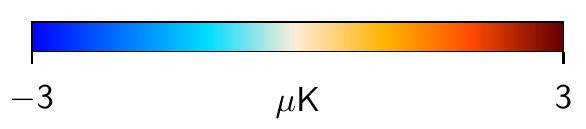}
  \caption{Posterior mean total data-minus-model residual maps $\r_{\nu} = \d_{\nu}-\s_{\nu}$ for \BP\ LFI 30 (\emph{top}), 44 (\emph{middle}), and 70\,GHz (\emph{bottom}). All maps are smoothed to a common angular resolution of $2^{\circ}$ FWHM.}\label{fig:LFI_residuals}

\end{figure*}

Starting from the top, we see that the \BP\ maps adopt a HEALPix\footnote{\url{https://healpix.jpl.nasa.gov}} \citep{gorski2005}
resolution of $N_{\mathrm{side}}=512$ for the 30 and 44\,GHz channels
and $N_{\mathrm{side}}=1024$ for the 70\,GHz, while the \Planck\ maps use
$N_{\mathrm{side}}=1024$ for all three channels. In this respect, it
is worth noting that the beam width of the three channels are
$30\arcm$, $27\arcm$, and $14\arcm$ \citep{planck2016-l02},
respectively, while the HEALPix pixel size is $7\arcm$ at
$N_{\mathrm{side}}=512$ and $3.4\arcm$ at $N_{\mathrm{side}}=1024$. A
useful rule-of-thumb for the HEALPix grid is that maps should have at
least 2.5--3 pixels per beam FWHM to be proper bandwidth limited at
$\ell_{\mathrm{max}}\approx 2.5\,N_{\mathrm{side}}$, above which
HEALPix spherical harmonic transforms gradually become more
non-orthogonal \citep{gorski2005}. With 3.9~pixels per beam FWHM, this
rule is more than satisfied at $N_{\mathrm{side}}=512$ for the 30 and
44\,GHz beams, while it would not be satisfied at 70\,GHz, with only
two pixels per beam FWHM. As essentially all modern component
separation and CMB power spectrum methods today are anyway required to
operate with multi-resolution data, simply to combine data from LFI,
HFI and \WMAP, there is little compelling justification for
maintaining identical pixel resolution among the three LFI channels,
and we instead choose to optimize CPU and memory requirements for the
30 and 44\,GHz channels.

The second entry in Table~\ref{tab:overview} lists the nominal center
frequency of each band. As discussed by \citet{bp09}, the \BP\ model
includes the 30\,GHz center frequency as a free parameter, but fixes
the 44 and 70\,GHz center frequencies at their default values. The
overall shift for the 30\,GHz channel is 0.2\,GHz, with
corresponding implications in terms of unit conversions, color
corrections, and bandpass integration. In addition to this 30\,GHz
correction, the \BP\ processing also includes
entirely reprocessed bandpasses for all detectors, mitigating artefacts
from standing waves which affected bandpass measurements 
in the pre-launch calibration campaign \citep{zonca2009,bp09}, 
and we highly recommend using these improved
bandpass profiles for any future processing of either the original
\Planck\ or the new \BP\ LFI maps.

The four next entries in Table~\ref{tab:overview} list either
``typical'' (for \Planck\ 2018; \citealp{planck2016-l02}) or
frequency- and time-averaged (for \BP; \citealp{bp06}) noise PSD
parameters. Overall, these agree well between the two
analyses. However, it is important to emphasize that whereas the
\Planck\ analysis assumes that the instrumental parameters listed in
lines 3--6 of Table~\ref{tab:overview} to be constant throughout the
full mission during mapmaking, the \BP\ processing allows for hourly
variations in all these parameters. The \BP\ method leads as such both
to more optimal noise weighting in the final map, and also to more
accurate full-mission white noise estimates. We also see that the
additional log-normal correlated noise contribution accounts for as
much as 53\,\% of the white noise level at intermediate frequencies in
the 30\,GHz channel, and 28\,\% at 44\,GHz \citep{bp06}. This term is
neither included in the \Planck\ noise model nor simulations.

\begin{figure}[t]
  \center
  \includegraphics[width=\linewidth]{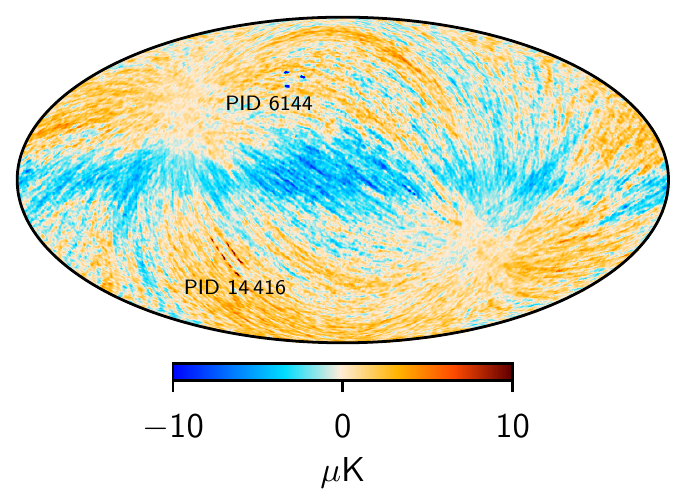}
  \includegraphics[width=\linewidth]{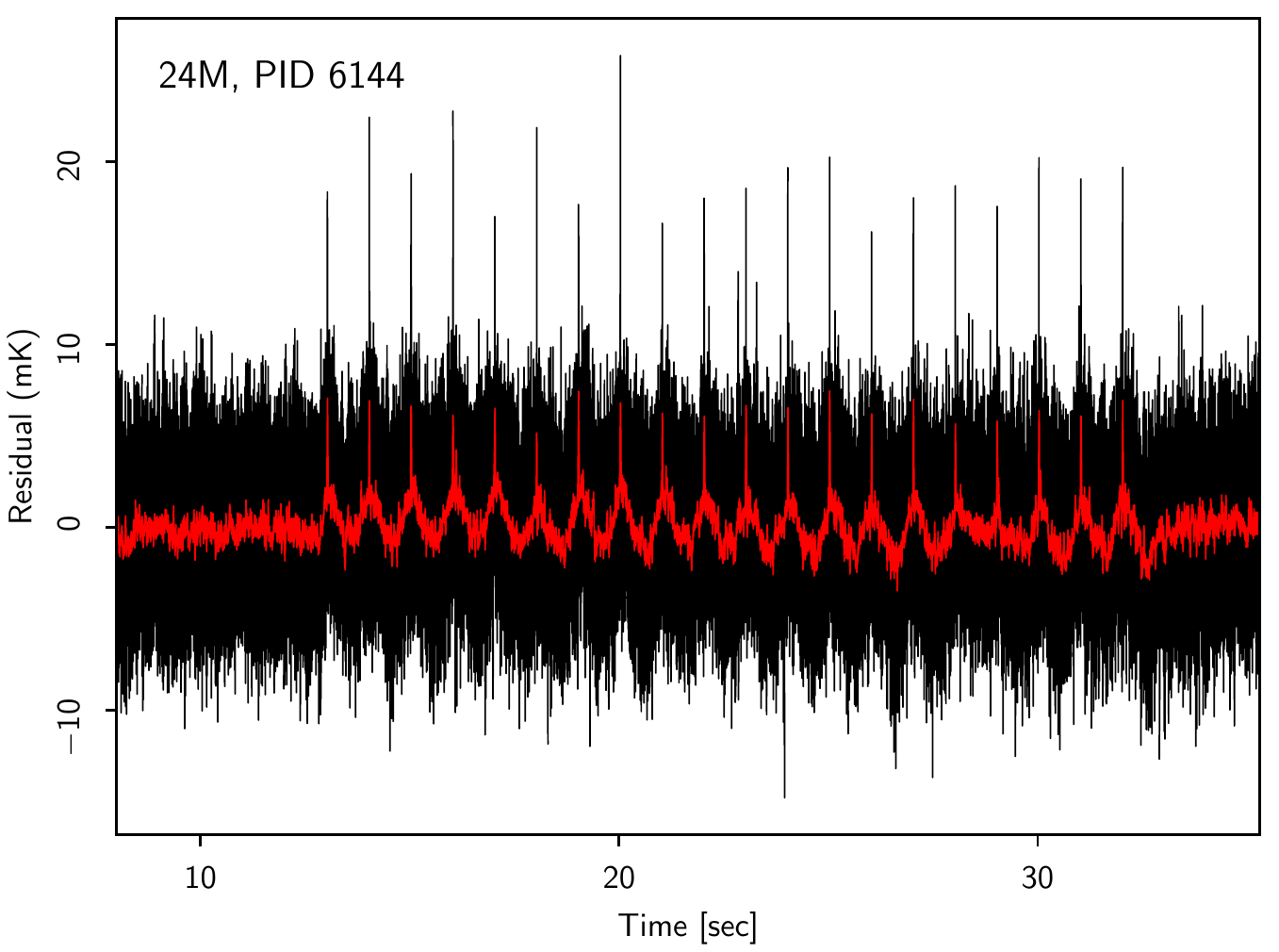}
  \includegraphics[width=\linewidth]{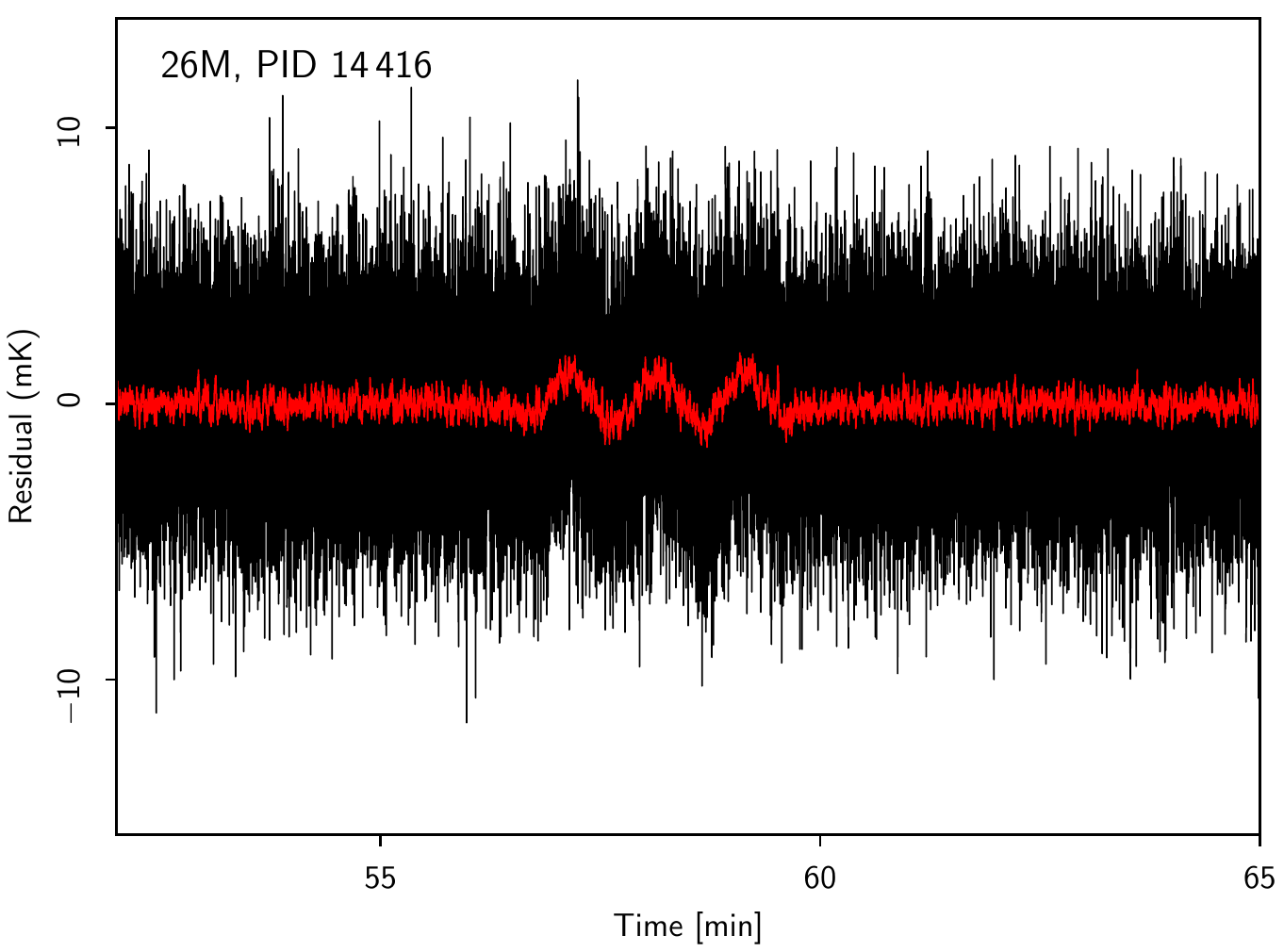}
  \caption{Previously undetected TOD artifacts found by visual inspection of the (preliminary) correlated noise map for the 44 \,GHz channel shown in the top panel. The middle and lower panels show a zoom-in of the source PIDs, namely PID 6144 (for the ``triangle feature'') and PID 14\,416 (for the ``dashed stripe feature''). In the two latter, the black curve shows the TOD residual, $\d-\s^{\mathrm{tot}}$, for each sample, and the red curve shows the same after boxcar averaging with a 1\,sec window.}
  \label{fig:features}
\end{figure}

The next entry in Table~\ref{tab:overview} lists the total data volume
that is actually co-added into the final maps, as measured in units of
detector-hours. In general, the accepted \BP\ data volume is about
9\,\% larger than the \Planck\ 2018 data volume, and this is
because the current processing also includes the so-called
``re-pointing'' periods, which are short intervals of about 5--8
minutes between each \Planck\ pointing period. These data were
originally excluded by an abundance of caution from the official
\Planck\ processing because the pointing model could not be
demonstrated to satisfy the arcsecond pointing requirements defined at
the beginning of the mission. However, \citet{npipe} have subsequently
demonstrated that these data perform fully equivalently in terms of
null maps and power spectra as the regular mission data, and the
potential additional arcsecond-level pointing uncertainties are
negligible compared to the overall instrument noise level. We
therefore follow \Planck\ PR4, and include these observations.

The last entry in Table~\ref{tab:overview} compares the nominal
absolute calibration uncertainties of \Planck\ 2018 and
\BP\ \citep{bp09}, and this may very well quantify the single most
important difference between the two sets of products. For the 70\,GHz
the nominal absolute \BP\ calibration uncertainty is as much as 40
times smaller than the \Planck\ 2018 uncertainty, corresponding to a
fractional uncertainty of $\Delta g_0/g_0 = 5\cdot10^{-5}$. For the
30\,GHz channel, the corresponding ratio is 24.

Clearly, when faced with such large uncertainty differences, two
questions must be immediately addressed, namely, 1) ``are the quoted
uncertainties reasonable?'', and 2) if they are, ``what is the
physical and algorithmic origin of these large differences?'' Starting
with the first question, it is useful to adopt the CMB Solar dipole
amplitude as a reference. This amplitude depends directly, but by no
means exclusively, on the 70\,GHz absolute calibration; other
important (and, in fact, larger) sources of dipole amplitude
uncertainties include foreground and analysis mask marginalization and
the uncertainty in the CMB monopole, $T_{\mathrm{CMB}} =
2.7255\pm0.0005\,\mathrm{mK}$ \citep{fixsen2009,bp11}. For \BP, the
total CMB Solar dipole posterior standard deviation is $1.4\,\muK$,
while the conditional uncertainty predicted by the 70\,GHz absolute
calibration only is $3.36\,\mathrm{mK}\,\times\, 5\cdot10^{-5} =
0.2\,\muK$, which is at least theoretically consistent with a total
dipole uncertainty of $1.4\,\muK$, given the other sources of
uncertainty. For \Planck\ 2018, the corresponding predicted
conditional dipole uncertainty is $3.36\,\mathrm{mK}\, \times\,
200\cdot10^{-5} = 7\,\muK$, which is more than twice as large as the
corresponding full Solar dipole uncertainty. The quoted LFI absolute
gain uncertainties thus appear to be overly conservative, and cannot
be taken as a realistic absolute gain uncertainty at face value; the
true uncertainty is most likely lower by at least a factor of two, and
possibly as much as an order of magnitude.

Even after accounting for this factor, it is clear that the
\BP\ uncertainty is still significantly lower. To understand the
origin of this difference, we recall the summary of the
\BP\ calibration approach provided by \citet{bp07}: Whereas
\Planck\ 2018 processes each frequency channel independently and
assumes the large-scale CMB polarization signal to be zero during
calibration, \BP\ processes all channels jointly and uses external
information from \emph{WMAP} to constrain the large-scale CMB
polarization modes that are poorly measured by \Planck\ alone. Thus,
the \BP\ gain model contains fewer degrees of freedom than the
corresponding \Planck\ 2018 model, and it uses more data to constrain
these degrees of freedom. The net result is a significantly more
well-constrained calibration model.

\subsection{Posterior mean maps and uncertainties}

Figures~\ref{fig:freq_maps30}--\ref{fig:freq_maps70} show the
posterior mean Stokes $T$, $Q$, and $U$ maps for each of the 30, 44,
and 70\,GHz channels. Note that the \BP\ intensity sky maps retain the
CMB Solar dipole. This is similar to \Planck\ PR4 \citep{npipe}, but
different from \Planck\ 2018. For the remainder of the paper, we add
the \Planck\ 2018 Solar dipole back into the \Planck\ 2018 frequency
maps using the parameters listed by \citet{planck2016-l01} when needed.

Figure~\ref{fig:freq_stddev} shows the posterior standard deviation
evaluated pixel-by-pixel directly from the sample set, while
Fig.~\ref{fig:freq_rmswn} shows the diagonals of the white noise
covariance matrices. The first point to notice is that the intensity
scale of the white noise matrices is $75\,\muK$, while the range for
the posterior standard deviation is $2\,\muK$. The \Planck\ LFI maps
are thus strongly dominated by white noise on a pixel level. However,
the fundamental difference between these two sets of matrices is that
while the white noise standard deviations scale proportionally with
the HEALPix pixel resolution, $N_{\mathrm{side}}$, the posterior
standard deviation does not, and the spatial correlations in the
latter therefore dominate on large angular scales.

A visually striking example of such spatial correlations is directly
visible in the 30\,GHz intensity panel in Fig.~\ref{fig:freq_stddev},
which actually appears almost spatially homogeneous, and with an
amplitude that is more than three times larger than the other
channels. The reason for this homogeneous structure is that the
per-pixel variance of the 30\,GHz channel is strongly dominated by the
same monopole variations seen in the bottom row of
Fig.~\ref{fig:param_corr_local}, and only small hints of the scanning
pattern variations (from gain and correlated noise fluctuations) may
be seen at high Galactic latitudes. Consequently, degrading the
30\,GHz map to very low $N_{\mathrm{side}}$'s will not change the per
pixel posterior standard deviation at all, while the white noise level
will eventually drop to sub-$\muK$ values.

Qualitatively speaking, similar considerations also hold for all the
other effects seen in these plots, whether it is the synchrotron
emission coupling seen in the 44\,GHz intensity map, the
\Planck\ scanning induced gain and correlated noise features seen in
the 30 and 44\,GHz polarization maps, or the Galactic plane features
seen in all panels. These are all spatially correlated, and do not
average down with smaller $N_{\mathrm{side}}$ as expected for Gaussian
random noise.

It is also interesting to note that the \BP\ processing mask is
clearly seen in the 44\,GHz polarization standard deviations. This is
because the TOD-level correlated noise within this mask is only
estimated through a constrained Gaussian realization using
high-latitude information and assuming stationarity \citep{bp06},
and this is obviously less accurate than estimating the correlated
noise directly from the measurements. Still, we see that this effect
only increases the total marginal uncertainty by a modest 10--20\,\%.

The motivation for introducing the processing mask during correlated
noise and gain estimation is illustrated in
Fig.~\ref{fig:LFI_residuals}, which shows the TOD-level residual,
$\d_{\nu}-\s_{\nu}^{\mathrm{tot}}$, binned into sky maps. These plots
summarize \emph{everything} in the raw TOD that cannot be fitted by the model
in Eq.~\eqref{eq:model}. The clearly dominant features in these maps
are Galactic plane residuals, which are due to a simplistic foreground 
model, and these are slightly modulated by the \Planck\ scanning
strategy through gain variations. At high Galactic latitudes, the
dominant residuals are point sources and Gaussian noise. Regarding the
point source residuals, we note that the current \BP\ data model does
not account for time variability \citep{rocha:2022}, and this should
be added in a future extension. Without a processing mask, these
residuals would bias the estimated gain and correlated noise model,
and thereby also bias even the high-latitude sky. At the same time, it
is important to note that the intensity scale in these plots is very
modest, and by far most of the sky exhibits variations smaller than
$3\,\muK$.

\begin{figure*}[p]
    \vspace*{1cm}
  
  \center
  \includegraphics[width=0.33\linewidth]{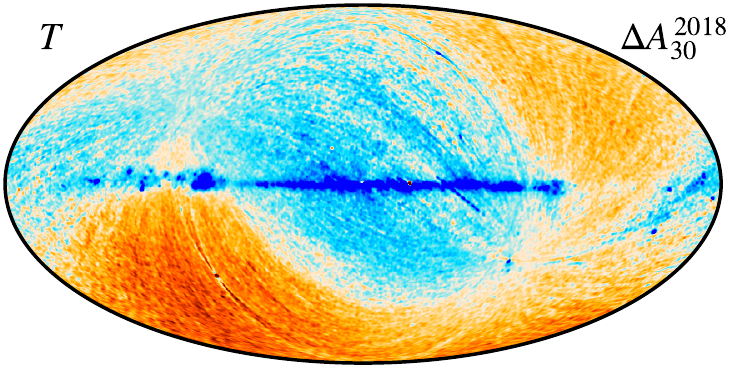}
  \includegraphics[width=0.33\linewidth]{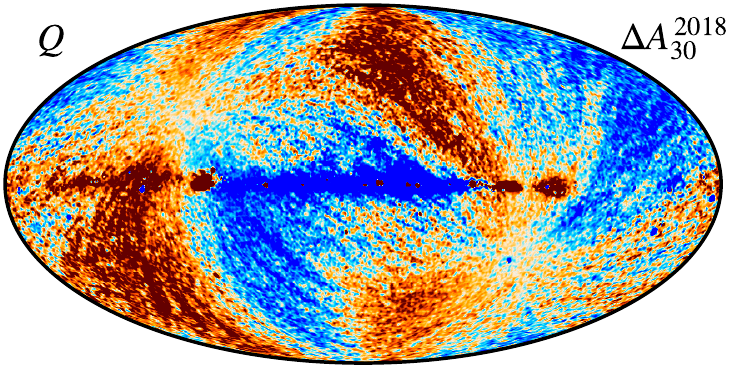}
  \includegraphics[width=0.33\linewidth]{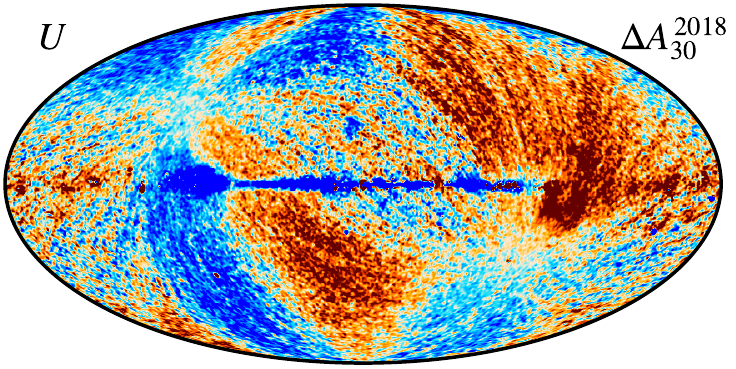}\\
  \includegraphics[width=0.33\linewidth]{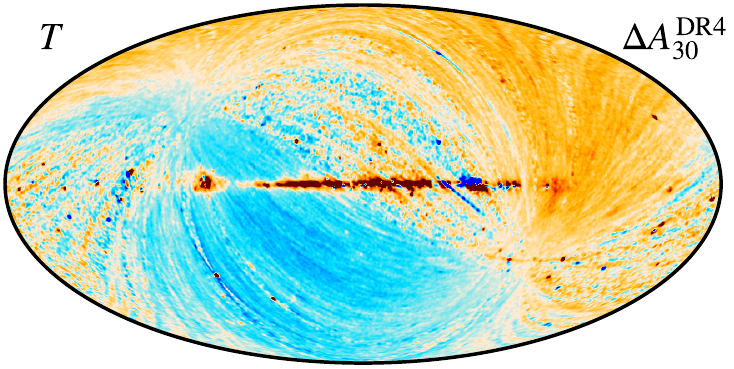}
  \includegraphics[width=0.33\linewidth]{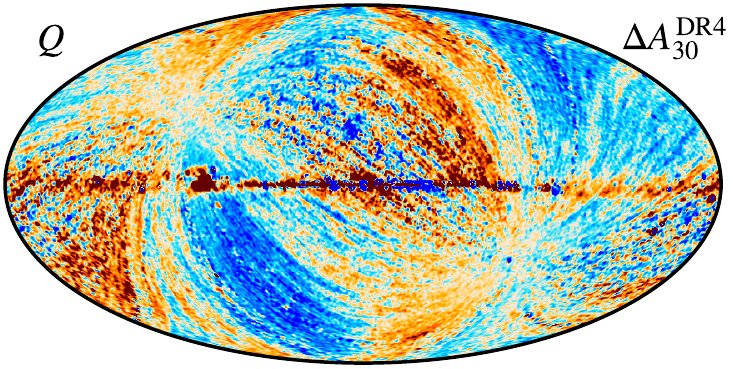}
  \includegraphics[width=0.33\linewidth]{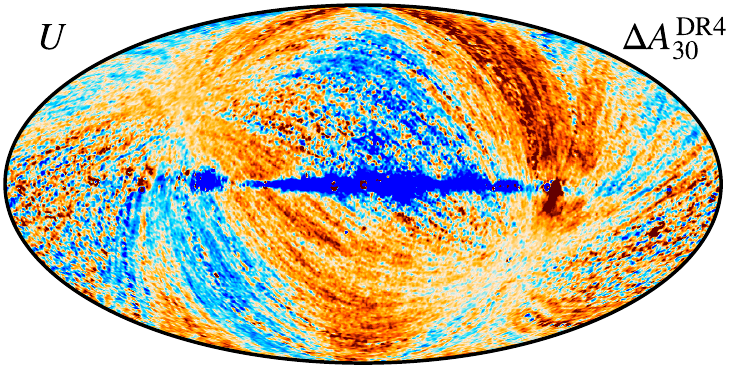}\\
  \vspace*{5mm}
  \includegraphics[width=0.33\linewidth]{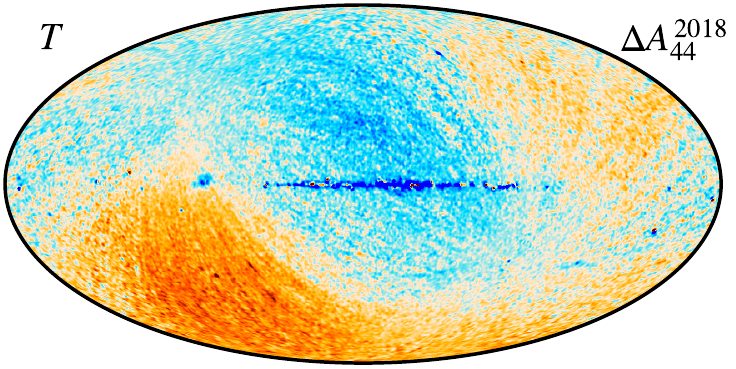}
  \includegraphics[width=0.33\linewidth]{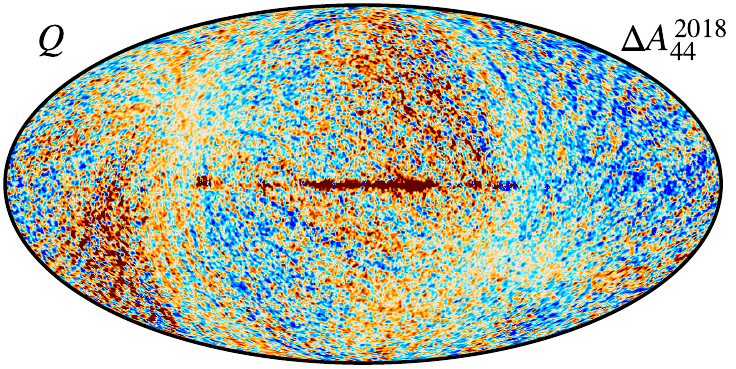}
  \includegraphics[width=0.33\linewidth]{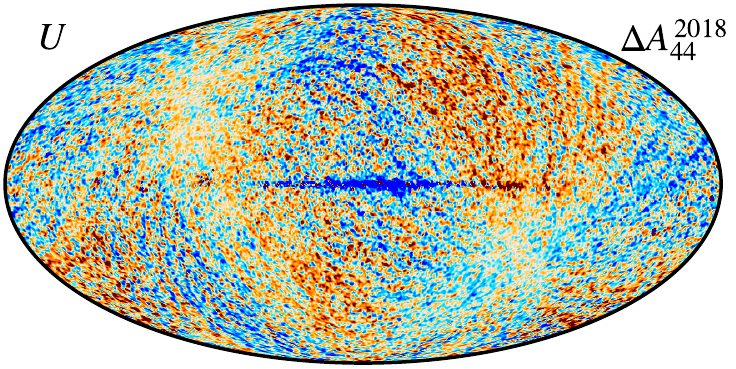}\\
  \includegraphics[width=0.33\linewidth]{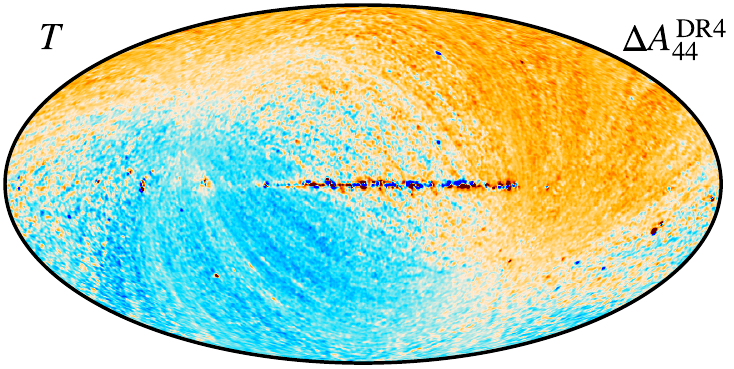}
  \includegraphics[width=0.33\linewidth]{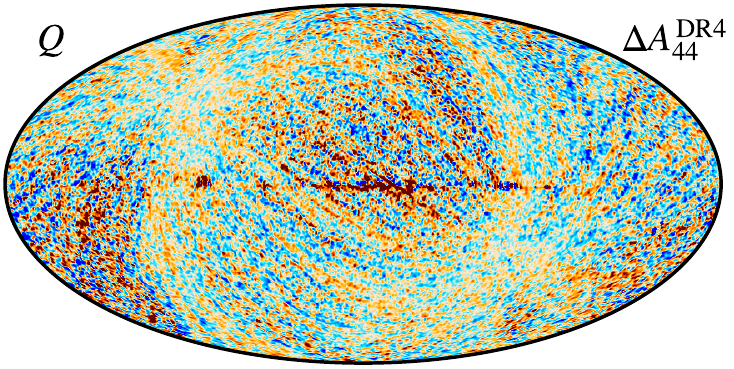}
  \includegraphics[width=0.33\linewidth]{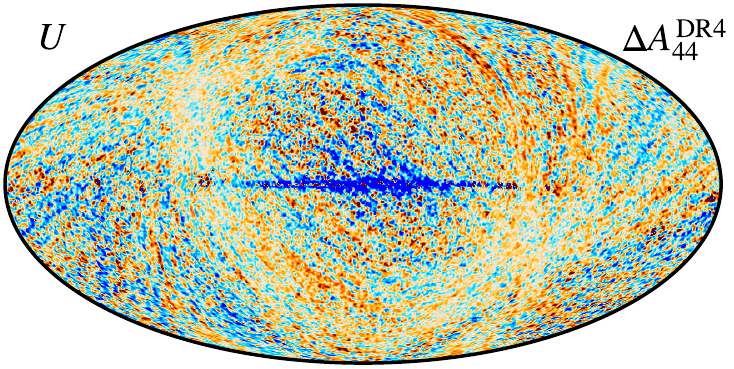}\\
  \vspace*{5mm}
  \includegraphics[width=0.33\linewidth]{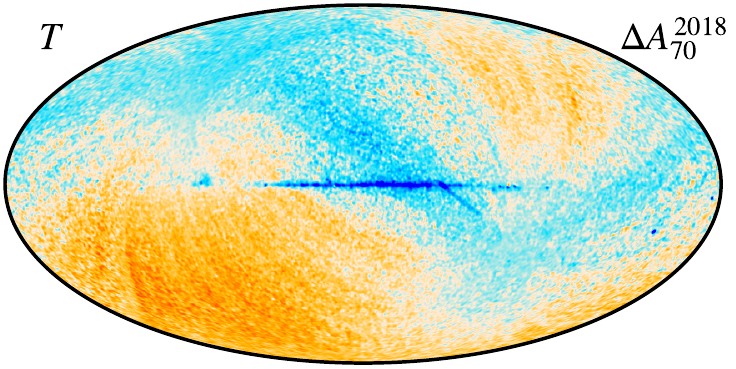}
  \includegraphics[width=0.33\linewidth]{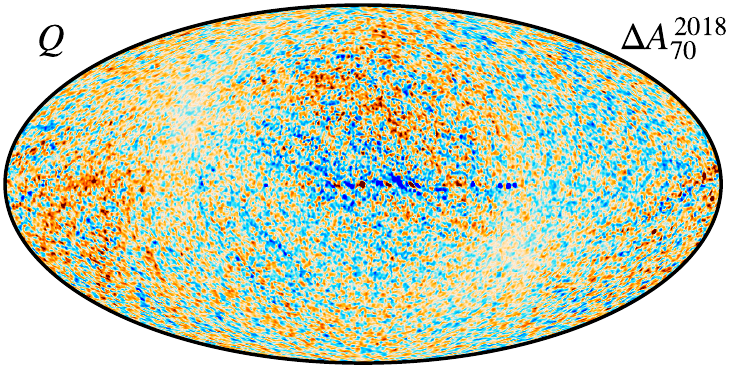}
  \includegraphics[width=0.33\linewidth]{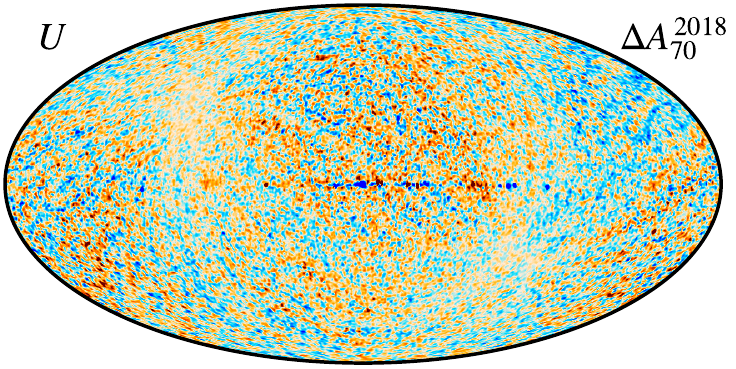}\\
  \includegraphics[width=0.33\linewidth]{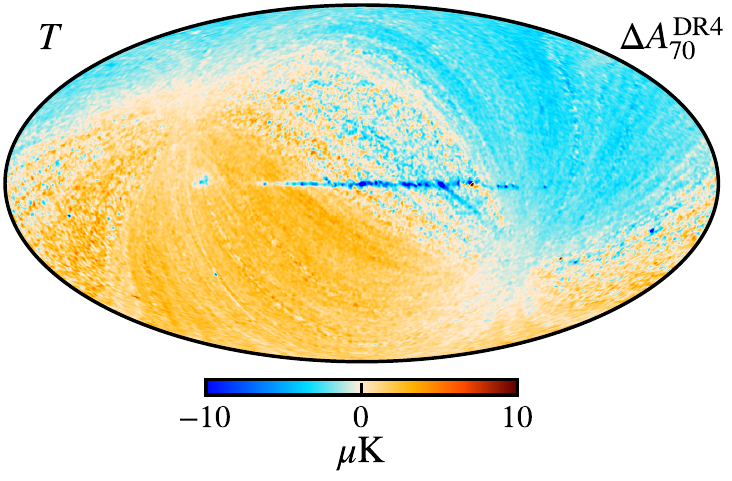}
  \includegraphics[width=0.33\linewidth]{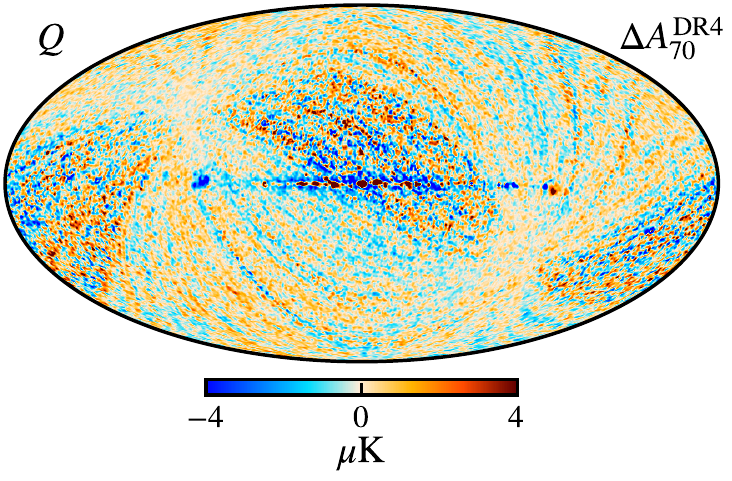}
  \includegraphics[width=0.33\linewidth]{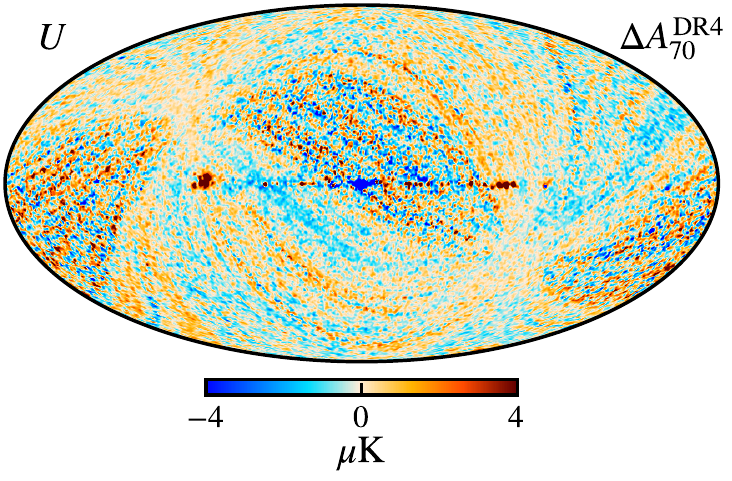}\\
  \caption{Differences between \BP\ and 2018 or PR4 frequency
    maps, smoothed to a common angular resolution of $2^{\circ}$
    FWHM. Columns show Stokes $T$, $Q$ and $U$ parameters,
    respectively, while rows show pair-wise differences with respect
    to the pipeline indicated in the panel labels. A constant offset
    has been removed from the temperature maps, while all other modes
    are retained. The 2018 maps have been scaled by their respective
    beam normalization prior to subtraction.
  }\label{fig:freqdiff}
\end{figure*}

\begin{figure*}[t]
  \center
  \includegraphics[width=\linewidth]{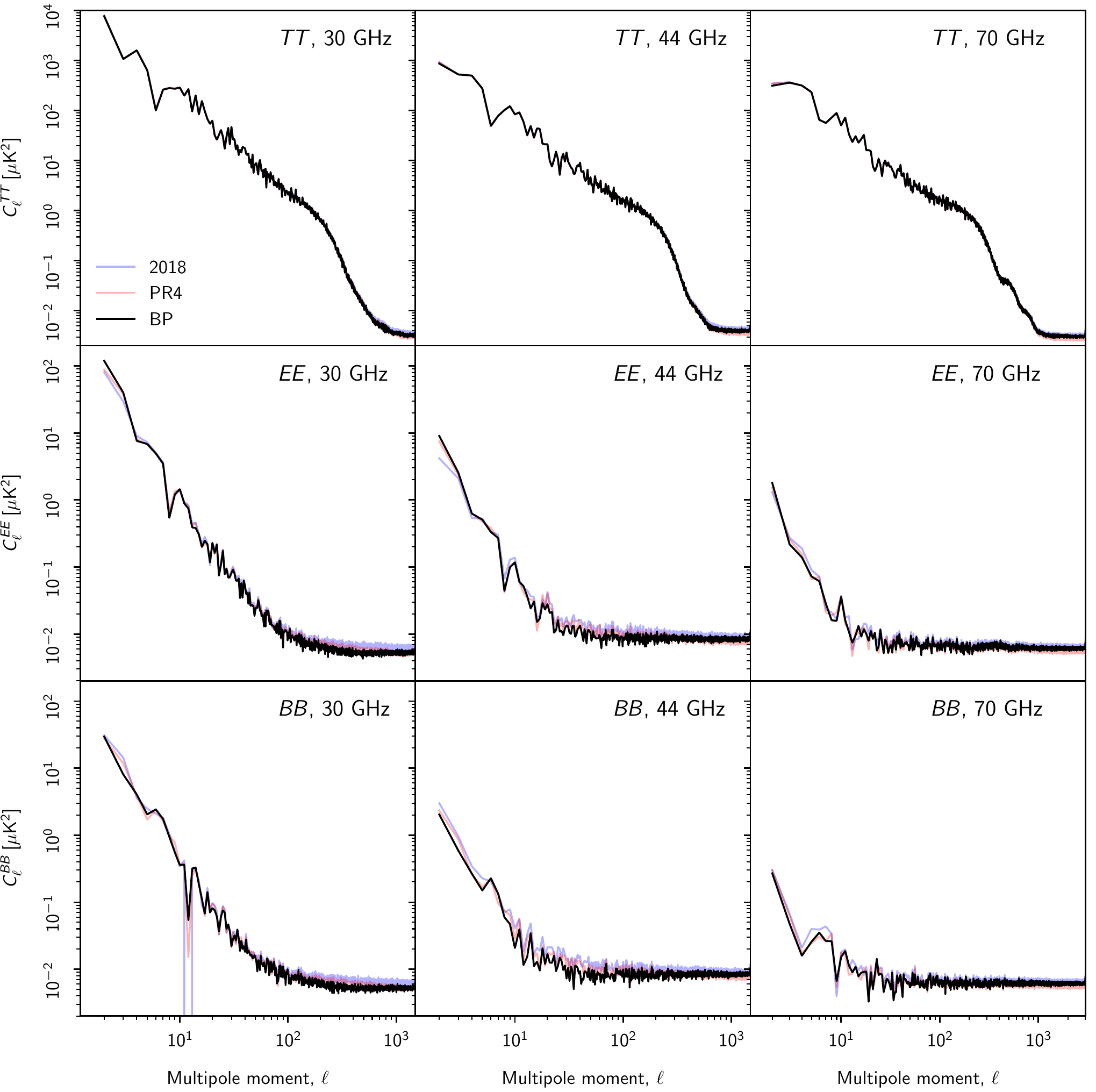}
  \caption{Comparison between angular auto-spectra computed from the \BP\
    (\emph{black}), \Planck\ 2018 (\emph{blue}), and PR4 (\emph{red})
    full-frequency maps. Rows show different frequencies, while columns show
    $TT$, $EE$, and $BB$ spectra. All spectra have been estimated with
    \texttt{PolSpice} using the \Planck\ 2018 common component separation confidence mask \citep{planck2016-l04}.}\label{fig:powspec_full}
\end{figure*}

\begin{figure*}[t]
  \center
  \includegraphics[width=\linewidth]{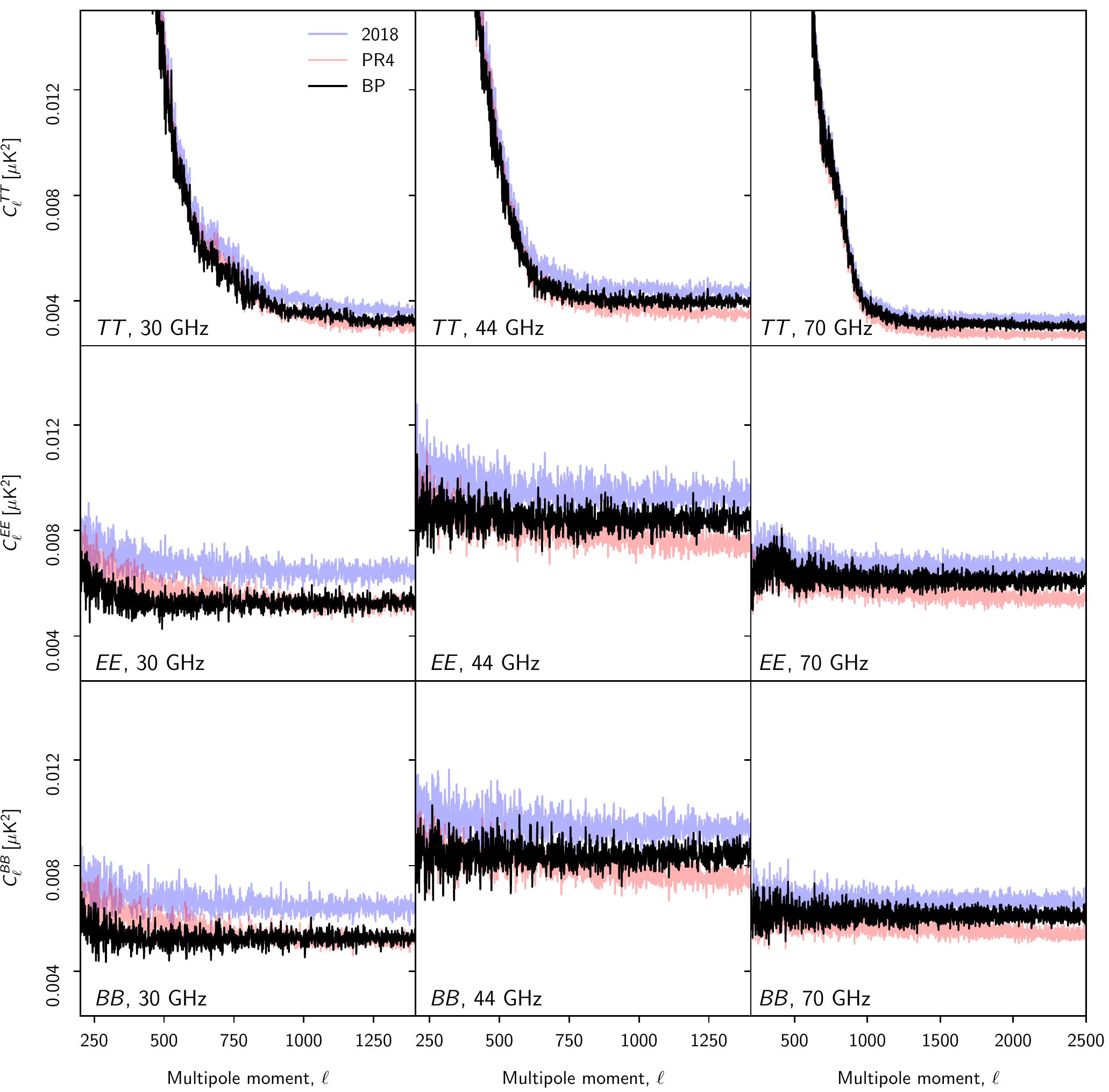}
  \caption{Same as Fig.~\ref{fig:powspec_full}, but zooming in on the noise-dominated high-$\ell$ multipole range.}\label{fig:powspec_full_zoom}
\end{figure*}

\begin{figure*}[t]
  \center
  \includegraphics[width=\linewidth]{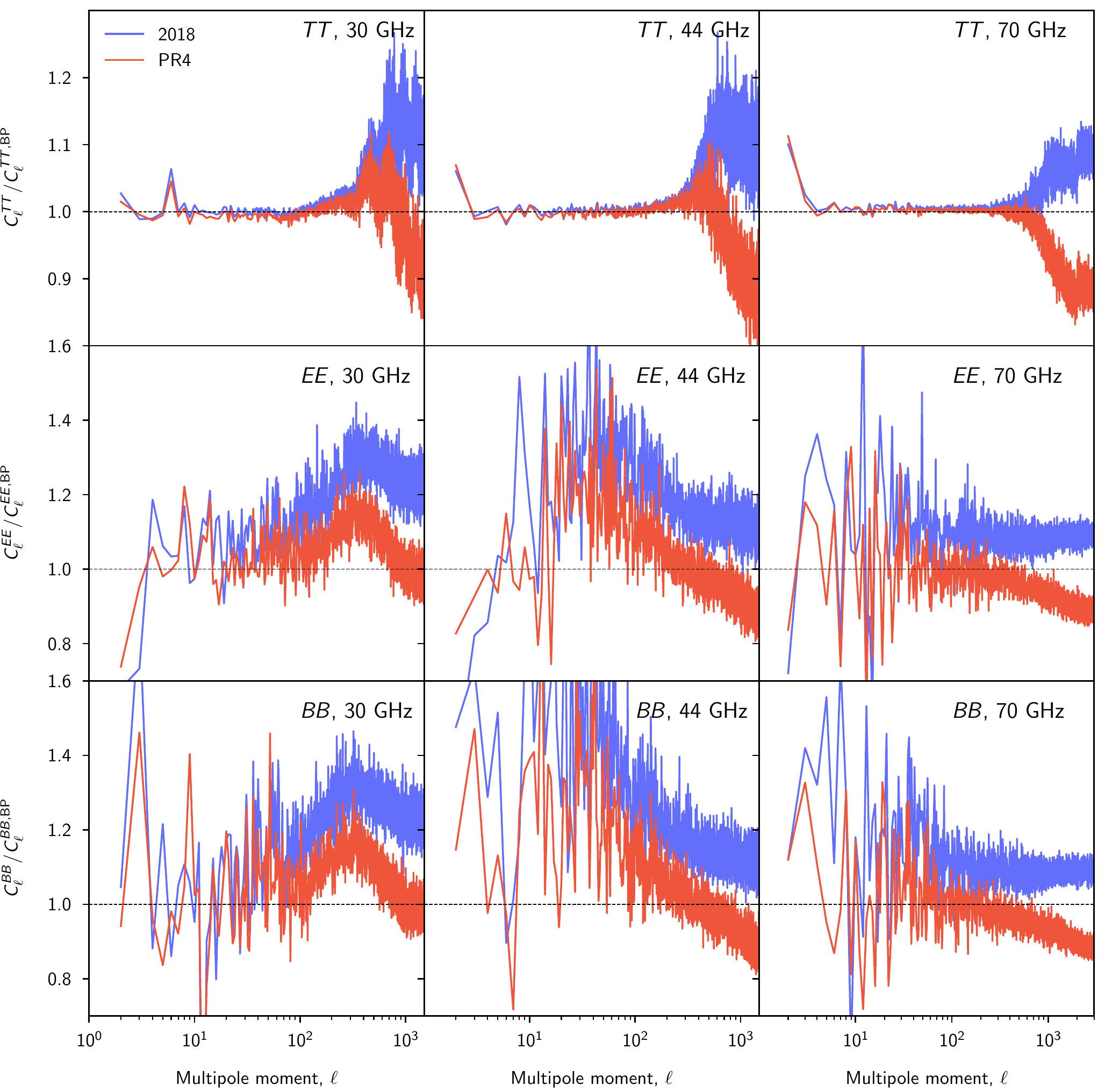}
  \caption{Ratios between the angular auto-spectra shown in
    Fig.~\ref{fig:powspec_full}, adopting the \BP\ spectra as reference.
    \Planck\ 2018 results are shown as blue lines, while \Planck\ PR4 results are
    shown as red lines. Values larger than unity imply that the respective map
    has more power than the corresponding \BP\ spectrum.}
\label{fig:powspec_ratio} 
\end{figure*}

\subsection{Peculiar artifacts removed by hand}

Before ending this section, we note that a few scanning periods are
removed by hand in the current analysis. These were identified by
projecting the estimated correlated noise into sky maps, and an
example of an older version of the 44\,GHz $\n_{\mathrm{corr}}$ map is
shown in the top panel of Fig.~\ref{fig:features}. Overall, this map
is dominated by coherent stripes, as expected for $1/f$-type
correlated noise, and there is also a negative imprint of the Galactic
plane, which is typical of significant foreground residuals outside
the processing mask. The latter effect was mitigated through better
foreground modelling and a more conservative processing mask in the
final production analysis. However, in addition to these well
understood features, there are also two localized and distinct
features in this plot. The first is a triangle, marked by PID 6144,
and the second consists of two extended stripes marked by PID 14\,416,
both identified through a series of bi-section searches. The
corresponding TOD residuals are shown in the bottom two panels; the
black curve shows the raw residual, while the red curve shows the same
after boxcar averaging with a 1-sec window. Here we see that the
triangle feature comprises a series of 20 strong spikes separated by
exactly one minute, which is identical to the \Planck\ spin frequency,
while the extended stripes are due to three slow oscillations.

An inspection of other detector residuals at the same time shows that
the spike event is present in all 44 and 70\,GHz
detectors. Furthermore, a similar event is also present in PID 6126,
and this is also spatially perfectly synchronized with the PID 6144
event, such that they appear superimposed in the top panel of
Fig.~\ref{fig:features}. The effect is thus quite puzzling, as it
appears naively to be stationary in the sky, and it was measured by
all detectors for two separate 20-minute periods; but if it were a
real signal, it would have to originate extremely close to the
\Planck\ satellite, since different detectors observe the signal in
widely separated directions. It is difficult to imagine any true
physical sources that could create such a signal, and we, therefore,
conclude that it most likely is due to an instrumental glitch. At the
same time, it is very difficult to explain why it appears as sky
stationary at two different times. We do not have any plausible
explanations for this signal, but it is obviously not a cosmological
signal signal, and we, therefore, remove PIDs 6126 and 6144 by hand
from the analysis.

In contrast, the PID 14\,416 event takes the form of three slow
oscillations and this looks very much like a thermal or electrical event on the
satellite. We therefore also exclude this by hand. We note, however, that neither of these artifacts were detected during the nominal
\Planck\ analysis phase and this demonstrates the usefulness of the
correlated noise and TOD residual maps in the current analysis as
catch-all systematics monitors. Indeed, low-level analysis in the
Bayesian framework may in fact be viewed as an iterative refinement
process in which coherent signals in these maps are gradually removed
through explicit parametric modelling, and assigned to well-understood
physical effects. The fact that the correlated noise maps in Fig.~9 in
\citet{bp06} appear visually clean of significant artifacts provides
some of the strongest evidence for low systematic contamination in the
\BP\ data products.

\section{Comparison with \Planck\ 2018 and PR4}
\label{sec:comparison}

Next, we compare the \BP\ posterior mean LFI frequency maps with the
corresponding \Planck\ 2018 and PR4 products, and we start by showing
difference maps in Fig.~\ref{fig:freqdiff}. All maps are smoothed to a
common angular resolution of $2^{\circ}$ FWHM, and a monopole has been
subtracted from the intensity maps. To ensure that this comparison is
well defined, the 2018 maps have been scaled by the uncorrected beam
efficiencies \citep{planck2016-l02}, and the best-fit \Planck\ 2018 
Solar CMB dipole has been added to each map, before computing the differences; 
both PR4 and \BP\ maps are intrinsically beam efficiency corrected 
and they retain the Solar dipole.

Overall, we see that the \BP\ maps agree with the other two pipelines
to $\lesssim 10\muK$ in temperature, and to $\lesssim 4\muK$ in
polarization. In temperature, we see that the dominant differences
between \Planck\ PR4 and \BP\ are dipoles aligned with the Solar dipole,
while differences with respect to the 2018 maps also exhibit a notable
quadrupolar pattern. The sign of the PR4 dipole differences
changes with frequency. This result is consistent with the original
characterization of the PR4 maps derived through multi-frequency
component separation in \citet{planck2020-LVII}; that paper reports a
relative calibration difference between the 44 and 70\,GHz channel of
0.31\,\%, which corresponds to 10\muK in the map domain. Overall, in
temperature \BP\ is thus morphologically similar to PR4, but it
improves a previously reported relative calibration uncertainty
between the various channels by performing joint analysis.

\begin{figure*}
  \includegraphics[width=0.33\linewidth]{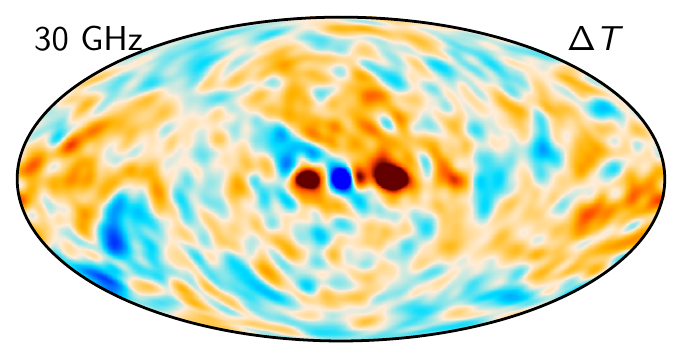}
  \includegraphics[width=0.33\linewidth]{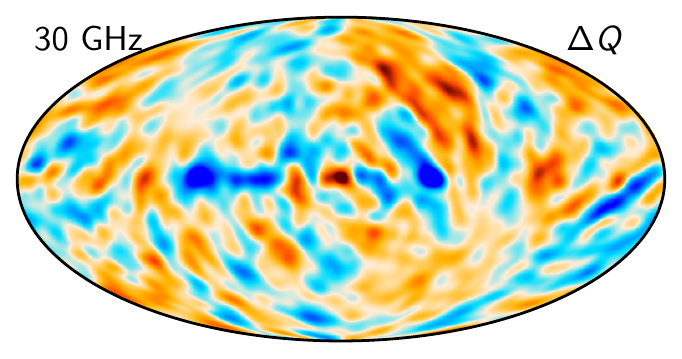}
  \includegraphics[width=0.33\linewidth]{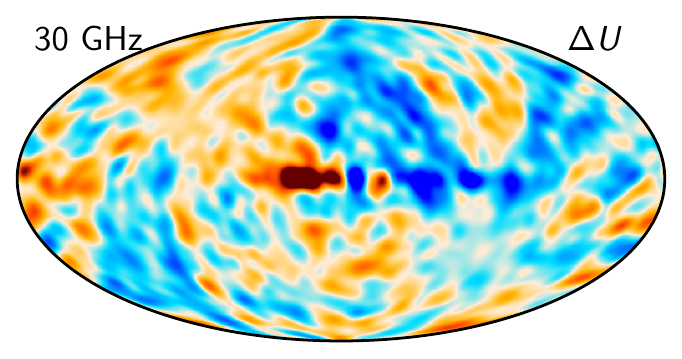}\\
  \includegraphics[width=0.33\linewidth]{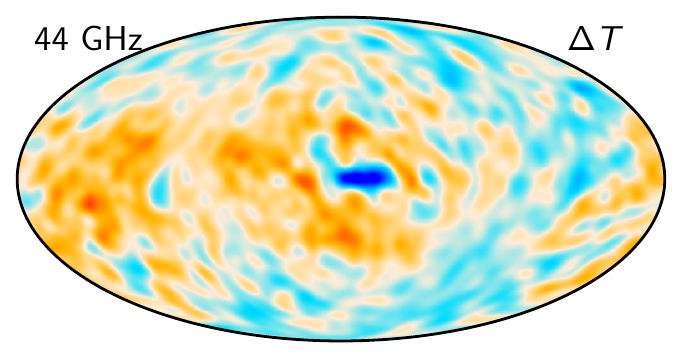}
  \includegraphics[width=0.33\linewidth]{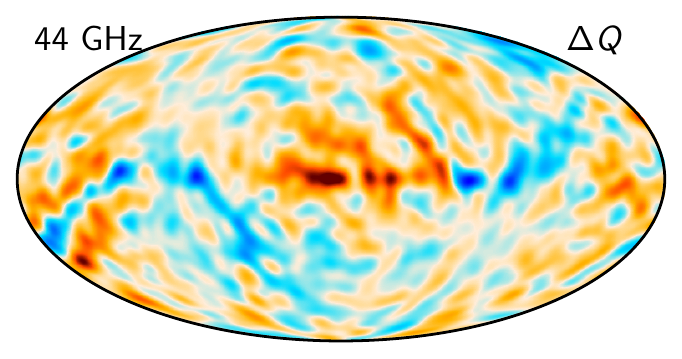}
  \includegraphics[width=0.33\linewidth]{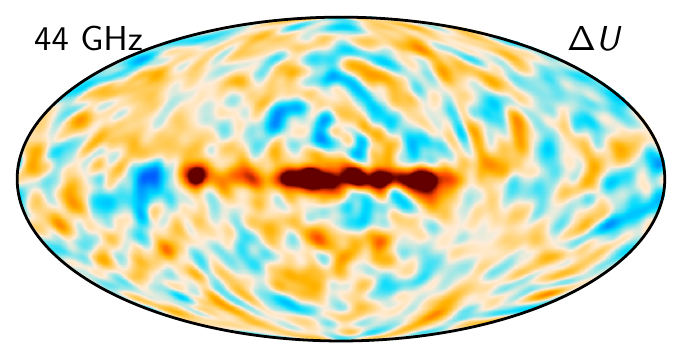}\\
  \includegraphics[width=0.33\linewidth]{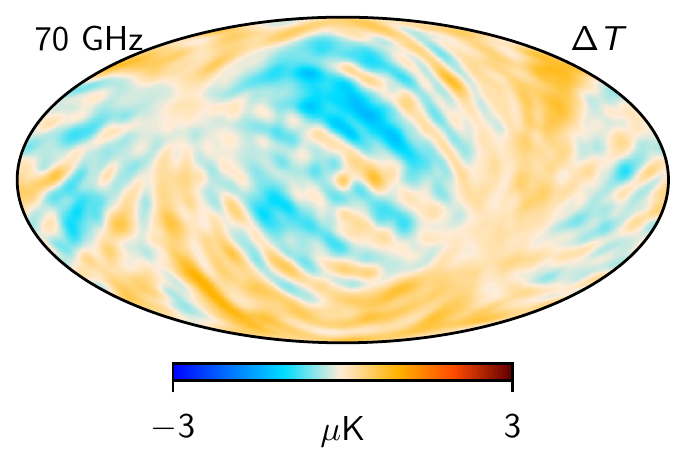}
  \includegraphics[width=0.33\linewidth]{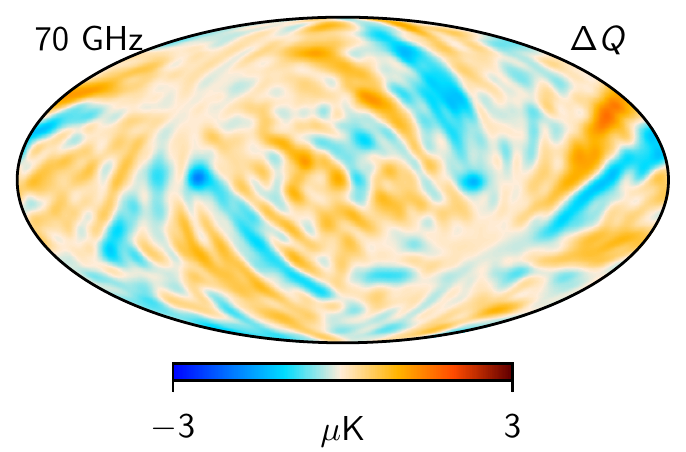}
  \includegraphics[width=0.33\linewidth]{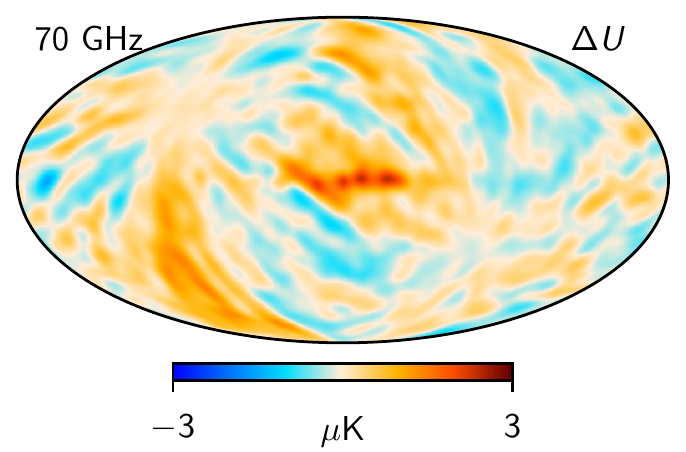}
  \caption{Difference maps between two frequency map samples, smoothed
    to a common angular resolution of $7^{\circ}$ FWHM. Rows show,
    from top to bottom, the 30, 44 and 70\,GHz frequency channels,
    while columns show, from left to right, the temperature and Stokes
    $Q$ and $U$ parameters. Monopoles of 2.8, 1.6, and $-0.5\muK$ have
    been subtracted from the three temperature components,
    respectively. }
  \label{fig:freq_sampdiff}  
\end{figure*}

In polarization, the dominant large-scale structures appear to be
spatially correlated with the \Planck\ gain residual template
presented by \citet{planck2016-l02}, and discussed by \citet{bp07}.
These patterns are thus plausible associated with the large nominal
calibration uncertainties between \Planck\ 2018 and \BP\ discussed in
Sect.~\ref{sec:general_char}. It is additionally worth noting that the
overall morphology of these difference maps is similar between
frequencies, and that the amplitude of the differences falls with
frequency. This strongly suggests that different foreground modelling
plays a crucial role. In this respect, two observations are
particularly noteworthy: First, while both the \Planck\ 2018 and PR4
pipelines incorporate component separation as an external input as
defined by the \Planck\ 2015 data release \citep{planck2014-a12} in
their calibration steps, \BP\ performs a joint fit of both
astrophysical foregrounds and instrumental parameters. Second, both
the LFI DPC and the PR4 pipelines consider only \Planck\ observations
alone, while \BP\ also exploits \WMAP\ information to establish the
sky model, which is particularly important to break scanning-induced
degeneracies in polarization \citep{bp07}.

We now turn our attention to the angular power spectrum properties of
the \BP\ frequency maps. Figure~\ref{fig:powspec_full} shows
auto-correlation spectra as computed with \texttt{PolSpice}
\citep{chon2004} outside the \Planck\ 2018 common component separation
confidence mask \citep{planck2016-l04}, which accepts a sky fraction
of 80\,\%. All these spectra are clearly signal-dominated at large
angular scales (as seen by the rapidly decreasing parts of the spectra
at low $\ell$'s), and noise-dominated at small angular scales (as seen
by the flat parts of the spectra at high $\ell$'s); note that the
``signal'' in these maps includes both CMB and astrophysical
foregrounds. Overall, the three pipelines agree well at the level of
precision supported by the logarithmic scale used here; the most striking
differences appear to be variations in the high-$\ell$ plateau,
suggesting notably different noise properties between the three
different pipelines.

We therefore zoom in on the high-$\ell$ parts of the spectra in
Fig.~\ref{fig:powspec_full_zoom}. Here the differences become much
more clear, and easier to interpret. In general we note two
different trends. First, we see that the overall noise levels of the
\BP\ maps are lower than in the \Planck\ 2018 maps, but also slightly
higher than PR4, although the latter holds less true for intensity
than polarization. Second, we also note that the \BP\ spectra are
notably flatter than the other two pipelines, and in particular than
PR4, which shows a clearly decreasing trend toward high multipoles.

These differences are further elucidated in
Fig.~\ref{fig:powspec_ratio}, which shows the power spectrum ratios
between \Planck\ 2018 and PR4, respectively, and \BP. Again, we see
that the three codes generally agree to well within 1\,\% in $TT$ in
the signal-dominated regimes of the spectra, but diverge in the
noise-dominated regimes. Indeed, at the highest multipoles PR4
typically exhibits about 10\,\% less noise than \BP, while
\BP\ exhibits 10\,\% less noise than \Planck\ 2018. As discussed in
\citet{planck2020-LVII}, \Planck\ PR4 achieves lower noise than
\Planck\ 2018 primarily through three changes. First, PR4 exploits the
data acquired during repointing periods, as explained in
Sect.~\ref{sec:markov_chains}, which account for about 8\,\% of the
total data volume. Second, PR4 smooths the LFI reference load data
prior to TOD differencing, and this results in a similar noise
reduction. Third, PR4 includes data from the so-called ``ninth
survey'' at the end of the \Planck\ mission, which accounts for about
3\,\% of the total data volume. In contrast, \BP\ currently uses the
repointing data, but neither smooths the reference load (essentially
only because of limited time for implementation and analysis), nor
includes the ninth survey. The reason for the latter is that we find
that the TOD $\chi^2$ statistics during this part of the mission show
greater variation from PID to PID, suggesting decreased stability of
the instrument.

These effects explain the different white noise levels. However, they
do not necessarily explain the different slopes of the spectra seen in Fig.~\ref{fig:powspec_full_zoom},
which instead indicate that the level of correlated noise is
significantly lower in the \BP\ maps as compared to the other two
pipelines. The main reason for this is as follows: While \Planck\ 2018
and PR4 both destripe each frequency map independently,
\BP\ effectively performs joint correlated noise estimation using all
available frequencies at once, as described by \citet{bp01} and \citet{bp06}. This
is implemented in practice by conditioning on the current sky model
during the correlated noise estimation phase in the Gibbs loop in
Eqs.~\eqref{eq:gain}--\eqref{eq:cl}, which may be compared to the
application of the destriping projection operator $\Z$ in the
traditional pipelines that is applied independently to each
channel. Intuitively, in the \BP\ approach the 30\,GHz channel is in
effect helped by the 70\,GHz channel to separate true CMB fluctuations
from its correlated noise, while the 70\,GHz channel is helped by the
30\,GHz channel to separate synchrotron and free-free emission from
its correlated noise. And both 30 and 70\,GHz are helped by both
\WMAP\ and HFI to separate thermal and spinning dust from correlated
noise. Of course, this also means that the correlated noise component
are correlated between frequency channels, as discussed in
Sect.~\ref{sec:markov_chains}, and it is therefore imperative to
actually use the Monte Carlo samples themselves to propagate
uncertainties faithfully throughout the system.\footnote{It must of
  course be noted that the traditional pipelines also exhibit a
  correlated noise component between different frequencies, simply
  because they use the same foreground sky model to estimate bandpass
  and gain corrections at different frequencies. This, however, is
  very difficult to both quantify or propagate, because of the
  substantial cost of including full component separation within a
  forward simulation pipeline.}

\section{Sample-based error propagation}
\label{sec:error_propagation}

As discussed in \citet{bp01}, one of the main motivations underlying
the current Bayesian end-to-end analysis framework is to support
robust and complete error propagation for current and next-generation
CMB experiments, and we use the LFI data as a worked example. In this
section, we discuss three qualitatively different manners in which the
posterior samples may be used for this purpose, namely 1)
post-processing of individual samples; 2) derivation of a
low-resolution dense covariance matrix, and 3) half-mission
half-difference maps. We discuss all three methods, and we start with
an intuitive comparison of the Bayesian and the traditional
approaches.

\subsection{Bayesian posterior sampling versus frequentist simulations}

Traditionally, CMB frequency maps are published in terms of a single
maximum-likelihood map with an error description that typically takes
one of two forms. We may call the first type ``analytical'', and this
usually takes the form of either a full-resolution but diagonal (i.e.,
``white noise'') or a low-resolution but dense covariance matrix,
coupled with a single overall calibration factor and a limited number
of multiplicative template-based corrections. This error description
aims to approximate the covariance matrix defined by the inverse
coupling matrix in Eq.~\eqref{eq:mapmaking}, within the bounds of
computational feasibility. The main advantage of this type of error
estimate is that it may both be computed and propagated into
higher-level analyses analytically.

The second type of classical error estimate may be referred to as
``simulations'', and is defined in terms of an ensemble of end-to-end
forward simulations. Each realization in this ensemble represents,
ideally, one statistically independent set of specifications for all
the astrophysical\footnote{In practice, the astrophysical sky model is
  often kept fixed between realizations for pragmatic reasons.} and
instrumental (noise, systematics, scanning strategy) parameters; then,
the simulated time-ordered data corresponding to this specification
set is processed in an identical manner as the real data
\citep[e.g.,][]{planck2014-a14}.  To actually propagate uncertainties
based on these simulations into higher-level products, one can either
analyze each realization with the same statistics as the real data,
and form a full histogram of test statistics, or use the simulations
to first generate a covariance matrix, and then propagate that
analytically through higher-level codes.

The Bayesian posterior sampling approach looks very similar to the
traditional simulation approach in terms of output products: Both
methods produce a large ensemble of sky map realizations that may be
analyzed with whatever higher-level analysis statistics the end-user
prefers. However, the statistical foundation and interpretation -- and
therefore their applicability -- of the two methods are in fact very
different: While the forward frequentist simulation approach considers
a set of \emph{random} instruments in a set of \emph{random}
universes, the Bayesian approach considers \emph{our} instrument in
\emph{our} universe; for an in-depth discussion regarding this issue,
see \citet{bp04}.

The relevance and importance of this difference for actual CMB
analysis applications may be illustrated by the following series of
observations. Suppose we are tasked with generating an optimal
simulation suite to analyze a given frequency map. One important
question we have to address is, ``what CMB dipole parameters should we
adopt?'' The importance of this question is highlighted in
Fig.~\ref{fig:freqdiff}, which shows the differences between the
\BP\ and \Planck\ maps; if completely random CMB Solar dipole
parameters were used, the large coherent structures in these plots
would move around randomly from realization to realization, to the
extent that the simulated uncertainties would not actually be a useful
description of \emph{our} sky map. For this reason, all CMB analysis
pipelines to date have actually used dipole parameters close to the
real sky.

A second important question is, ``what Galactic model should we use?''
The same observations hold in this case; if one were to use random
Galactic models for each simulation, the Galactic plane's morphology
in Fig.~\ref{fig:freqdiff} would change from realization to
realization, and no longer even be aligned with our Milky Way, and the
simulated ensemble would obviously not be useful to describe the
uncertainties in the real sky map. For this reason, all CMB pipelines
to data have also adopted a Galactic model close to the real sky.

So far, the discussion has been rather straightforward. However, the
next question that emerges is more controversial, namely ``what CMB
fluctuations should we use to generate the simulations?'' In this
case, most traditional pipelines actually adopt statistically
independent $\Lambda$CDM realizations that have no coupling to the real
sky. This component is thus treated logically very differently from
the CMB dipole and the Galaxy, and this has important consequences for
any correlation structures that result from the simulated
ensemble; the simulations are no longer useful to describe the
features seen in Fig.~\ref{fig:freqdiff}, simply because the
coherence between the small-scale CMB fluctuations and the gain
variations, the correlated noise, the foreground fluctuations etc.,
has been broken.

This issue does not only apply to the CMB fluctuations, but also to
the instrument parameters. Of course, the basic knowledge of
instrument parameters comes from hardware calibration
measurements. However, these inevitably come with an uncertainty, and
in the traditional approach these uncertainties are extremely
difficult to propagate through the data analysis. So in some cases
(e.g., bandpasses) the parameters are treated as fixed values in the
pipeline, and the potential bias arising from the uncertainty in the
hardware measurement is evaluated separately (see, e.g., Figs.~24, 25,
and 26 of \citealp{planck2014-a14}).  This approach is able to
quantify the potential impact of the systematic effect compared to the
signal power spectra, but it does not control the correlation of the
effect with other parameters (in this case, most notably,
astrophysical foreground parameters) thus preventing a rigorous error
analysis. In other cases, such as $1/f$ noise, the relevant parameters
(e.g., gain change as a function of time, or ``baseline'') are not
derived from hardware measurements, but are chosen randomly from some
hyper-distribution, independent from the real data. However, gain
fluctuations correlate with the CMB dipole and fluctuations and the
Galactic model, and consequently they may generate coherent features
at very specific positions in the final frequency map.

\begin{figure*}[p]
  \center
  \includegraphics[width=0.33\linewidth]{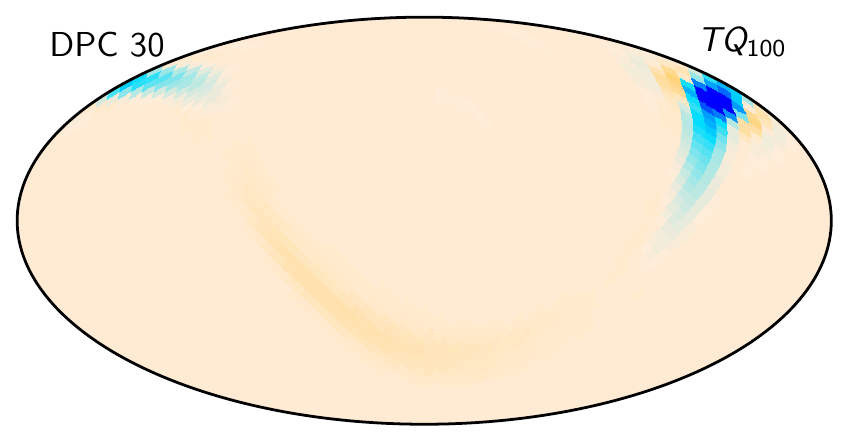}
  \includegraphics[width=0.33\linewidth]{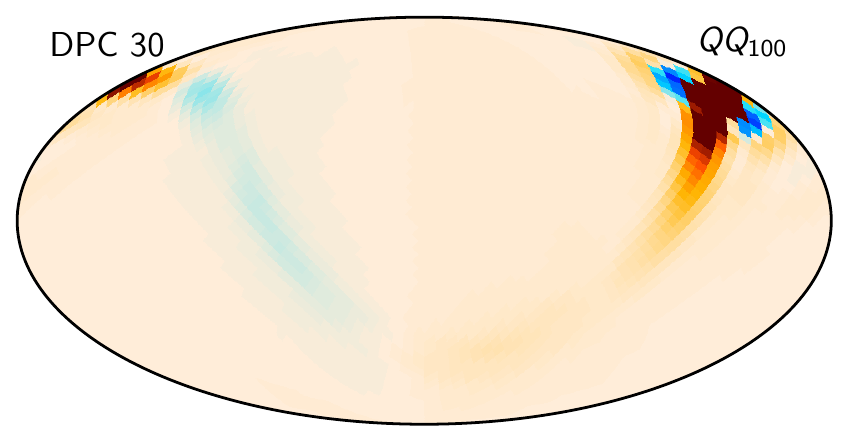}
  \includegraphics[width=0.33\linewidth]{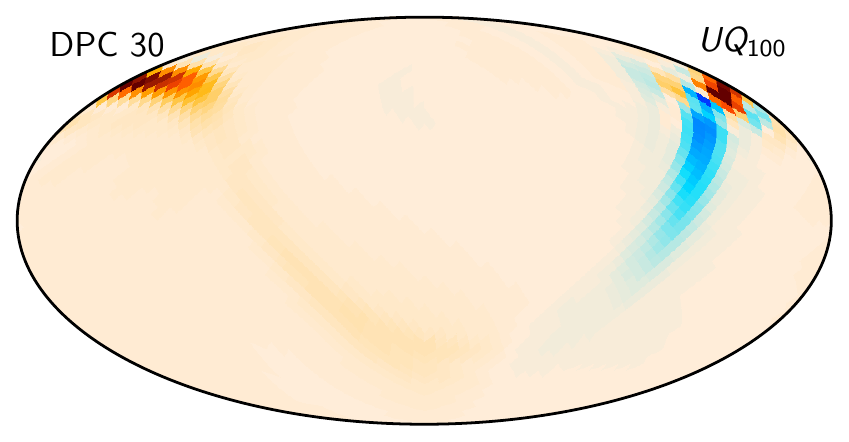}\\
  \includegraphics[width=0.33\linewidth]{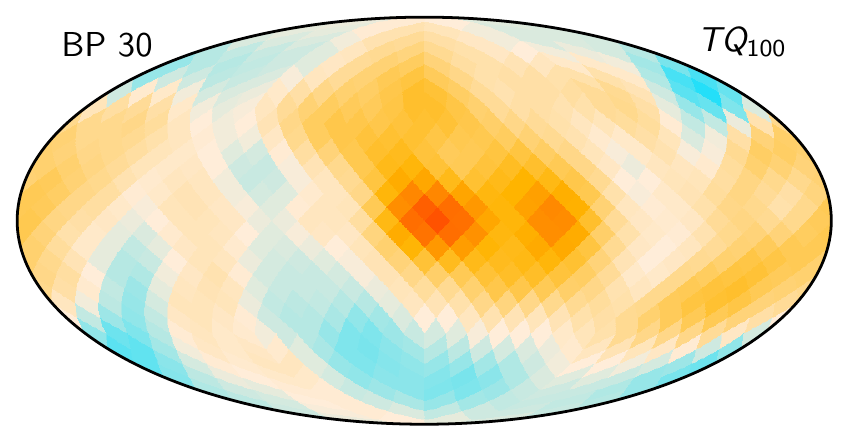}
  \includegraphics[width=0.33\linewidth]{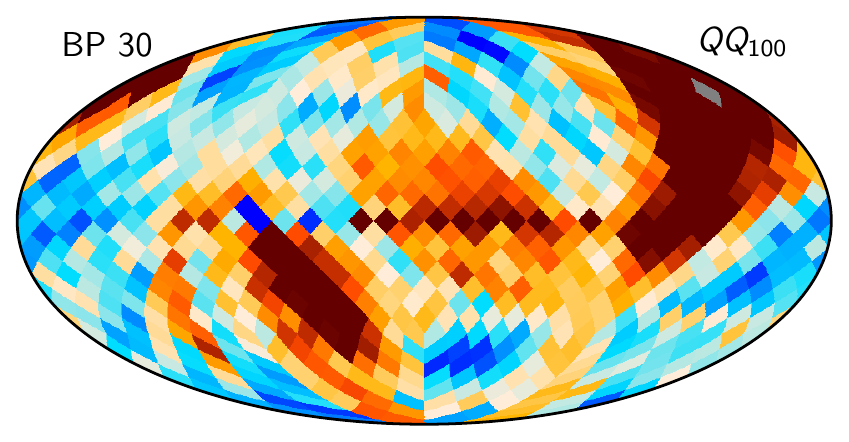}
  \includegraphics[width=0.33\linewidth]{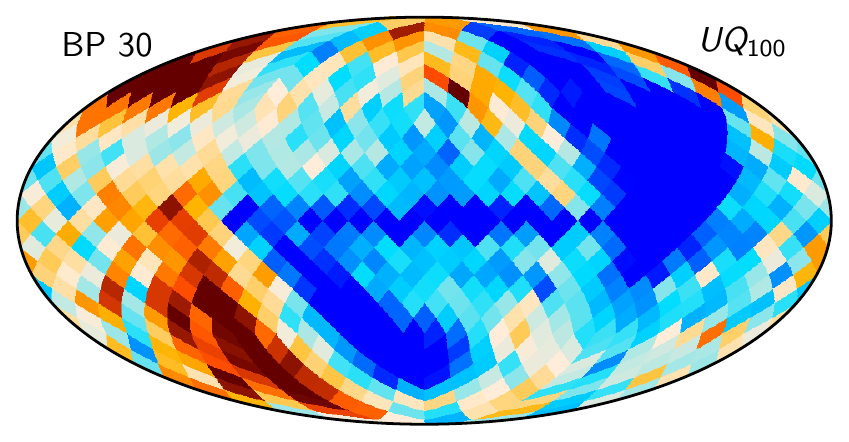}\\\vspace*{5mm}
  \includegraphics[width=0.33\linewidth]{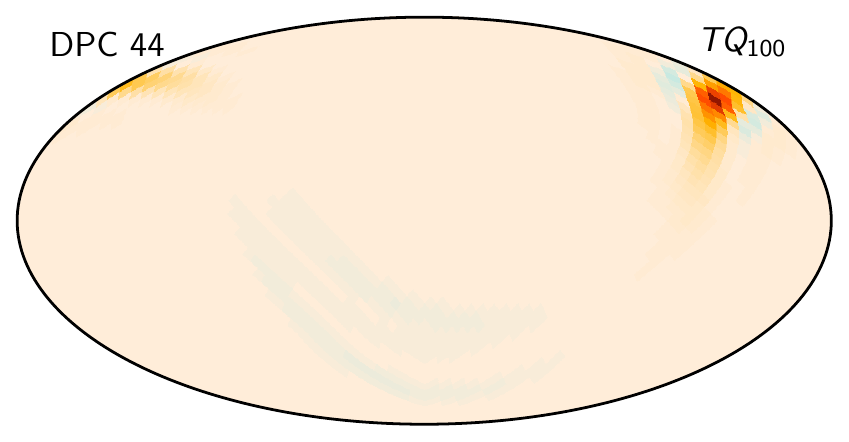}
  \includegraphics[width=0.33\linewidth]{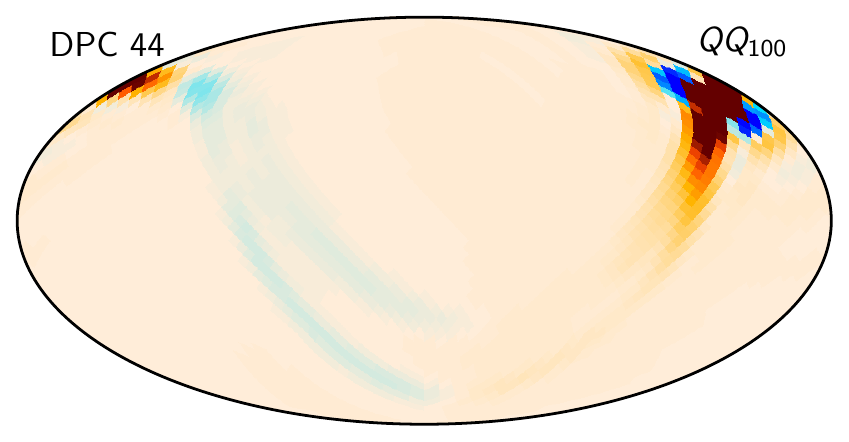}
  \includegraphics[width=0.33\linewidth]{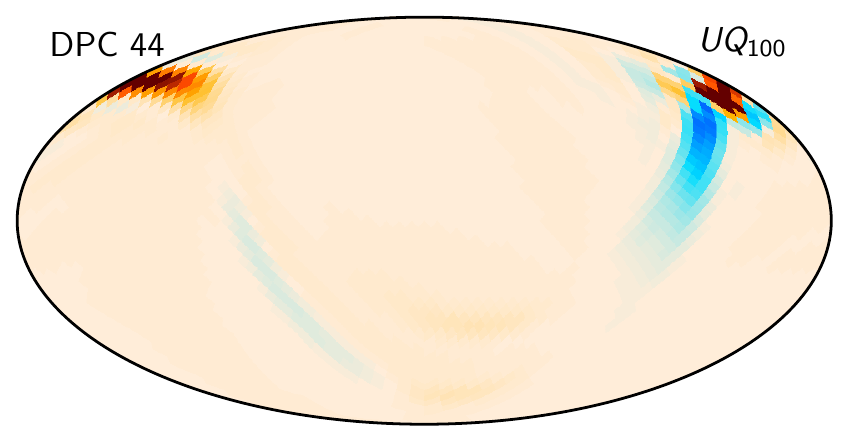}\\
  \includegraphics[width=0.33\linewidth]{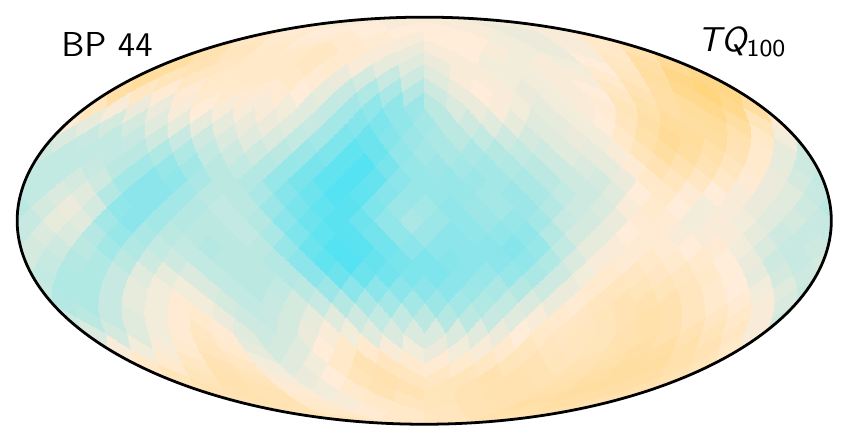}
  \includegraphics[width=0.33\linewidth]{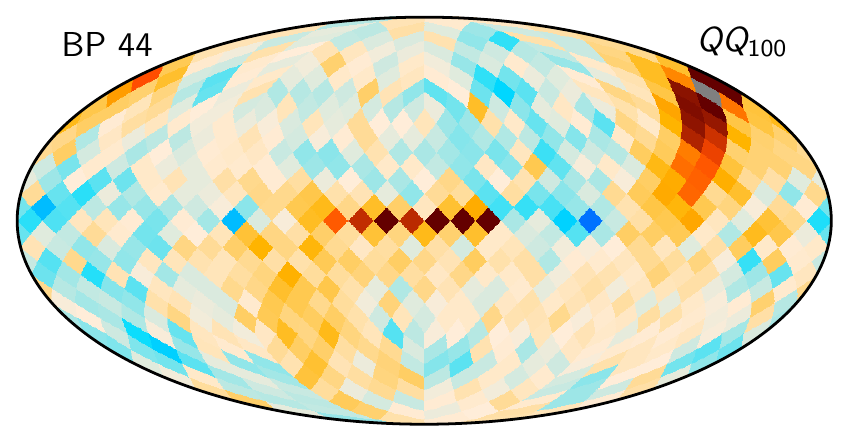}
  \includegraphics[width=0.33\linewidth]{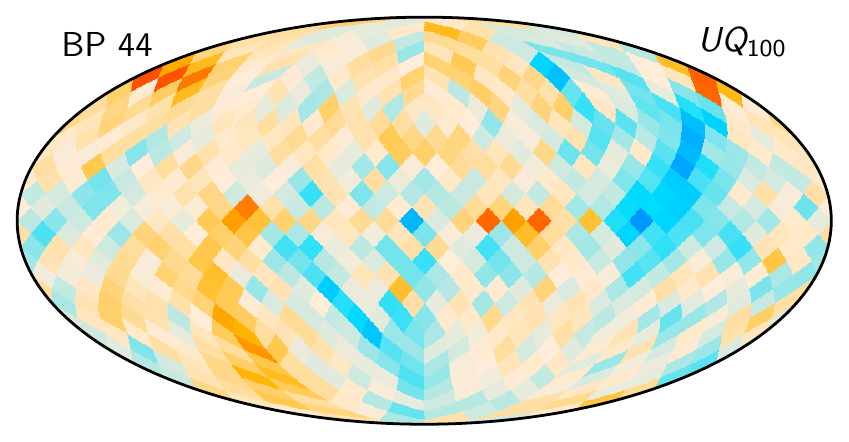}\\\vspace*{5mm}
  \includegraphics[width=0.33\linewidth]{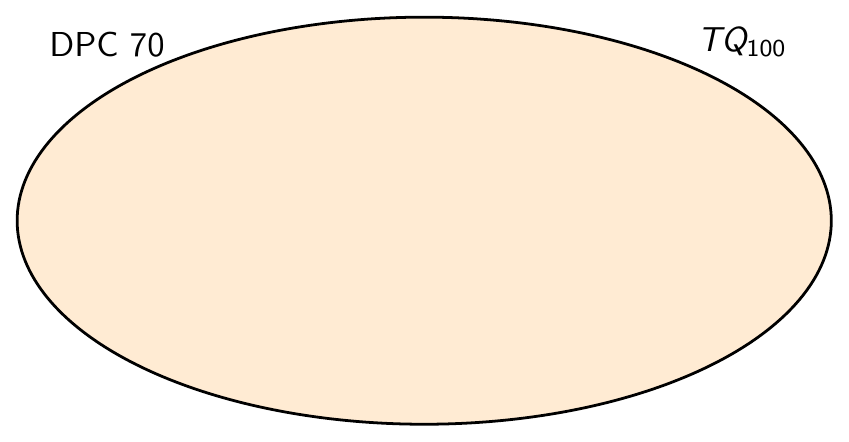}
  \includegraphics[width=0.33\linewidth]{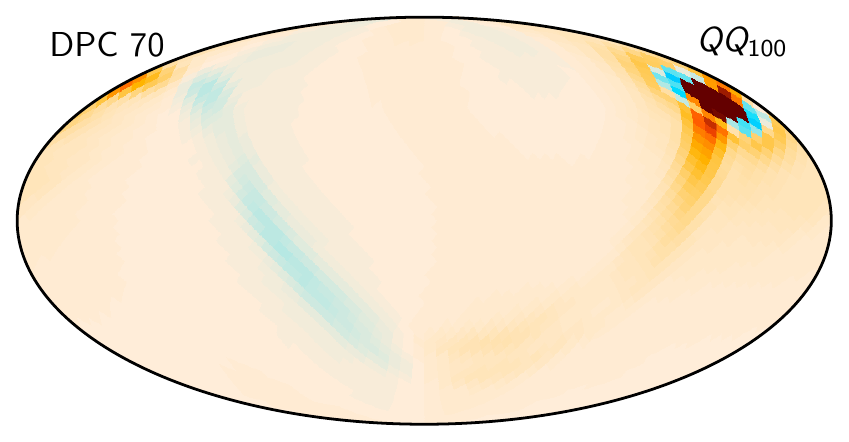}
  \includegraphics[width=0.33\linewidth]{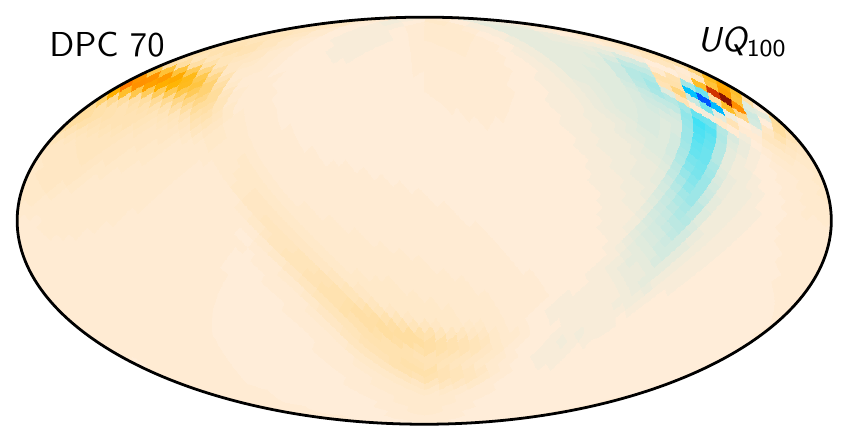}\\
  \includegraphics[width=0.33\linewidth]{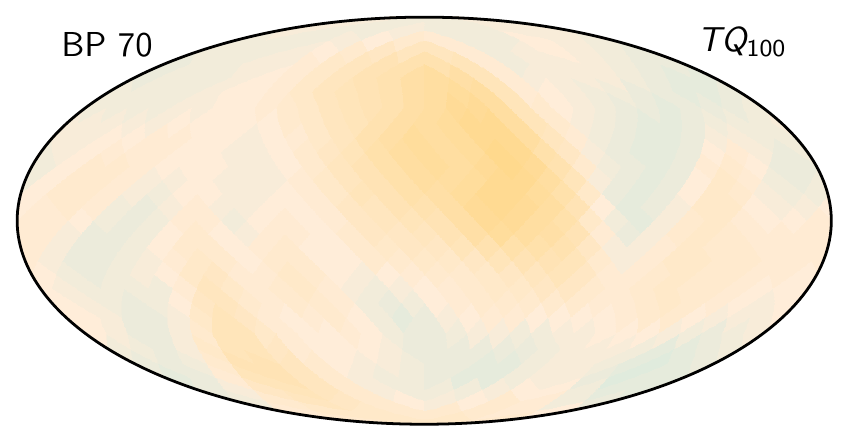}
  \includegraphics[width=0.33\linewidth]{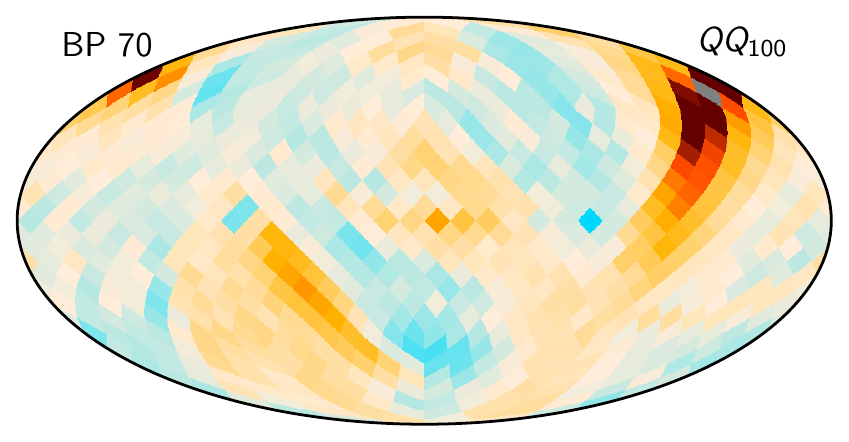}
  \includegraphics[width=0.33\linewidth]{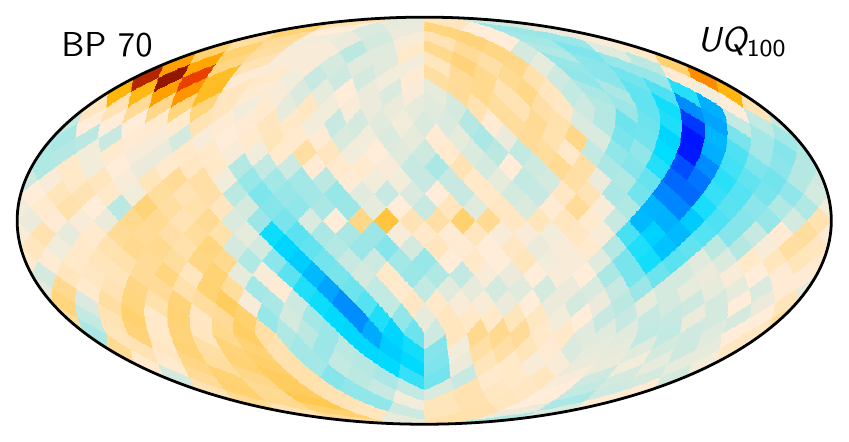}\\
  \includegraphics[width=0.35\linewidth]{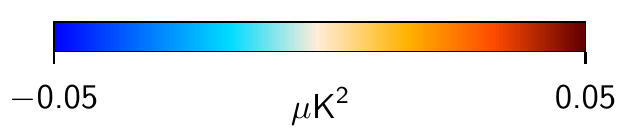}
  \caption{Single column of the low-resolution 30 (\emph{top
      section}), 44 (\emph{middle section}), and 70\,GHz (\emph{bottom
      section}) frequency channel covariance matrix, as estimated
    analytically by the LFI DPC (\emph{top rows}) and by posterior
    sampling in \BP\ (\emph{bottom rows}). The selected column
    corresponds to the Stokes $Q$ pixel marked in gray, which is
    located in the top right quadrant in the \BP\ maps. Note that the
    DPC covariance matrix is constructed at $N_{\mathrm{side}}=16$ and
    includes a cosine apodization filter, while the \BP\ covariance
    matrix is constructed at $N_{\mathrm{side}}=8$ with no additional
    filter. The temperature components are smoothed to $20^{\circ}$
    FWHM in both cases, and \Planck\ 2018 additionally sets the $TQ$
    and $TU$ elements by hand to zero for the 70\,GHz channel. }
  \label{fig:ncov_highlat}
\end{figure*}

\begin{figure*}
	\center
	\includegraphics[width=\linewidth]{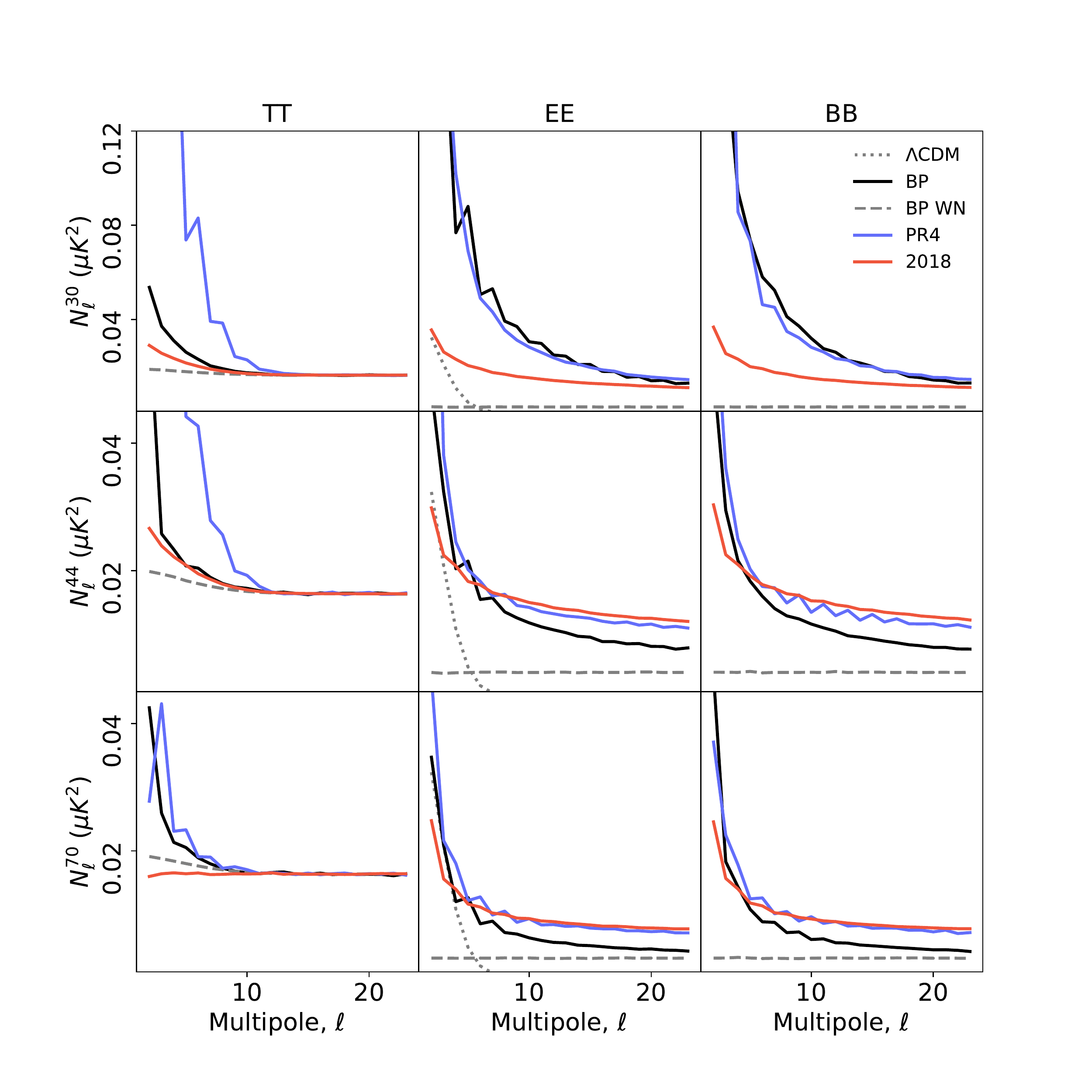}
	\caption{Comparison of noise power spectra from  \BP\
	(\emph{black}), \Planck\ 2018 (\emph{red}) and PR4 (\emph{blue});
	We report the \Planck\ 2018 best-fit with a dotted line, and 
	the \BP\ white noise contribution as a dashed line. Each dataset
	includes a regularization noise of $2\;\mu\rm{K}/\rm{pixel}$
	for temperature, and $0.02\;\mu\rm{K}/\rm{pixel}$ for 
	polarization.}
	\label{fig:noise_ps}
\end{figure*}

In contrast to the frequentist forward simulation approach that uses
random simulations to describe uncertainties, the Bayesian posterior
sampling approach couples every single parameter to the actual data
set in question, and treats the CMB fluctuations and instrument
parameters on the same statistical footing as the CMB dipole and
Galactic signal. Intuitively speaking, the Bayesian sampling approach
is identical to the simulation approach, with the one important
difference that the Bayesian method adopts a constrained realization
to generate the end-to-end simulation for all parameters, while the
frequentist approach uses a random realization (or a mixture of random
and constrained realizations).

The conclusion from this set of observations is that the Bayesian and
frequentist approaches actually addresses two fundamentally different
questions \citep{bp04}. The Bayesian samples are optimally tuned to
answer questions like ``what are the best-fit $\Lambda$CDM parameters
of our specific data set?'', while the frequentist simulations are
optimally tuned to answer questions like ``is our data set consistent
with the $\Lambda$CDM model?'' The fundamental difference lies in
whether the correlation structures between realizations are tuned to
describe uncertainties in our specific instrument and universe, or
uncertainties in a random instrument in a random universe.

It is, however, clear that both the Bayesian posterior sampling and
the frequentist simulation approaches have decisive practical
advantages over the analytical method in terms of modelling complex
systematic effects and their interplay. As a practical demonstration
of this, Fig.~\ref{fig:freq_sampdiff} shows the difference between two
\BP\ frequency map samples for each frequency channel, smoothed to a
common angular resolution of $7^{\circ}$ FWHM. Here we clearly see
correlated noise stripes along the \Planck\ scan direction in all
three frequency channels, as well as coherent structures along the
Galactic plane. Clearly, modelling such correlated fluctuations in
terms of a single standard deviation per pixel is inadequate,
and, as we will see in the next section, these structures also cannot
be described as simply correlated $1/f$ noise. End-to-end analysis
that simultaneously accounts for all sources of uncertainties is 
key for fully describing these uncertainties.

\subsection{Sample-based covariance matrix evaluation}

For several important applications, including low-$\ell$ CMB
likelihood evaluation \citep[e.g.,][]{page2007,planck2016-l05,bp12},
it is important to have access to full pixel-pixel covariance
matrices. Since the memory required to store these scale as
$\mathcal{O}(N_{\mathrm{side}}^4)$, and the CPU time required to
invert them scale as $\mathcal{O}(N_{\mathrm{side}}^6)$ \citep{sherman1949adjustment}, these are
typically only evaluated at very low angular resolution; \Planck\ used
$N_{\mathrm{side}}=16$ \citep{planck2016-l05}, while \BP\ uses
$N_{\mathrm{side}}=8$ \citep{bp12}.

\begin{figure*}[t]
  \center
    \center
  \includegraphics[width=0.33\linewidth]{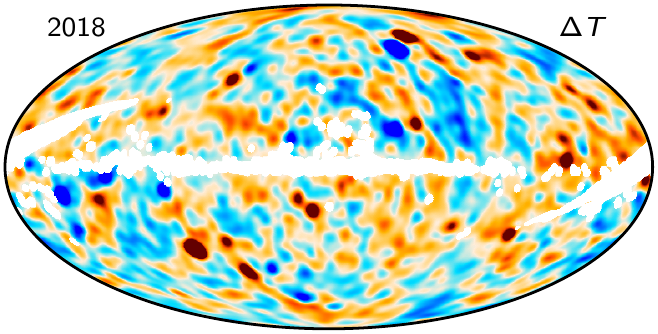}
  \includegraphics[width=0.33\linewidth]{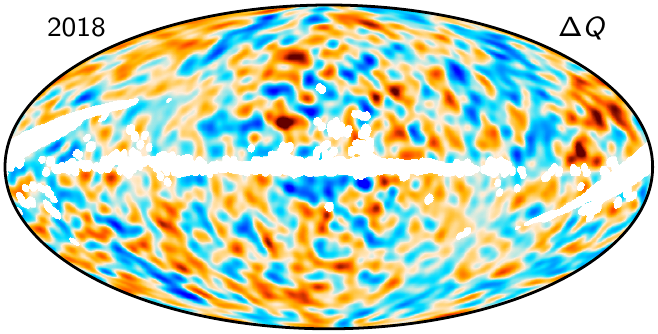}
  \includegraphics[width=0.33\linewidth]{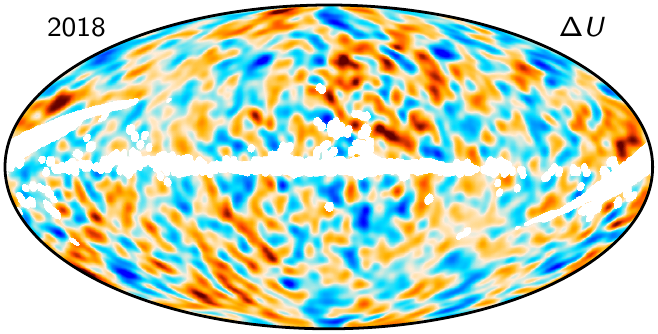}\\
  \includegraphics[width=0.33\linewidth]{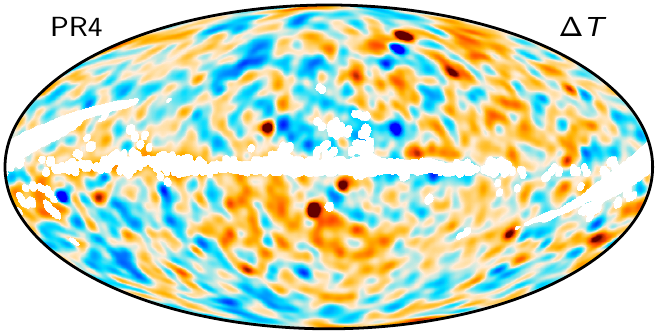}
  \includegraphics[width=0.33\linewidth]{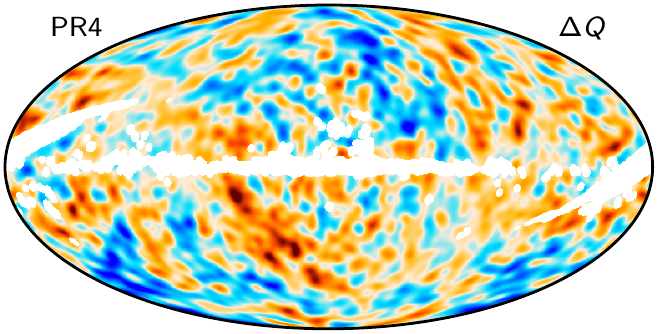}
  \includegraphics[width=0.33\linewidth]{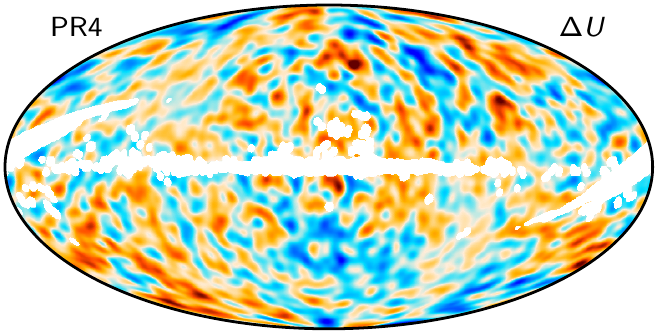}\\
  \includegraphics[width=0.33\linewidth]{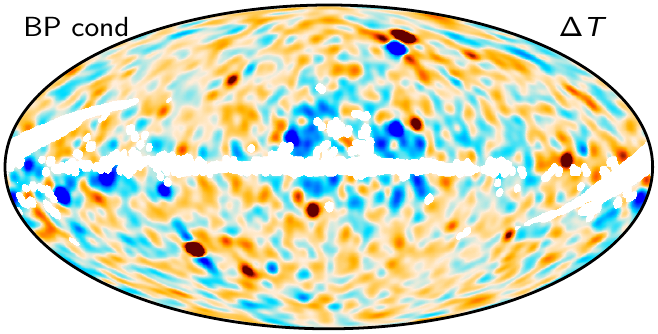}
  \includegraphics[width=0.33\linewidth]{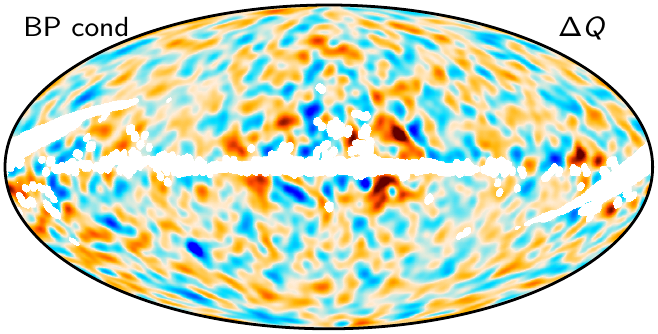}
  \includegraphics[width=0.33\linewidth]{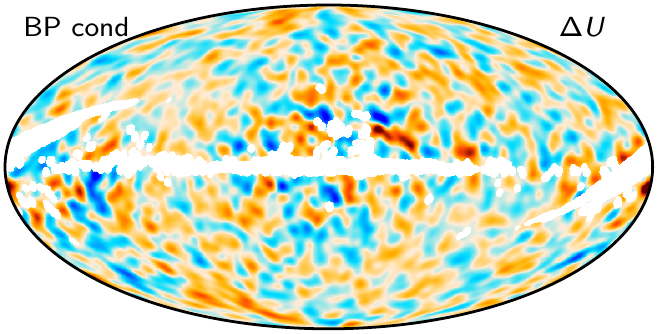}\\
  \includegraphics[width=0.33\linewidth]{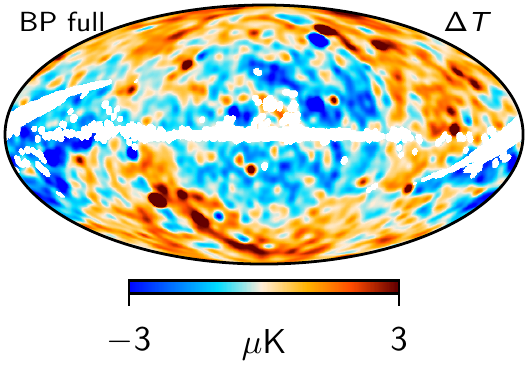}
  \includegraphics[width=0.33\linewidth]{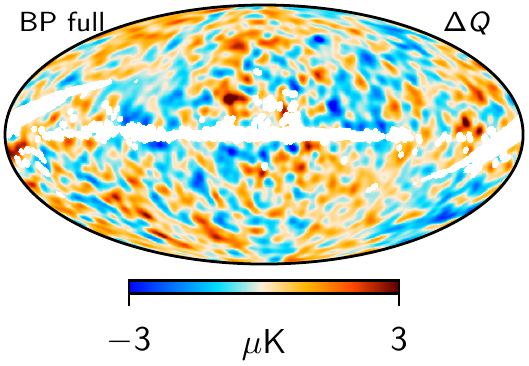}
  \includegraphics[width=0.33\linewidth]{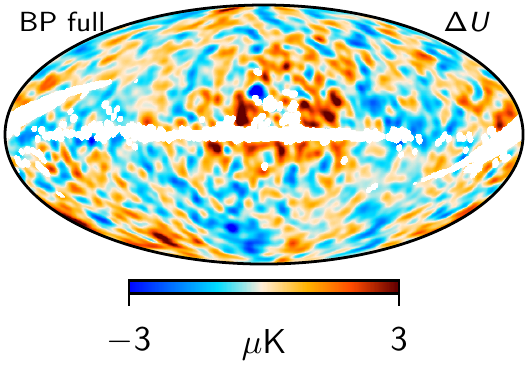}\\
  \caption{Half mission difference maps for the 30\,GHz channel. The top row shows the \Planck\ data release 3 maps, while the second top row shows the difference for the \Planck\ data release 4 maps. The bottom two rows show BP maps.
  The third row shows the first sample of a conditional run on astrophysical and instrumental parameters. The bottom row shows the difference between the half mission maps made from two independent runs that sample over the full model. The latter maps are averaged over 50 samples.}
  \label{fig:halfdiff-30}
\end{figure*}

In the traditional approach, it is most common to primarily account
for $1/f$-type correlated noise, and the matrix may then be
constructed by summing up the temporal two-point correlation function
between any two sample pairs separated by some maximum correlation
length. This calculation is implementationally straightforward, but
expensive. In addition, an overall multiplicative calibration
uncertainty may be added trivially, and spatially fixed template
corrections (for instance for foregrounds or gain uncertainties;
\citealp{planck2016-l02}) may be accounted for through the
Sherman-Morrison-Woodbury formula.

This analytical approach, however, is not able to account for more
complex sources of uncertainty, perhaps most notably gain
uncertainties that vary with detector and time \citep{bp07}, but also
more subtle effects such as sidelobe \citep{bp08} or bandpass
uncertainties \citep{bp09}, all of which contribute to the final total
uncertainty budget. To account for these, end-to-end simulations --
whether using random or constrained realizations -- are
essential. The resulting covariance matrix may then be constructed
simply by averaging the pixel-pixel outer product over all samples,
\begin{align}
  \hat{\m}_{\nu} &= \langle\m^i_{\nu}\rangle \\
  \N_{\nu} &= \left\langle\left( \m^i_{\nu}-\hat{\m}_{\nu} \right)
  \left( \m^i_{\nu}-\hat{m}_{\nu} \right)^t\right\rangle.
  \label{eq:covmat}
\end{align}
In practice, we also follow \Planck\ 2018, and add $2\,\muK$
($0.02\,\muK$) Gaussian regularization noise to each pixel in
temperature (polarization), in order to stabilize nearly singular
modes that otherwise prevent stable matrix inversion and determinant
evaluations.

\begin{figure*}

  \includegraphics[width=\linewidth]{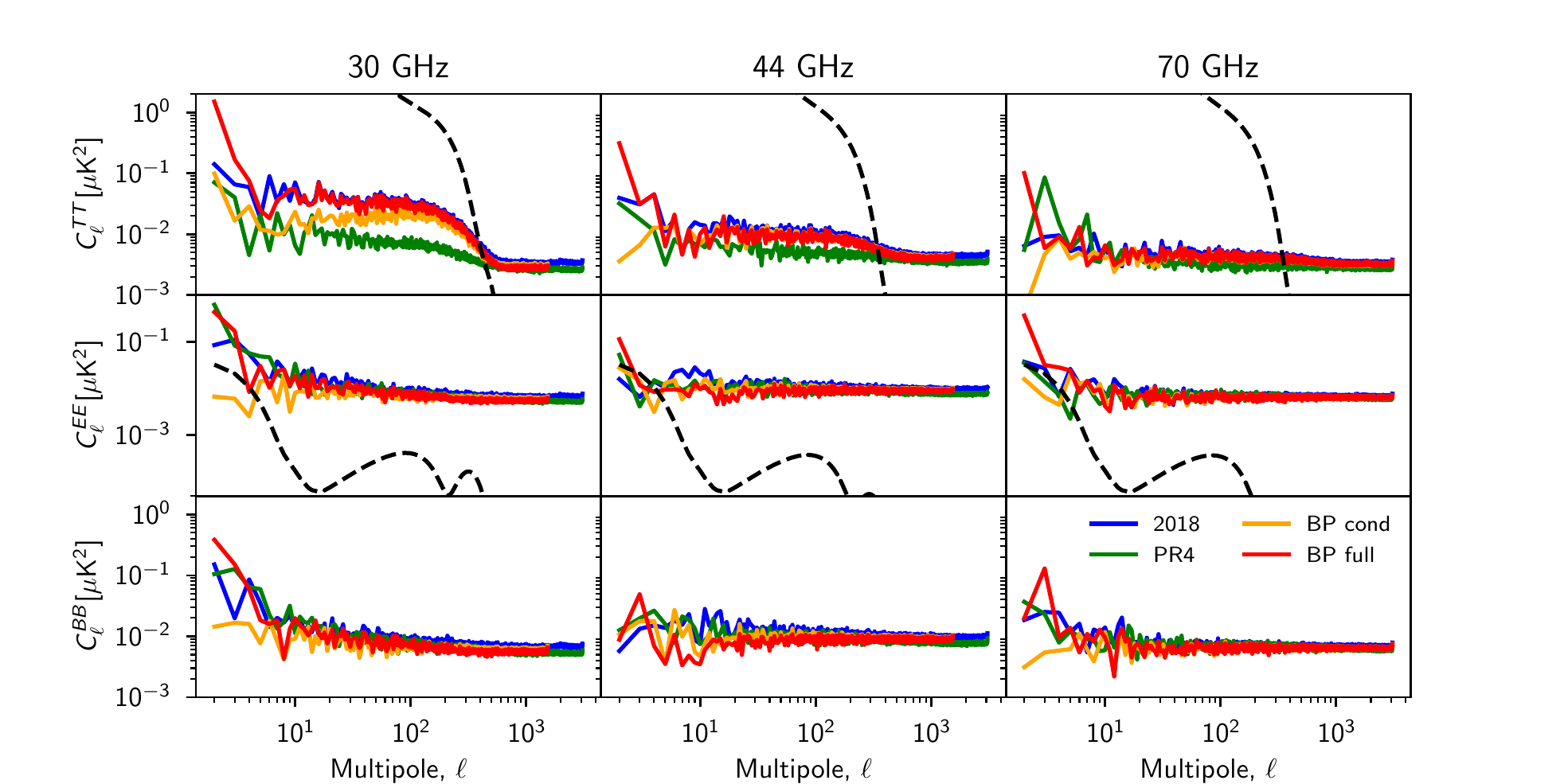}
  \caption{Half-mission half-difference power spectra estimated by \Planck\ 2018 (\emph{blue}), \Planck\ PR4 (\emph{green}), and \BP. In the case of \BP, orange curves show spectra obtained from maps that are calibrated jointly (or ``conditionally'' with respect to the signal model), while red curves show spectra from maps that are processed completely independently. Columns show 30, 44, and 70\,GHz results, while rows show $TT$, $EE$, and $BB$ spectra. The dashed black line shows the best-fit \Planck\ 2018 $\Lambda$CDM spectrum convolved with the instrument beam.  }
  \label{fig:halfdiff_powspec}  
\end{figure*}

The importance of complete error propagation is visually illustrated
in Fig.~\ref{fig:ncov_highlat}, which compares an arbitrarily selected
slice through the \Planck\ 2018 LFI low-resolution covariance matrix
that accounts only for $1/f$ correlated noise with the corresponding
\BP\ covariance matrix \citep{planck2016-l05}. (Note that the two
matrices are evaluated at different pixel resolution, and that the
\Planck\ 2018 matrix applies a cosine apodization filter that is not
used in \BP. Also, the $TP$ cross-terms at 70\,GHz are set to zero by
hand in the \Planck\ covariance matrices.)

When comparing these slices, two obvious qualitative differences stand
out immediately. First, the \BP\ matrices appear noisy, while the
\Planck\ appear smooth; this is because the former are constructed by
Monte Carlo sampling, while the latter are constructed
analytically. In practice, this means that any application of the
\BP\ covariance matrices must be accompanied with a convergence
analysis that shows that the number of Monte Carlo samples is
sufficient to reach robust results for the final statistic in
question. How many this is will depend on the statistic in question;
for an example of this as applied to estimation of the optical depth
of reionization, see \citet{bp12}. Second, and even more importantly,
we also see that the \BP\ matrices are far more feature rich than the
\Planck\ 2018 matrices, and this is precisely because they account for
a full stochastic data model, and not just $1/f$ correlated noise. 

This is most striking for the 30\,GHz channel, for which it is in fact
nearly impossible to see the $1/f$ imprint at all. Rather, the slice
is dominated by a large-scale red-blue quadrupolar pattern aligned
with the Solar dipole, and this is an archetypal signature of
inter-detector calibration differences \citep[see,
  e.g.,][]{planck2016-l02,bp07}. A second visually striking effect is
the Galactic plane, which is due to bandpass mismatch uncertainties
\citep{bp09}. There are of course numerous other effects also present
in these maps, although these are generally harder to identify
visually.

Figure~\ref{fig:noise_ps} shows the angular noise power spectra
corresponding to these matrices (black for \BP, and red for
\Planck\ 2018) in both temperature and polarization. For comparison,
this plot also includes similar spectra computed from the end-to-end
\Planck\ PR4 simulations (blue). The dashed line shows the pure white
noise contribution in the \BP\ covariances, while the dotted line
shows the best-fit \Planck\ 2018 $EE$ $\Lambda$CDM spectrum.

We first note that the temperature component of these covariance
matrices is heavily processed by first convolving to $20^{\circ}$
FWHM, and then adding uncorrelated regularization noise of
$2\,\muK$/pixel; this explains why all curves converge above
$\ell\approx10$ in temperature. Secondly, we note that \Planck\ PR4
did not independently re-estimate the LFI $1/f$ parameters, but rather
adopted the \Planck\ 2018 values for this. The difference between the
blue and red curves (after taking into account the 15\,\% lower white
noise level in PR4) thus provides a direct estimate of the sum of other
effects than correlated noise in \Planck\ 2018, most notably gain
uncertainties.

Overall, the \BP\ polarization noise spectra agree well with
\Planck\ PR4 at 30\,GHz, while they are generally lower by as much as
20--50\,\% at 44 and 70\,GHz. In fact, we see that the \BP\ 70\,GHz
noise spectrum almost reaches the white noise floor around
$\ell\approx 20$, suggesting that the current processing has succeeded
in removing both excess correlated noise and gain fluctuations to an
unprecedented level.

\subsection{Half-mission split maps}

Before concluding this section, we also consider error propagation by
half-mission split maps, as such maps have been used extensively by
all generations of the \Planck\ pipelines
\citep{planck2013-p01,planck2014-a01,planck2016-l01}. These maps are
generated by dividing the full data set into two disjoint sets,
typically either by splitting each scanning period in two (resulting
in half-ring maps); or by dividing the entire mission into two halves
(resulting in half-mission maps); or by dividing detectors into
separate groups (resulting in detector maps). The goal of each of
these splits is the same, namely to establish maps with similar signal
content, but statistically independent noise realizations. The
resulting maps may then be used for various cross-correlation
analyses, including power spectrum estimation
\citep[e.g.,][]{planck2016-l05}.

In general, the usage of cross-correlation techniques is an implicit
admission that our understanding of the instrumental systematic
effects is incomplete, and the main goal of the current
\BP\ processing is precisely to establish a statistically adequate
error propagation model. As such, the primary CMB results presented by
\citet{bp11} and \citet{bp12} do not employ cross-correlation techniques at all,
but rather rely fully on statistically optimal auto-spectrum
estimation, in which all information is fully exploited. The current
section is therefore included primarily for comparison purposes with
respect to the original \Planck\ processing.

Before presenting half-mission maps from \BP, we note that
\Planck\ 2018 and PR4 adopted different conventions for how
to split the mission into two halves. Specifically, while \Planck\ 2018
simply divided the mission in two, and made separate maps for years
1+2 and 3+4, \Planck\ PR4 instead chose to co-add years 1+3 and
2+4. The main reason for this is that years 3+4 do not result in 
full-sky coverage for the LFI maps, but leaves a small hole, which is
awkward for cross-correlation analyses. The cost of this choice,
however, is slightly less independent halves, which may lead to
additional common modes. An important example of this is asymmetric
beams; as discussed by \citet{planck2014-a01}, the scanning phase of
the \Planck\ satellite was reversed between years 2 and 3, and this
adds additional beam symmetrization between the first and second half
of the mission. Therefore, the effect of beam asymmetries is
maximized in the \Planck\ 2018 split, but minimized in the
\Planck\ PR4 split. In this paper, we choose to follow \Planck\ 2018,
and focus on years 1+2 versus 3+4 splits.

Technically speaking, half-mission maps may be generated very
straightforwardly in the \BP\ pipeline at the parameter file level
\citep{bp03}, simply by restricting the start and end PID of the
analysis. In this case, all parameters, including the calibration and
foreground model, will be fitted independently in each data set. We
generate such half-mission samples for each frequency channel and
compute pairwise differences between these. This set of half-mission
maps are denoted ``full'' in the following, indicating that the full
model is fitted in each half-mission set.

However, it is important to note that this mode differs significantly
from the official \Planck\ implementation. Rather, in the
\Planck\ 2018 case, both the calibration and foreground model are
fixed at their full-mission values, while in the case of \Planck\ PR4,
the foreground model is fixed, and the calibration model is
re-estimated. To mimic the \Planck\ 2018 behaviour, we therefore
create a second set of \BP\ half-mission maps, in which all model
parameters except the correlated noise are fixed at their full mission
values. We refer to this set as ``conditional'' (or ``cond'') in the
following, indicating that most parameters are in fact conditioned
during generation.

Figure~\ref{fig:halfdiff-30} shows half-mission half-difference maps
($\m_{\mathrm{hmhd}} = (\m_{1}-\m_2)/2$) for the LFI 30\,GHz channel
for all four cases, all smoothed to a common angular resolution of
$5^{\circ}$. (Similar plots for the 44 and 70\,GHz channels look
qualitatively similar, and is omitted for brevity.) Starting with the
two bottom rows that show \BP\ ``cond'' and ``full'', respectively, we
see that conditioning on the calibration and foreground parameters
has a substantial impact in terms of overall variations. The typical
large-scale fluctuation level is at least a factor of two larger when
fitting all parameters freely, and there is significantly more
pronounced coherent large-scale features. We also see that these two
cases (at least in temperature) bound the \Planck\ 2018 and PR4 cases,
in the sense that the fluctuations generally increase as more and more
parameters are re-fitted; the main difference between \Planck\ PR4 and
\BP\ is a large-scale quadrupole pattern, which looks very much like
the familiar bandpass residuals are seen by both \Planck\ 2018 and PR4
\citep{npipe}, and may originate from the foreground and/or beam
differences.

In polarization, this trend is less obvious, as in this case the
absolute magnitude of the pipeline-specific residuals is comparable
with the level from parameter conditioning. In this case, the
\BP\ ``cond'' case shows clearly smaller differences than either of the
two \Planck\ implementations, while the ``full'' case shows comparable
levels. 

The sky maps shown in Fig.~\ref{fig:halfdiff-30} emphasize the very
largest angular scales. To obtain additional information regarding
intermediate and small scales, we show in
Fig.~\ref{fig:halfdiff_powspec} the auto-spectra of each case. For
comparison, the dashed curve shows the beam-convolved best-fit
$\Lambda$CDM spectrum for each channel. Starting with the
noise-dominated polarization spectra, we see that all four pipelines
result in roughly similar half-mission power, and this indicates that
their abilities to propagate and describe instrumental noise alone are
very similar. For temperature, the picture is qualitatively different,
and there is a strong excess from the sky signal, most notably at
30\,GHz, that appears to fall off with the beam smoothing. The fact
that \Planck\ 2018 and \BP\ appears nearly identical in this case,
while \Planck\ PR4 is substantially lower, strongly suggests that
asymmetric beams can account for at least some of this additional
power. At the same time, the fact that \BP\ ``full'' and ``cond''
differ shows that gain fluctuations also play a significant role in
explaining the full excess. However, the main conclusion from these
results is that \Planck\ 2018 and \BP\ appear to perform similarly in
terms of half-mission power on small scales, and this strongly
suggests that both pipelines are likely to be close to the theoretical
noise limit. 


\section{Systematic error corrections and uncertainties}
\label{sec:systematics}

A significant strength of the novel Bayesian end-to-end framework is
that all TOD contributions are modelled in terms of specific and
physically motivated parametric models, and there are very few ``black
boxes'' that can hide non-understood contaminants; any contribution is
either present in one of the parametric components, or it shows up in
the TOD residual maps. This physical foundation is useful both when
interpreting the results, and when debugging and tuning the analysis
configuration. In this section, we provide a concrete example of this
by presenting a complete survey of each systematic correction term,
both in the form of projected sky maps and as residual power spectra,
for all three LFI channels.

\subsection{Sky map corrections}

\begin{figure*}[p]
  \center
  \includegraphics[width=0.265\linewidth]{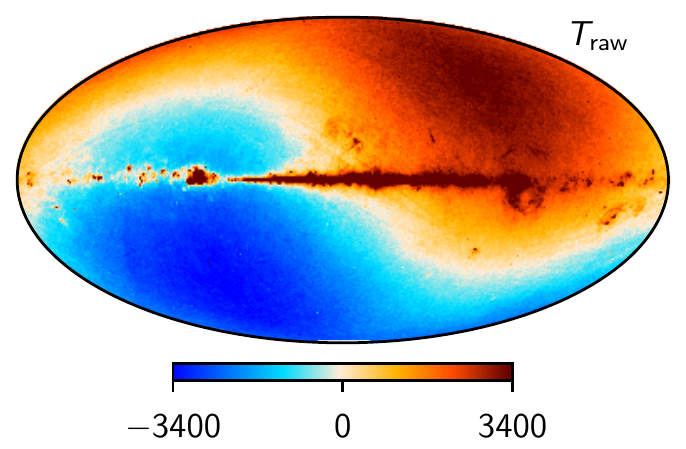}
  \includegraphics[width=0.265\linewidth]{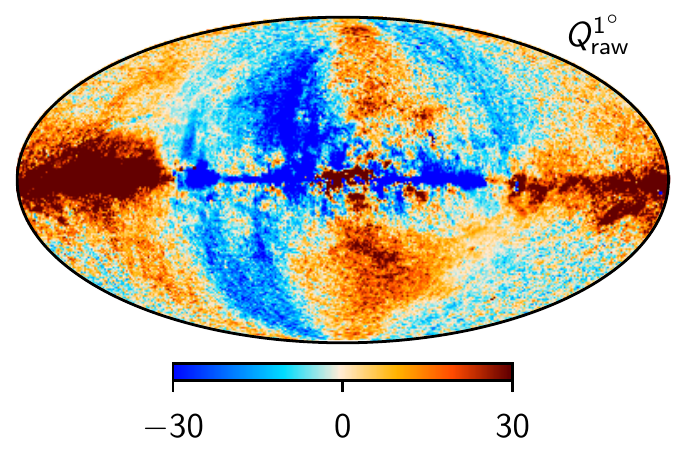}
  \includegraphics[width=0.265\linewidth]{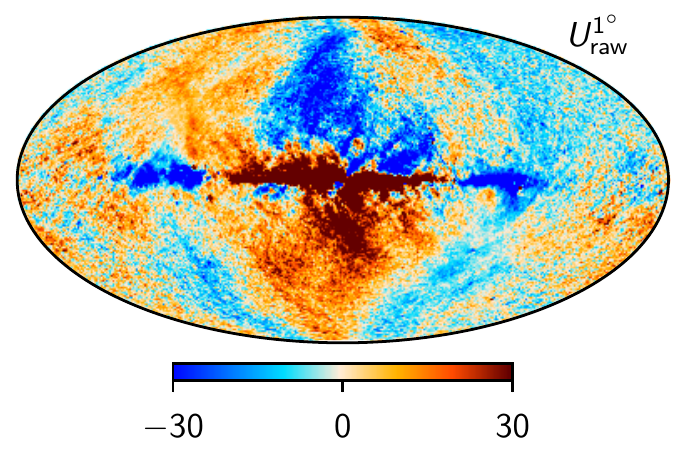}\\\vspace*{3mm}
  \includegraphics[width=0.265\linewidth]{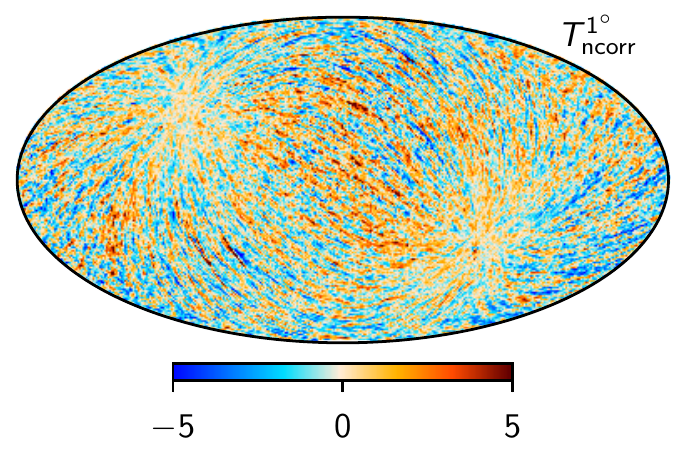}
  \includegraphics[width=0.265\linewidth]{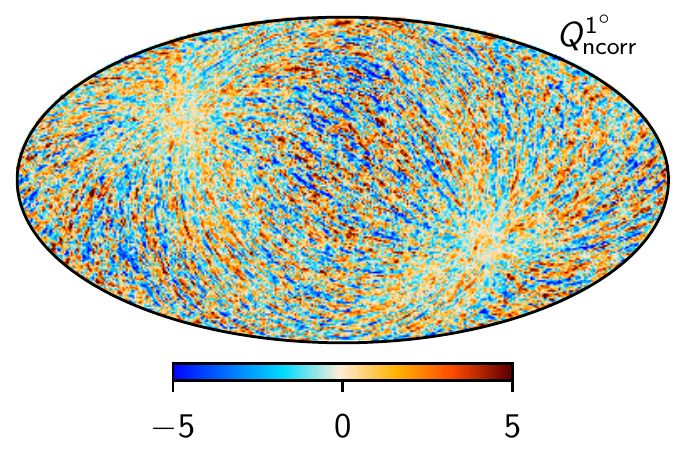}
  \includegraphics[width=0.265\linewidth]{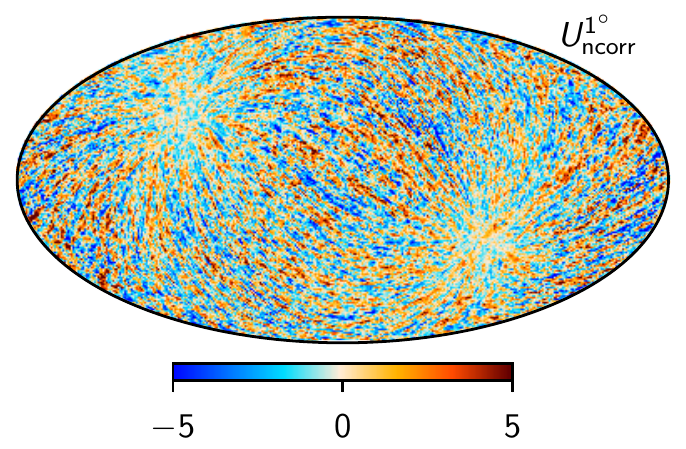}\\
  \includegraphics[width=0.265\linewidth]{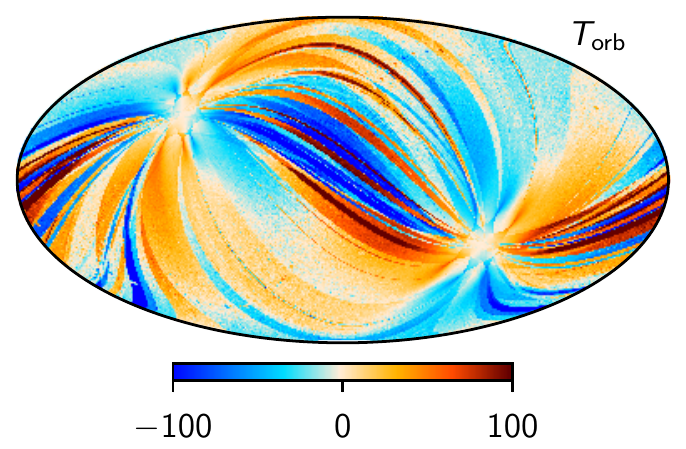}
  \includegraphics[width=0.265\linewidth]{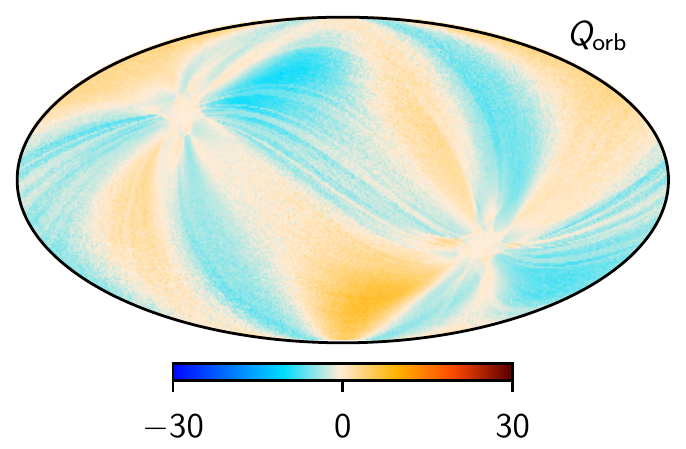}
  \includegraphics[width=0.265\linewidth]{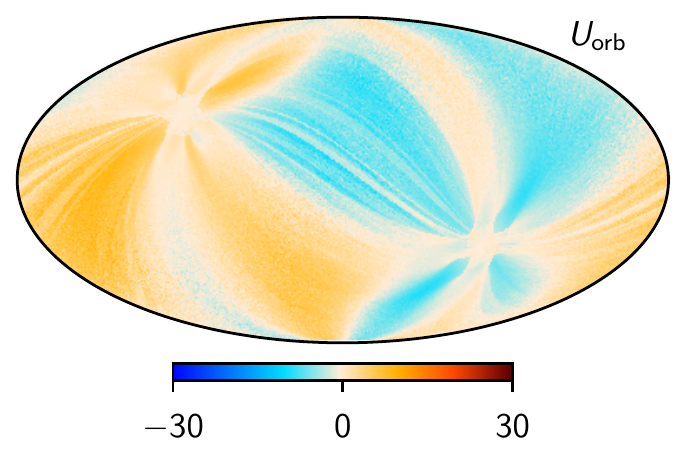}\\
  \includegraphics[width=0.265\linewidth]{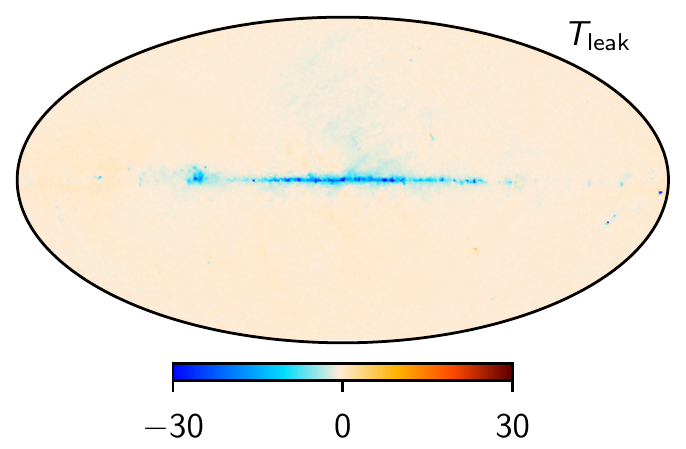}
  \includegraphics[width=0.265\linewidth]{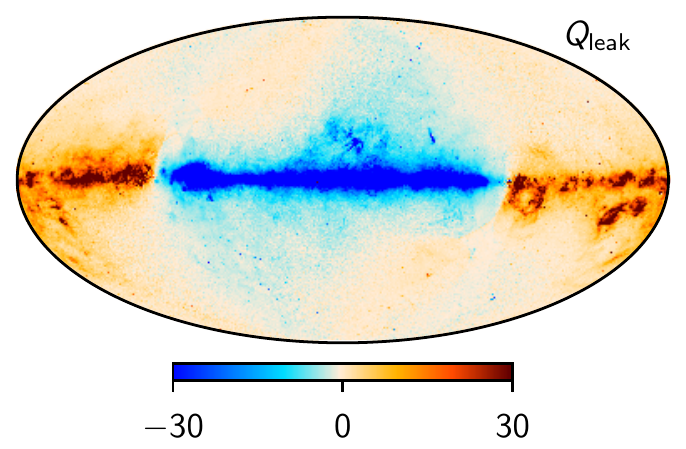}
  \includegraphics[width=0.265\linewidth]{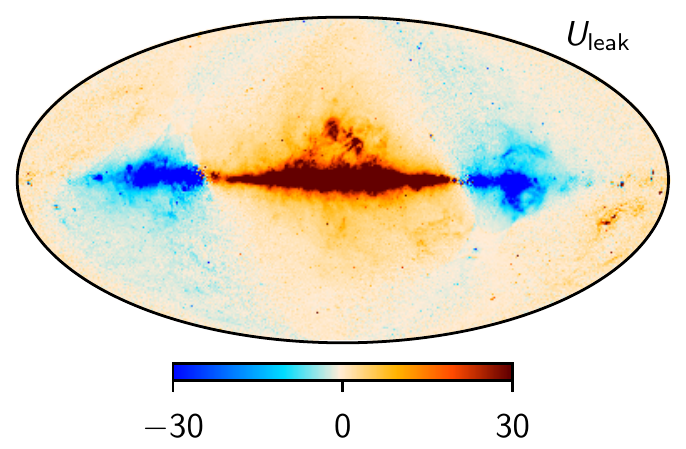}\\
  \includegraphics[width=0.265\linewidth]{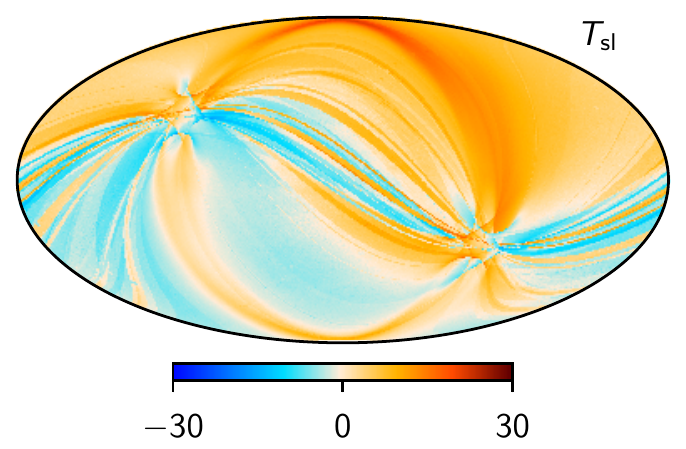}
  \includegraphics[width=0.265\linewidth]{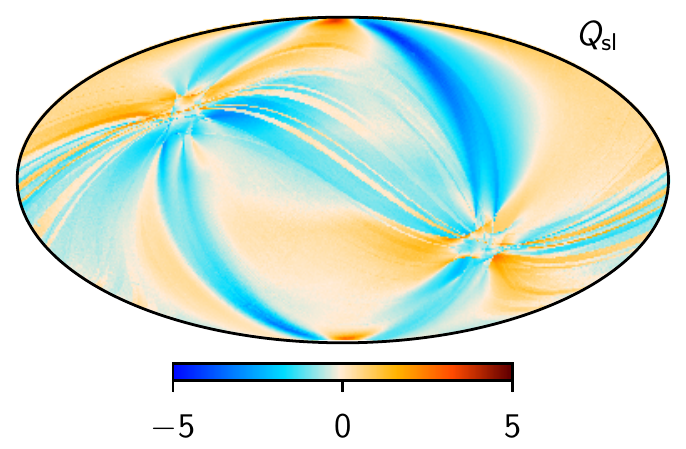}
  \includegraphics[width=0.265\linewidth]{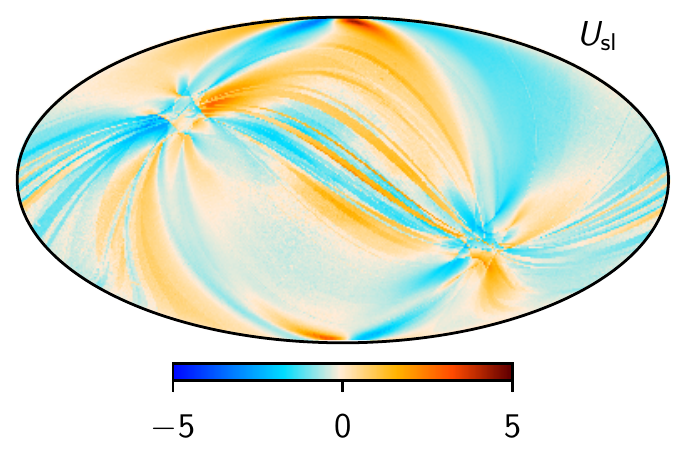}\\
  \includegraphics[width=0.265\linewidth]{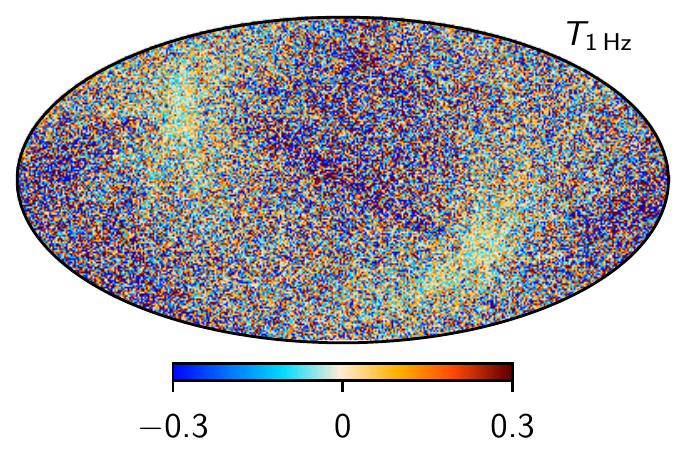}
  \includegraphics[width=0.265\linewidth]{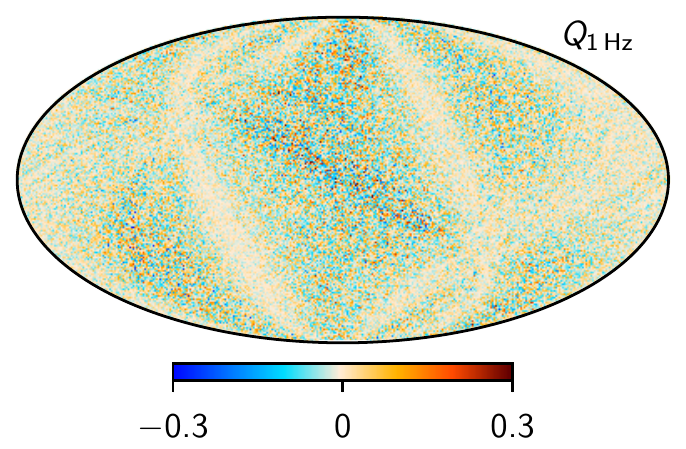}
  \includegraphics[width=0.265\linewidth]{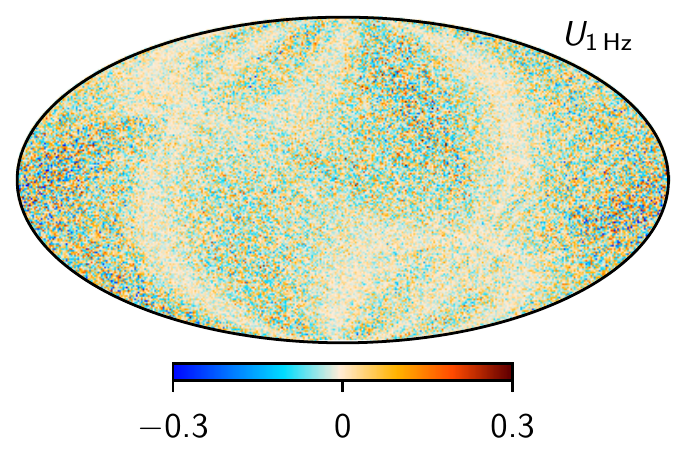}\\
  \includegraphics[width=0.265\linewidth]{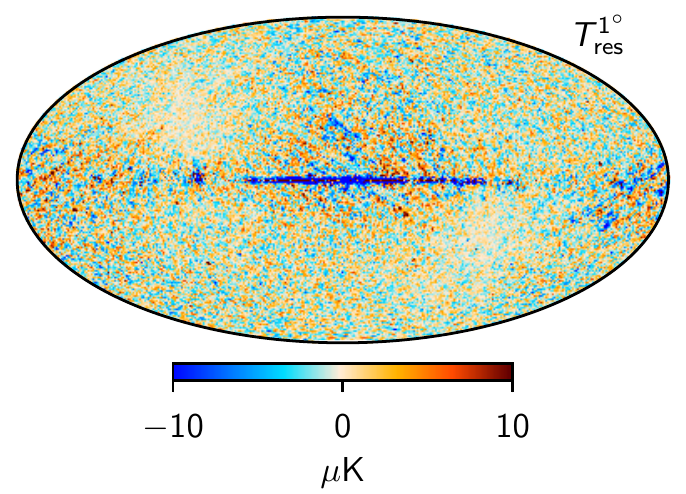}
  \includegraphics[width=0.265\linewidth]{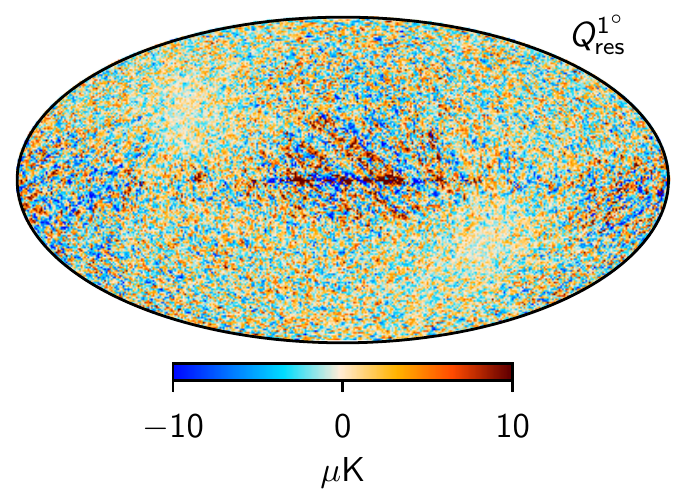}
  \includegraphics[width=0.265\linewidth]{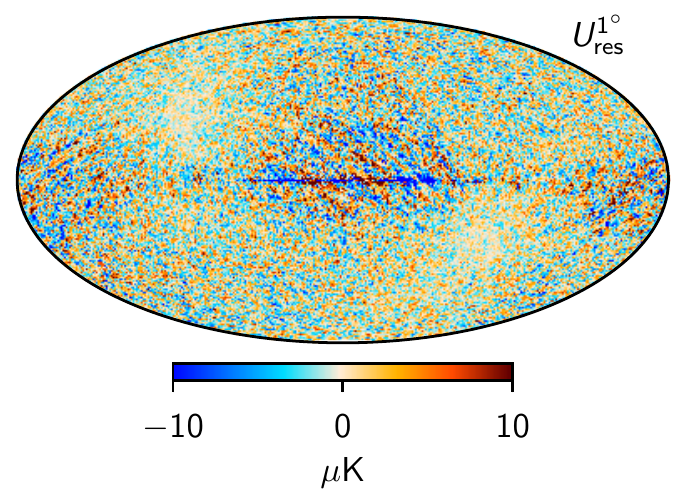}
  \caption{Comparison between TOD corrections for the 30\,GHz channel
    for a single Gibbs sample, projected into sky maps. Columns show
    Stokes $T$, $Q$, and $U$ parameters. Rows show, from top to
    bottom, 1) raw TOD; 2) correlated noise; 3) the orbital dipole; 4)
    bandpass and beam mismatch leakage; 5) sidelobe corrections; and
    6) 1\,Hz electronic spike correction. The bottom row shows the
    shows the residual obtained when binning the sky and systematics
    subtracted TOD into a sky map. Note that some components have been
    smoothed to an angular resolution of $1^{\circ}$ FWHM.  }\label{fig:corrmaps30}
\end{figure*}

We start by showing projected systematic effect sky maps at 30\,GHz, as
summarized in Fig.~\ref{fig:corrmaps30}. The top panel shows the raw
TOD binned into a sky map, and this provides intuition regarding the
overall quality of the data before applying any corrections. Indeed,
for the temperature component it is very difficult to spot major
artifacts of any kind; the most notable feature is a few correlated
noise stripes in the lower left quadrant. For polarization, the
dominant effect is the alternating sign along the Galactic plane as a
function of longitude, which is due to bandpass mismatch.

The second panel shows the correlated noise component. The most
notable features in this map are coherent stripes along the satellite
scanning path. It should also be noted that this component is the one
that is the least constrained from a priori considerations among all
TOD components, and therefore acts as a ``trash can'' for possible
unmodelled errors; this is the first place one expects to see
residuals from modelling errors. The fact that this appears
statistically clean, and has a morphology that is close to that
actually expected by $1/f$-type correlated noise provides very strong
evidence that the current processing has succeeded in cleaning the
30\,GHz channel with respect to most systematic effects.

\begin{figure*}[p]
  \center
  \includegraphics[width=0.265\linewidth]{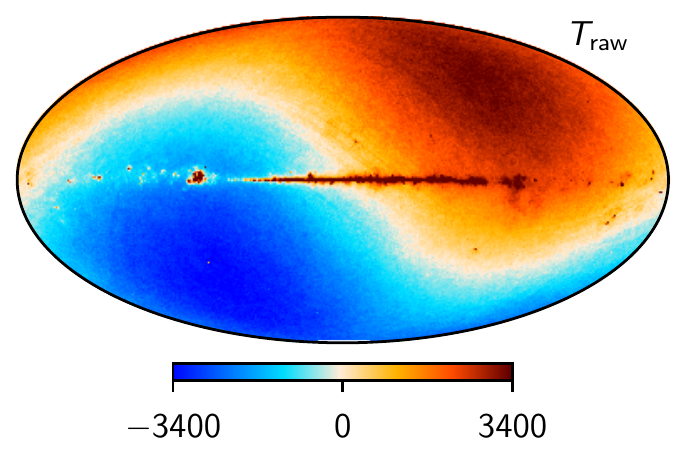}
  \includegraphics[width=0.265\linewidth]{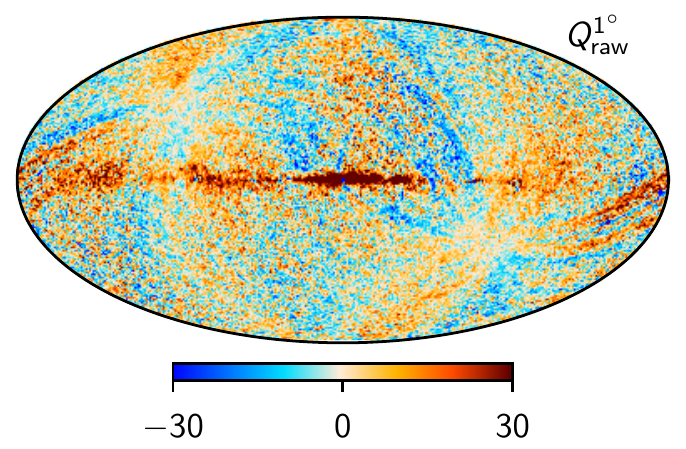}
  \includegraphics[width=0.265\linewidth]{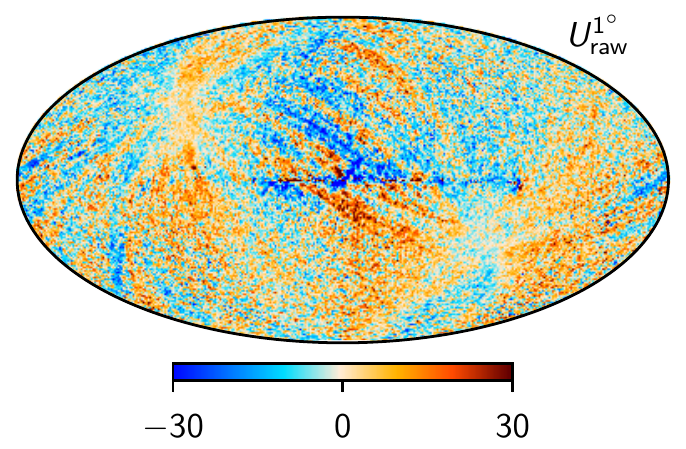}\\\vspace*{3mm}
  \includegraphics[width=0.265\linewidth]{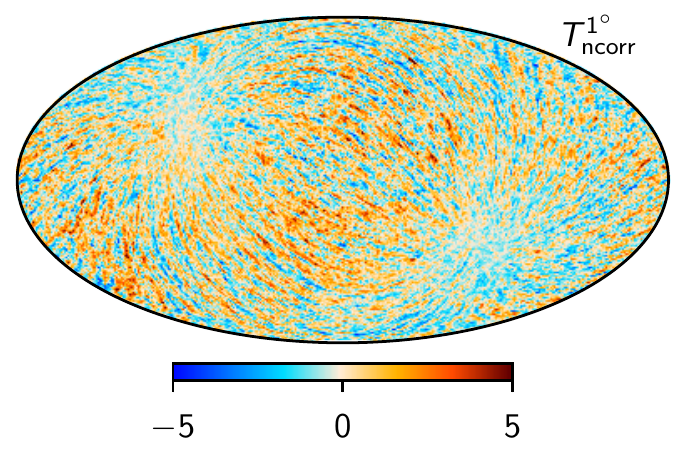}
  \includegraphics[width=0.265\linewidth]{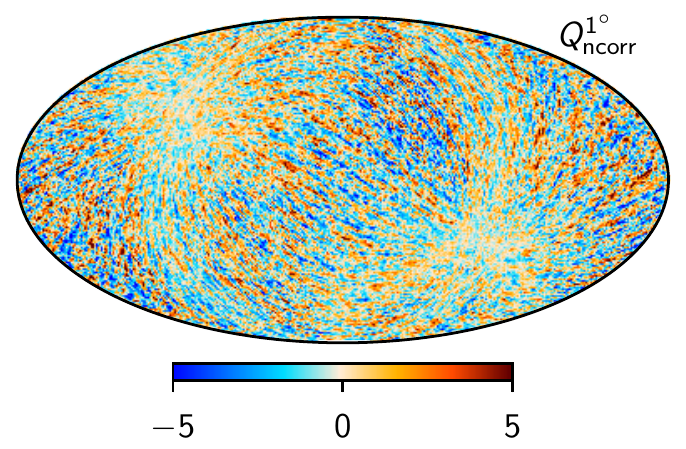}
  \includegraphics[width=0.265\linewidth]{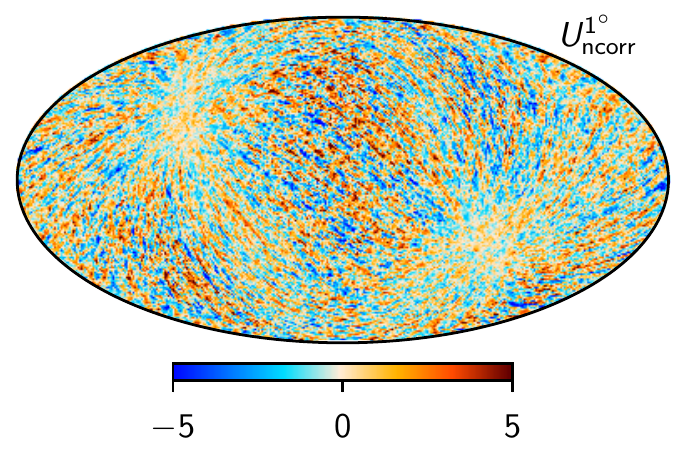}\\
  \includegraphics[width=0.265\linewidth]{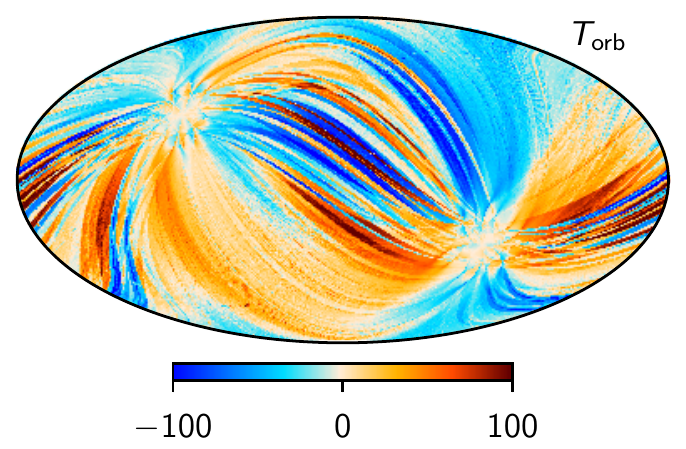}
  \includegraphics[width=0.265\linewidth]{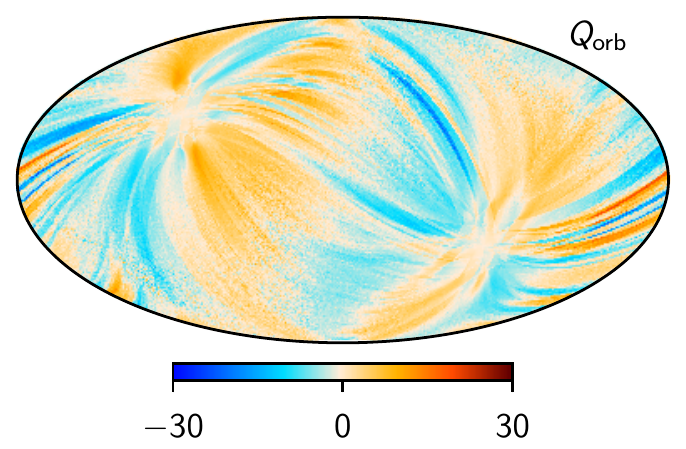}
  \includegraphics[width=0.265\linewidth]{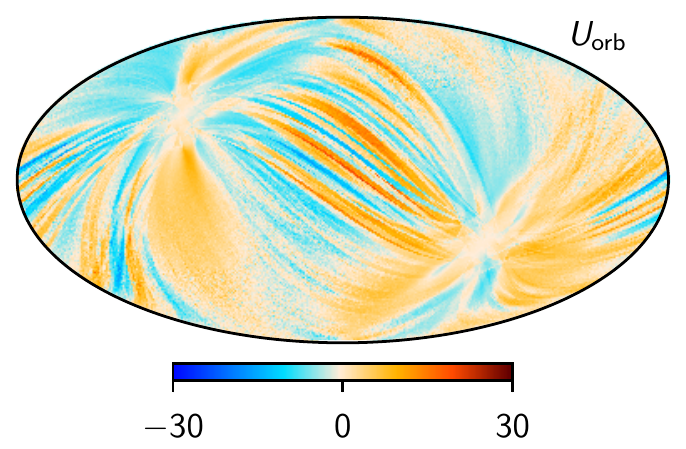}\\
  \includegraphics[width=0.265\linewidth]{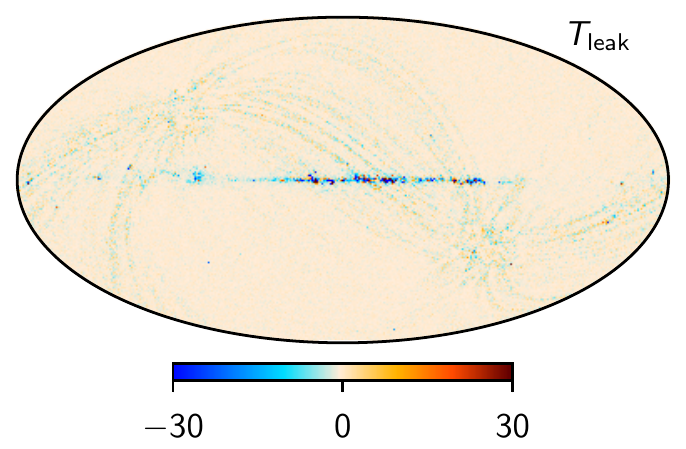}
  \includegraphics[width=0.265\linewidth]{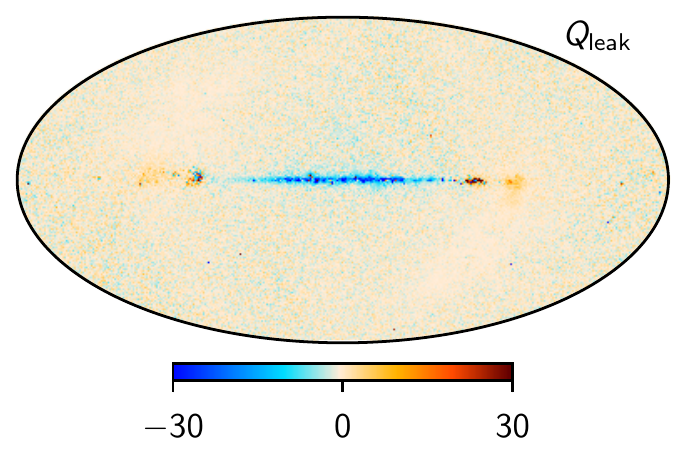}
  \includegraphics[width=0.265\linewidth]{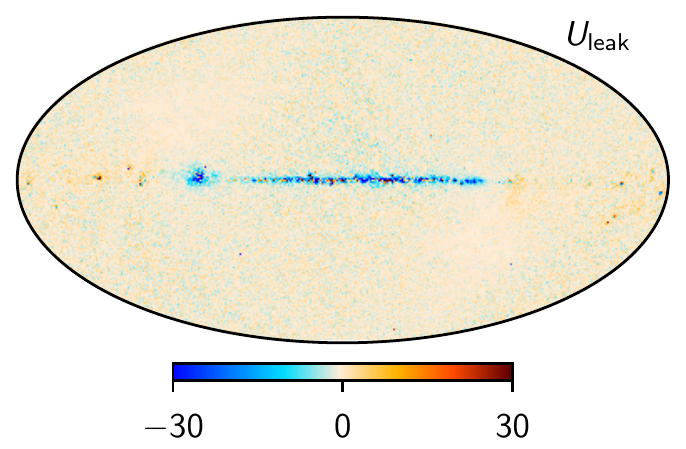}\\
  \includegraphics[width=0.265\linewidth]{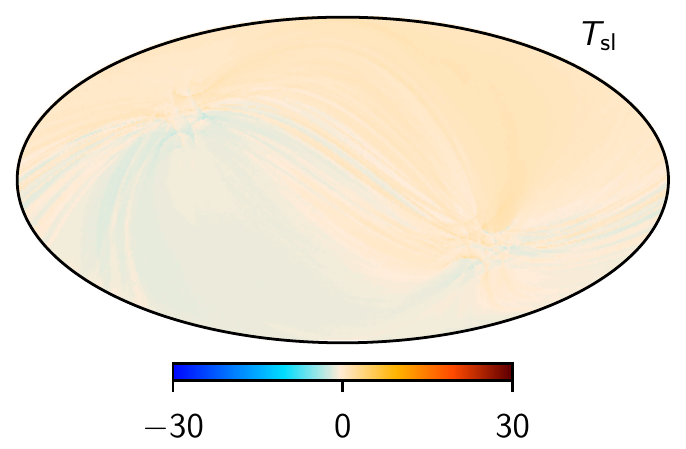}
  \includegraphics[width=0.265\linewidth]{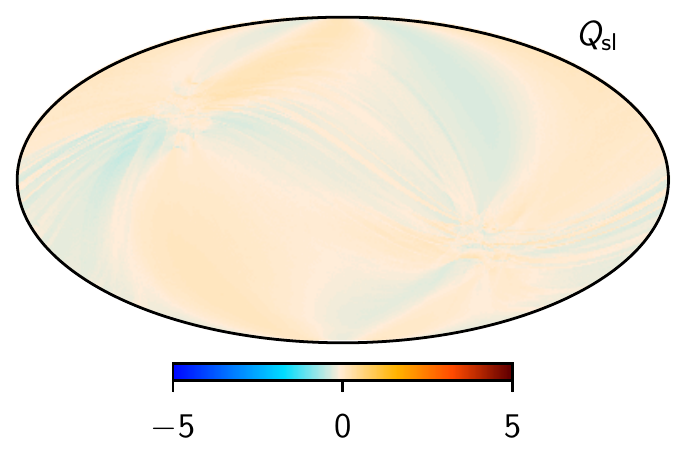}
  \includegraphics[width=0.265\linewidth]{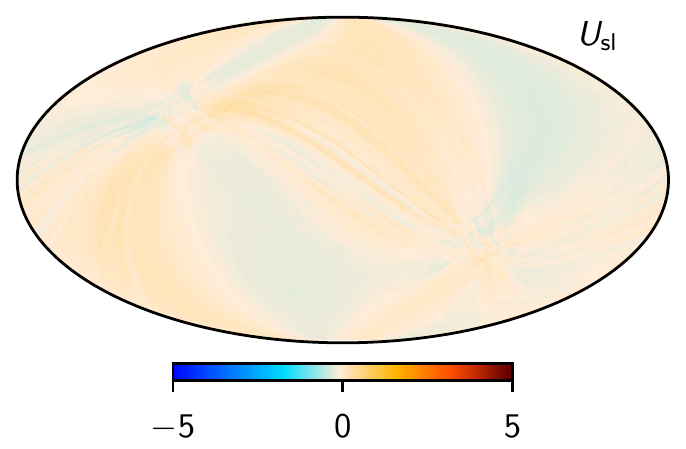}\\
  \includegraphics[width=0.265\linewidth]{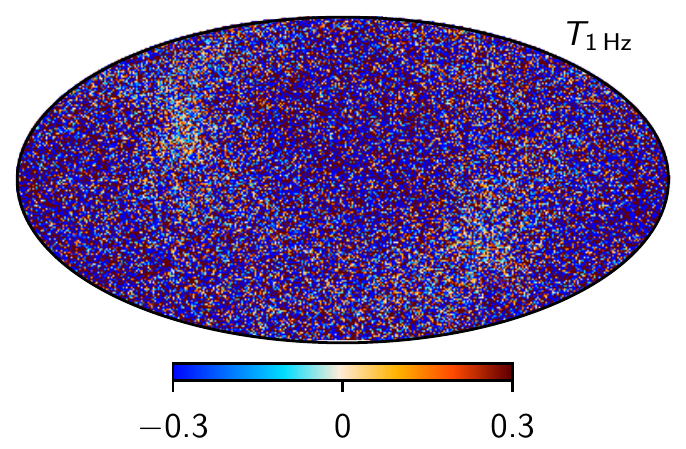}
  \includegraphics[width=0.265\linewidth]{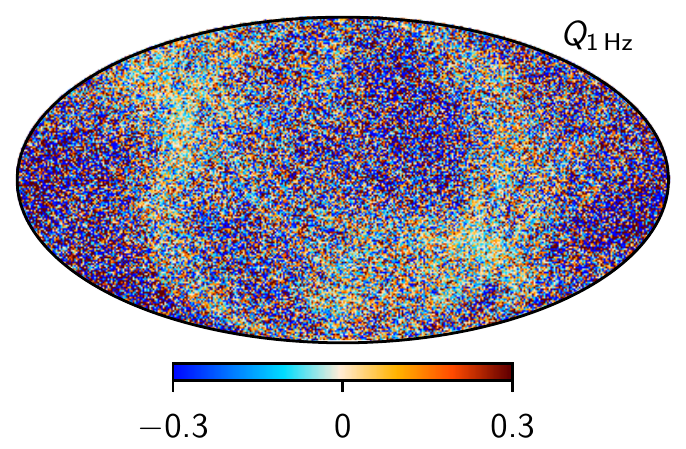}
  \includegraphics[width=0.265\linewidth]{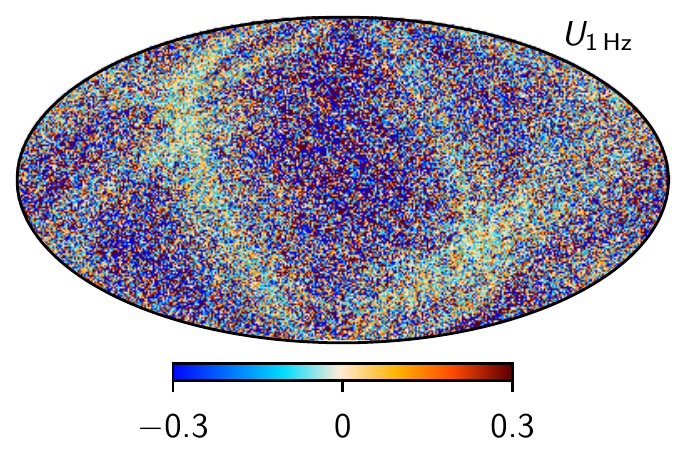}\\
  \includegraphics[width=0.265\linewidth]{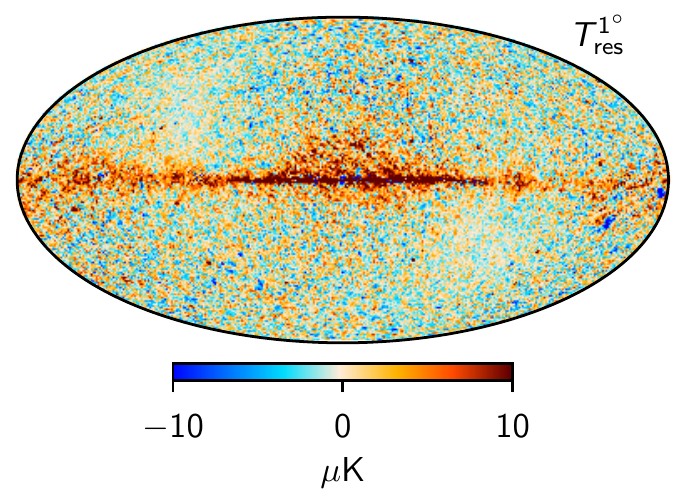}
  \includegraphics[width=0.265\linewidth]{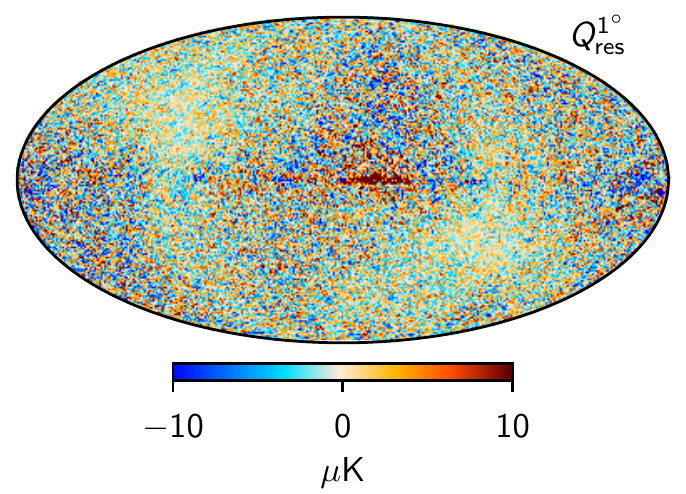}
  \includegraphics[width=0.265\linewidth]{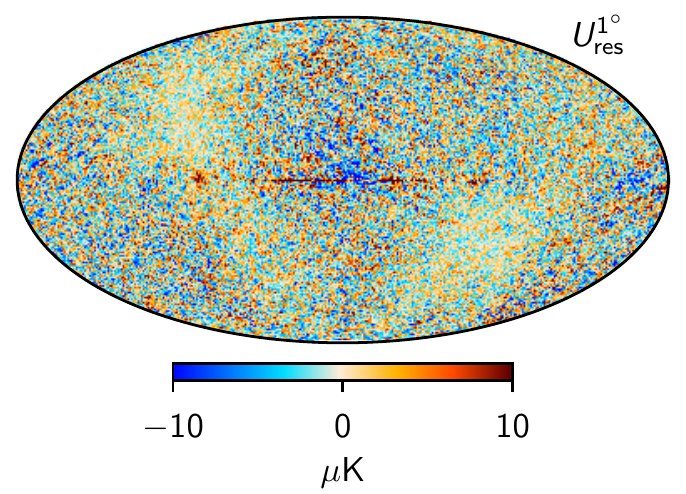}
  \caption{Same as Fig.~\ref{fig:corrmaps30}, but for the 44\,GHz channel.}
  \label{fig:corrmaps44}
\end{figure*}

\begin{figure*}[p]
  \center
  \includegraphics[width=0.265\linewidth]{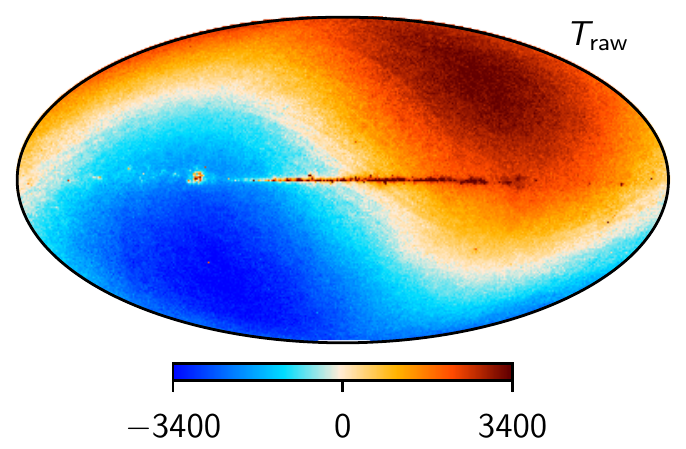}
  \includegraphics[width=0.265\linewidth]{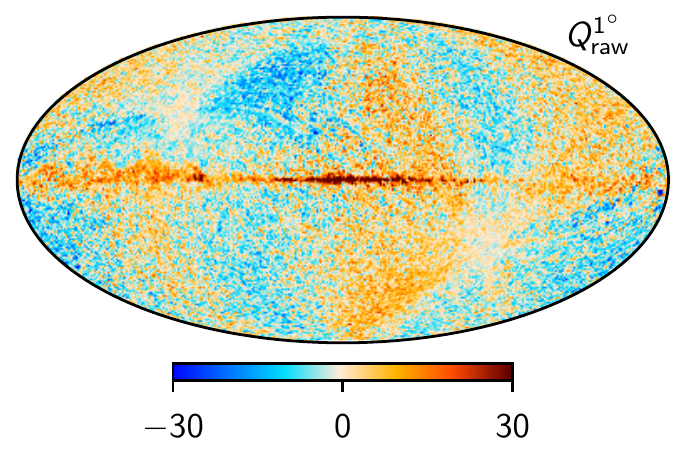}
  \includegraphics[width=0.265\linewidth]{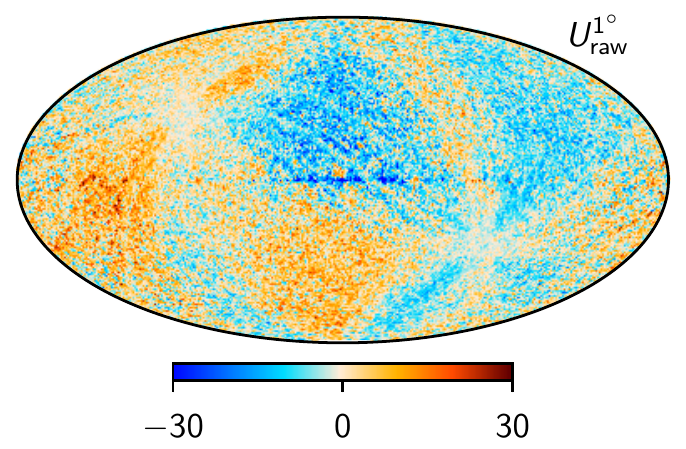}\\\vspace*{3mm}
  \includegraphics[width=0.265\linewidth]{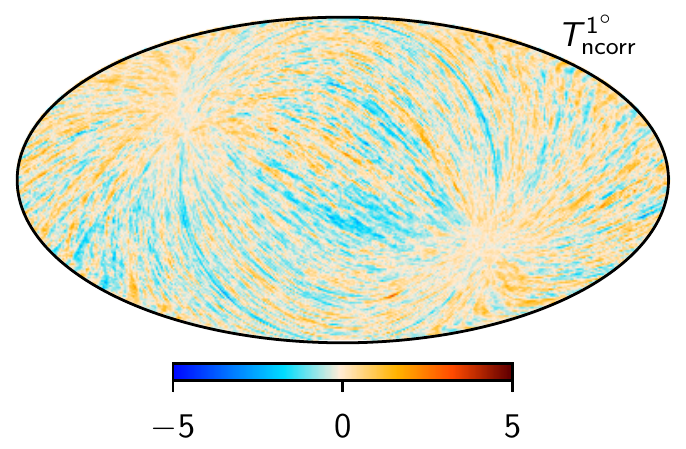}
  \includegraphics[width=0.265\linewidth]{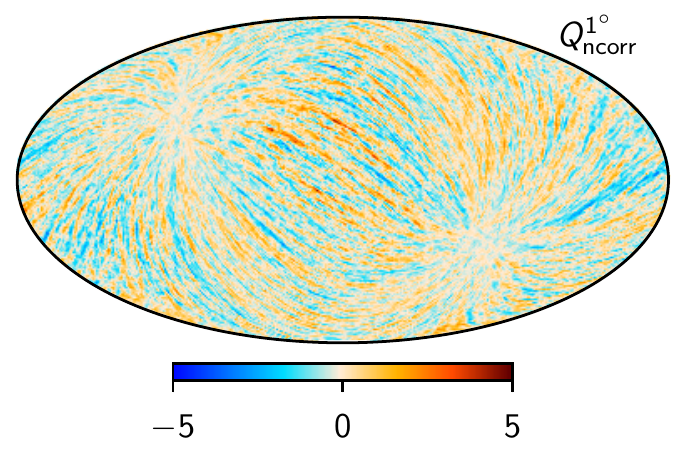}
  \includegraphics[width=0.265\linewidth]{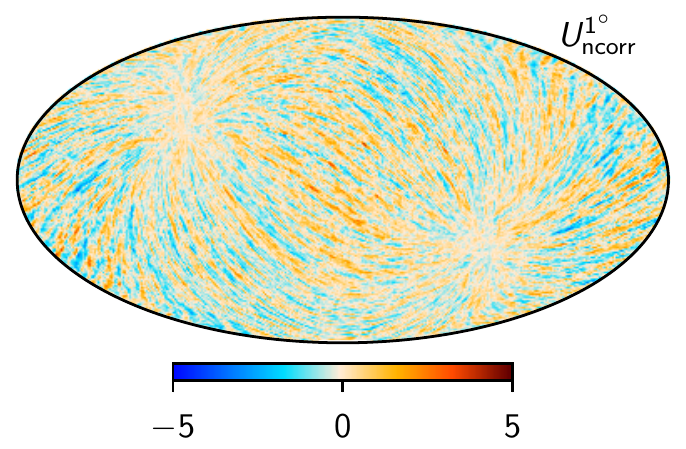}\\
  \includegraphics[width=0.265\linewidth]{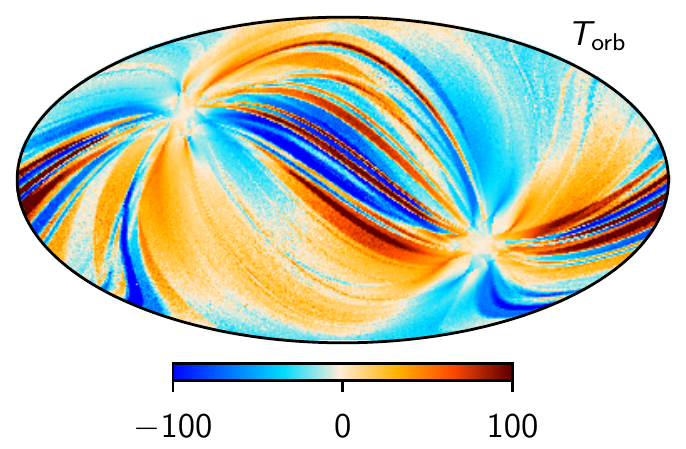}
  \includegraphics[width=0.265\linewidth]{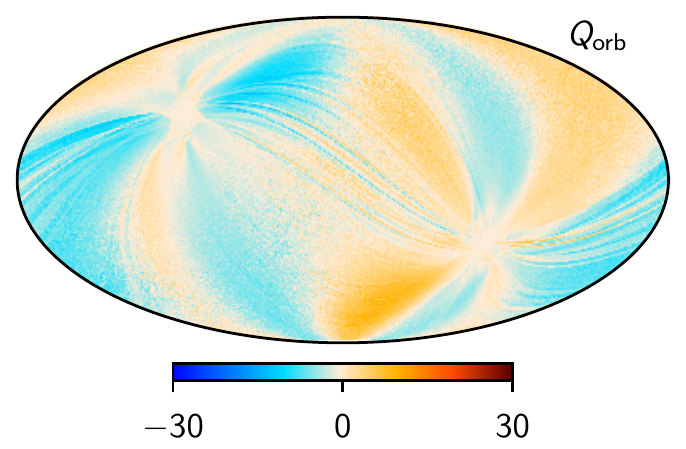}
  \includegraphics[width=0.265\linewidth]{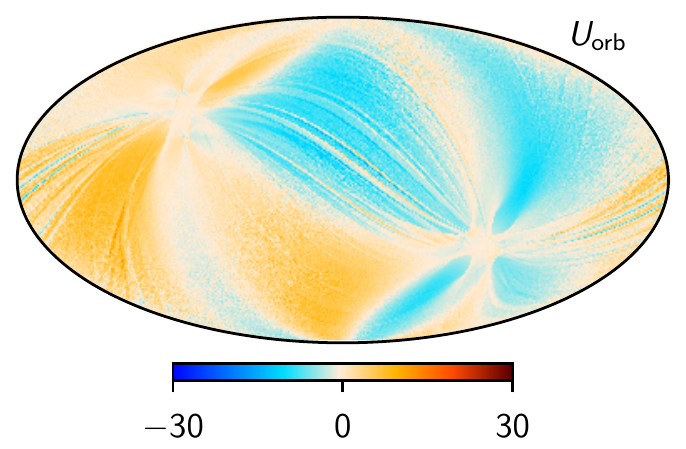}\\
  \includegraphics[width=0.265\linewidth]{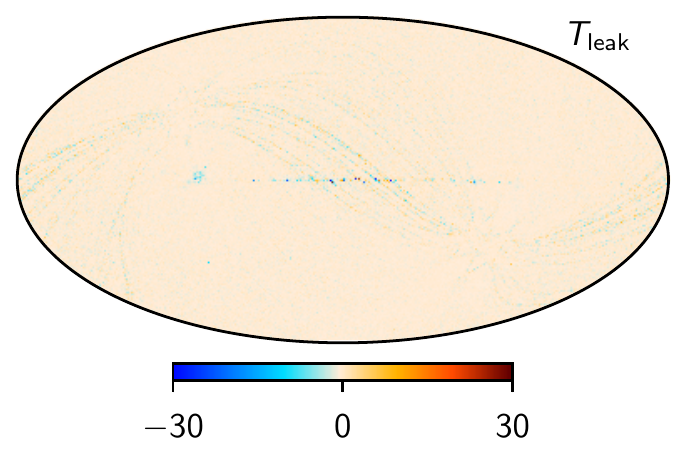}
  \includegraphics[width=0.265\linewidth]{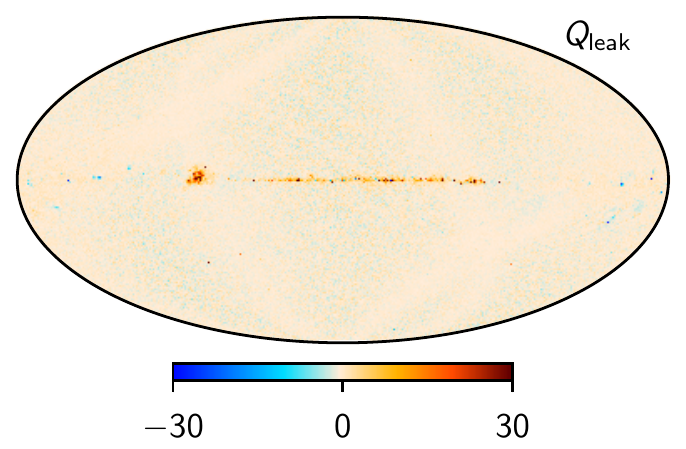}
  \includegraphics[width=0.265\linewidth]{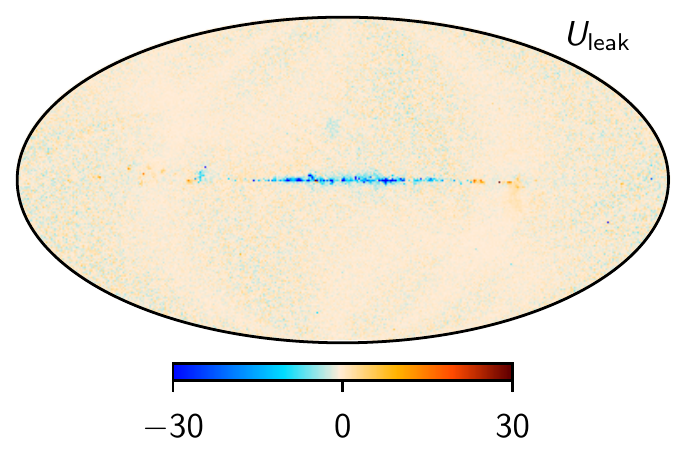}\\
  \includegraphics[width=0.265\linewidth]{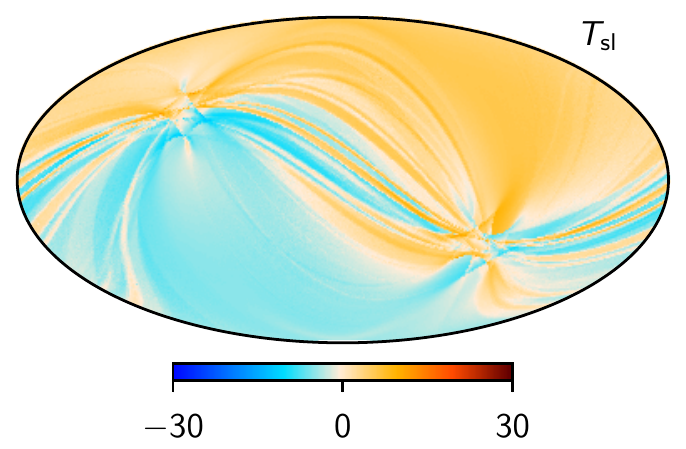}
  \includegraphics[width=0.265\linewidth]{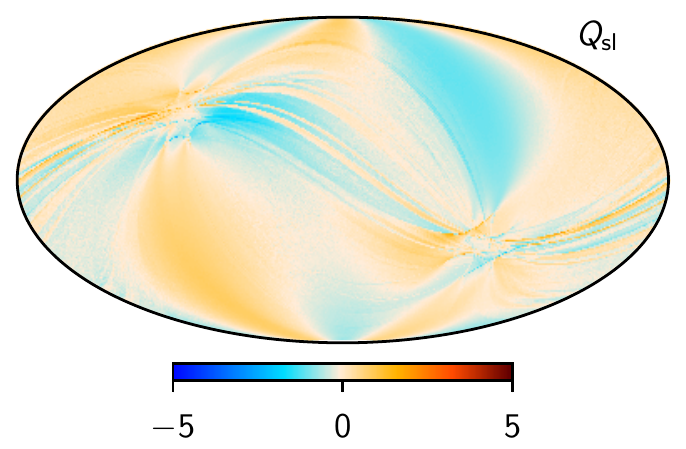}
  \includegraphics[width=0.265\linewidth]{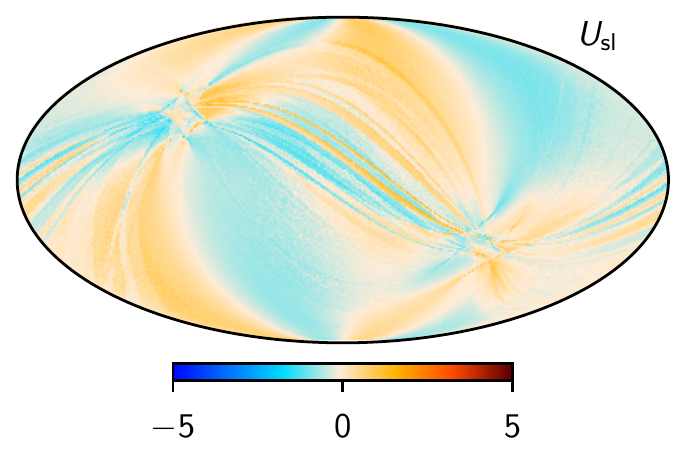}\\
  \includegraphics[width=0.265\linewidth]{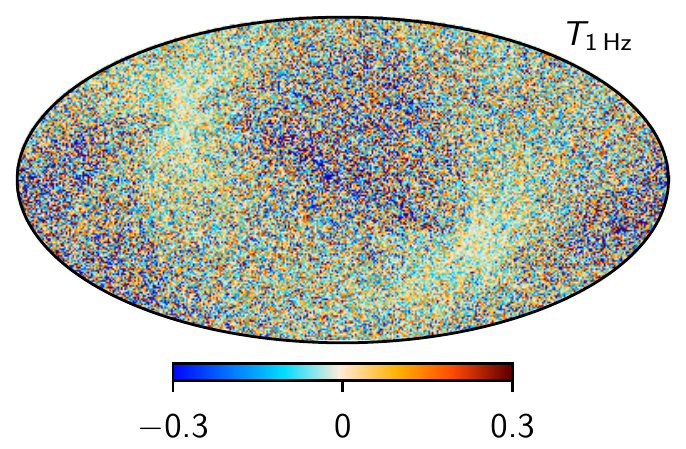}
  \includegraphics[width=0.265\linewidth]{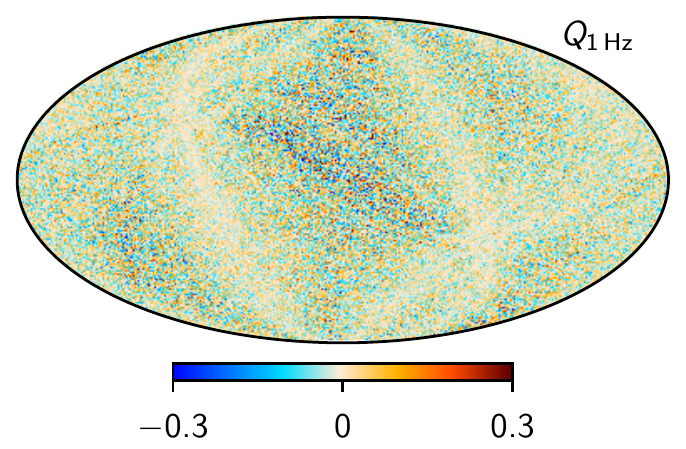}
  \includegraphics[width=0.265\linewidth]{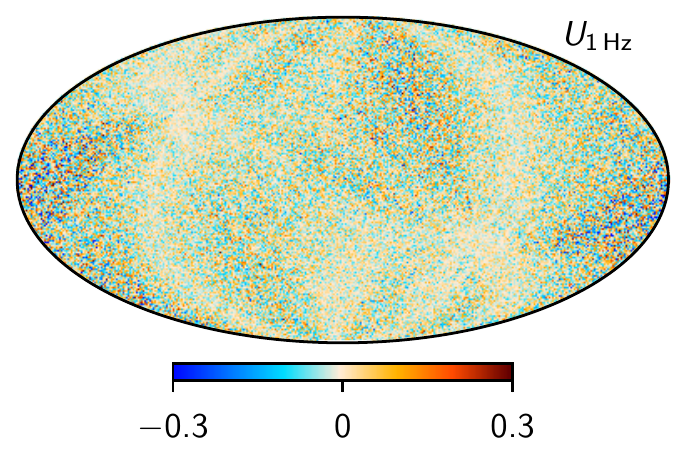}\\
  \includegraphics[width=0.265\linewidth]{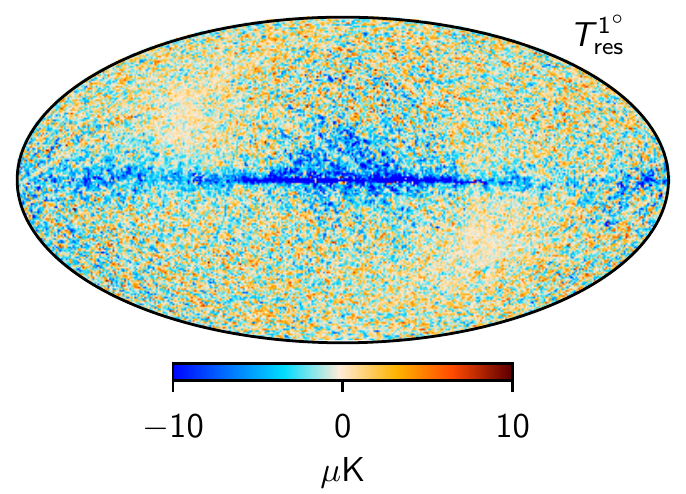}
  \includegraphics[width=0.265\linewidth]{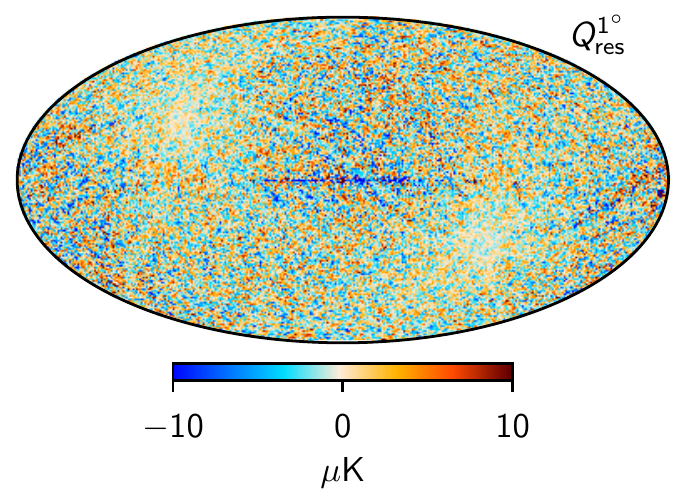}
  \includegraphics[width=0.265\linewidth]{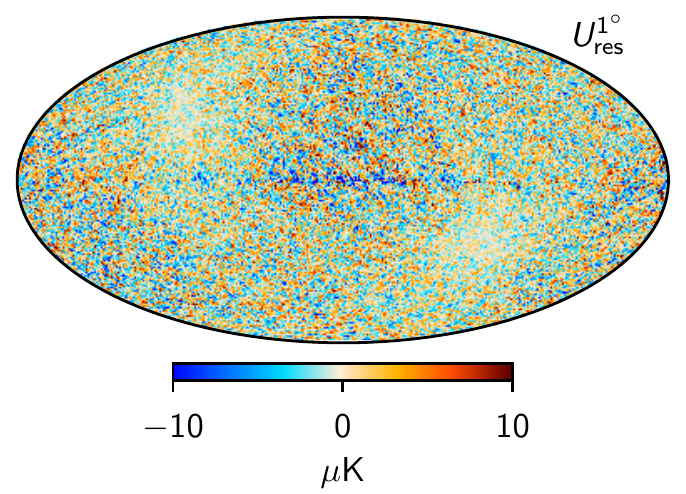}
  \caption{Same as Fig.~\ref{fig:corrmaps30}, but for the 70\,GHz channel.}
  \label{fig:corrmaps70}
\end{figure*}


\begin{figure*}
  \center	
  \includegraphics[width=0.98\linewidth]{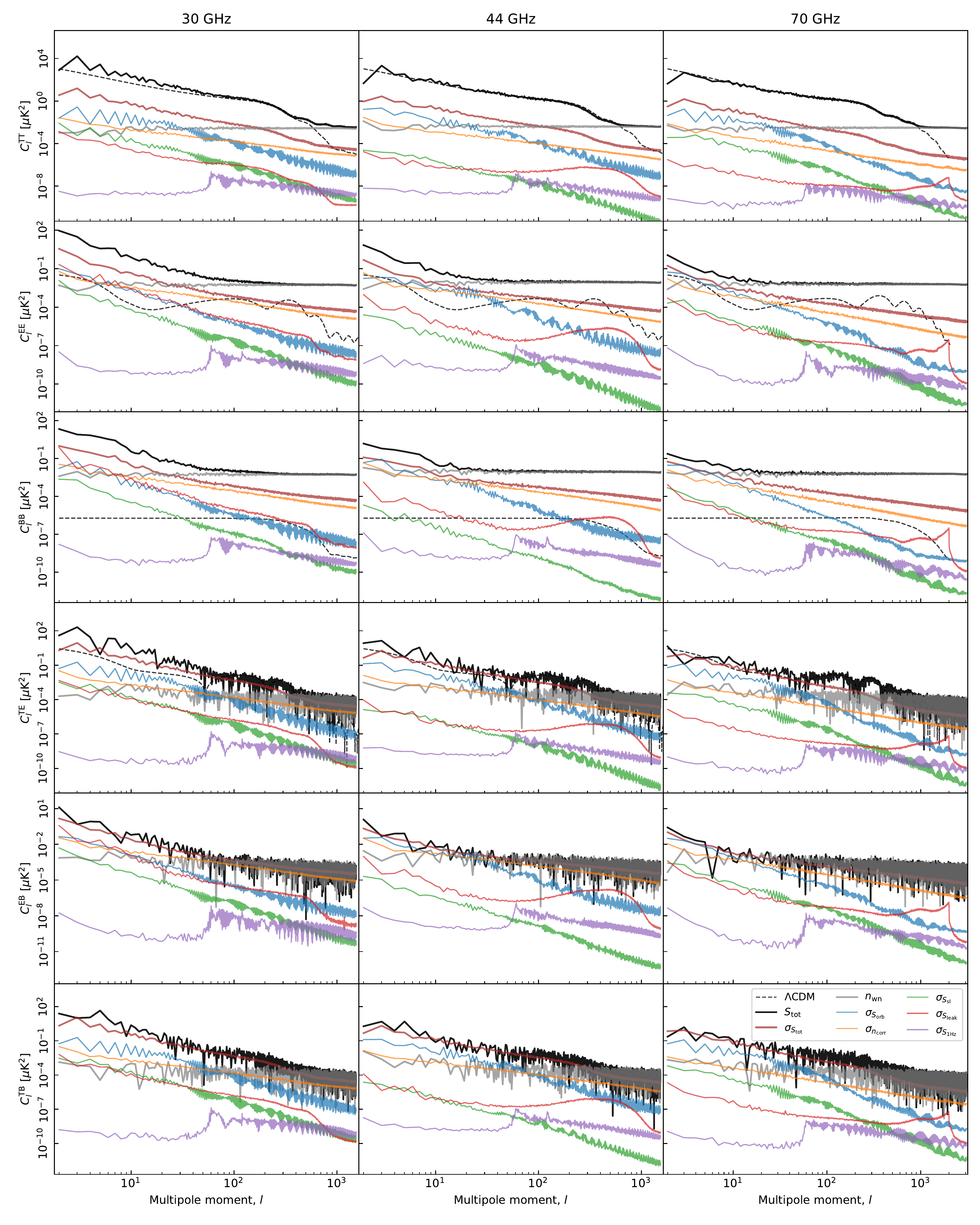}
  \caption{Pseudo-spectrum standard deviation for each instrumental
    systematic correction shown in
    Figs.~\ref{fig:corrmaps30}--\ref{fig:corrmaps70} (\emph{thin
      colored lines}). For comparison, thick black lines show spectra
    for the full co-added frequency map; thick red lines show the
    standard deviation of the same (i.e., the full systematic
    uncertainty); gray lines show white noise; and dashed black lines
    show the best-fit \Planck\ 2018 $\Lambda$CDM power spectrum
    convolved with the instrument beam. Columns show results for 30,
    44 and 70\,GHz, respectively, while rows show results for each of
    the six polarization states ($TT$, $EE$, $BB$, $TE$, $TB$, and
    $EB$). All spectra have been derived outside the CMB confidence
    mask presented by \citet{bp13} using the HEALPix \texttt{anafast}
    utility, correcting only for sky fraction and not for mask mode
    coupling. }
  \label{fig:corrmap_powspec_stddev}
\end{figure*}

The third row shows the orbital dipole. For a single PID, this signal
is defined by a perfect dipole along the scanning ring with an
amplitude of about 270\muK, convolved with the $4\pi$ LFI
beam. However, when the same ring is observed six months apart, the
phase of the signal is reversed, and the total gain- and
noise-weighted sum is then both smaller and difficult to
predict. Also, although the intrinsic signal is entirely unpolarized,
convolution with far-sidelobes algebraically couples this model to the
polarization sector as well. Because of the relatively large amplitude
of this signal, and the fact that it has no free parameters to be
fitted in the pipeline, the orbital dipole represents our best
available tracer of gain fluctuations.

The fourth row shows the bandpass and beam leakage correction. This
effect is clearly the strongest among all polarization corrections,
with amplitudes of many tens of $\mathrm{\mu K}$ in the
Galactic plane, while still being almost entirely negligible in
temperature. Morphologically speaking, the archetypal signature of
bandpass mismatch is a variable sign along the Galactic plane, tracing
the specific orientation of the detector polarization angles as the
different detectors observe at slightly different effective
frequencies. At high latitudes, this map is dominated by
temperature-to-polarization leakage resulting from different
radiometers observing the signal model with different beam FWHMs;
large angular scales are dominated by CMB dipole leakage, while small
angular scales are dominated by foregrounds and CMB temperature
fluctuations.

The fifth row shows the impact of sidelobe pickup. In temperature, the
two dominant features are, first, a large-scale pattern broadly
aligned with the solar CMB dipole resulting from interactions with the
intermediate sidelobes, and, second, individual rings created by the
far sidelobes hitting the Galactic plane. The same features are also
seen in polarization, but now a more complicated pattern arises due to
the additional modulation by the relative orientation of the
polarization angles at any given time.

The sixth row shows the contribution from electronic 1\,Hz spikes. Two
points are important to note regarding this signal. First and
foremost, the color bar only spans 0.3\,\muK, and this term is thus
very small in absolute magnitude compared to all other, 
in agreement with the \Planck\ 2018 analysis. Second, this 
signal is also primarily located on small
angular scales, and looks almost like white noise at the level of sky
maps.

The last row shows the TOD residuals binned into a sky map. For most
of the sky, this is consistent with white noise, but clear residuals
are seen in the Galactic plane, reflecting the structures seen in the
bottom panel of Fig.~\ref{fig:todplot}. This indicates that the
adopted foreground and/or instrument model is not statistically
adequate in these very bright regions of the sky, and all higher-level
CMB-oriented analyses should obviously mask these regions prior to
power spectrum or parameter estimation. However, the main conclusion
to be drawn from this plot is indeed that that the parametric data
model summarized in the above panels is very efficient at describing
the raw LFI TOD at 30\,GHz.

For completeness, Figs.~\ref{fig:corrmaps44} and \ref{fig:corrmaps70}
show corresponding surveys for the 44 and 70\,GHz channels. Overall,
these behave very similarly as the 30\,GHz channel, with minor
variations. For instance, we see that the bandpass corrections are
much smaller at the two higher frequencies, while the sidelobe
correction is particularly low at 44\,GHz, es expected from 
the optical analysis \citep{Sandri_2010}. On the other hand, the
electronic 1\,Hz spike signal is larger at 44\,GHz than either of the
other two channels, but still very small compared to all other
terms. Most importantly, however, we see once again that the
correlated noise and TOD residual maps appear visually clean of large
systematic effects at high Galactic latitudes, and this demonstrates
that the parametric model is able to describe also the 44 and 70\,GHz
TODs to a very high precision.

\subsection{Power spectrum residuals}

The maps shown in Figs.~\ref{fig:corrmaps30}--\ref{fig:corrmaps70}
provide useful intuition regarding the typical amplitude of each
systematic effect, but they do not provide an estimate of the residual
uncertainty associated with each effect. To quantify these residuals,
we first compute the mean of each effect across the full posterior
ensemble, and subtract this from each individual Gibbs sample. We then
compute the (pseudo-)power spectrum of each effect, adopting the CMB
analysis mask presented by \citet{bp11}, while correcting for the
masked sky fraction. We then plot the standard deviation of the
resulting power spectra in Fig.~\ref{fig:corrmap_powspec_stddev} for
each frequency and each polarization spectrum. For comparison, we also
plot the power spectrum of the full posterior mean map (thick black
curves); the standard deviation of the same across all samples (thick
red curves); the white noise level of each channel (thick gray
curves); and the best-fit \Planck\ 2018 $\Lambda$CDM spectrum (dashed
black curves). Thus, this plot provides a fairly comprehensive
description of the various systematic uncertainties modelled by the
\BP\ pipeline.

Starting with the 30\,GHz $TT$ spectrum, we first note that the
posterior mean sky map (thick black) is reasonably well modelled by
CMB (dashed black) and white noise (thick gray) above
$\ell\approx100$. At lower multipoles, there is a notable excess
compared to $\Lambda$CDM, and this may be explained in terms of
contributions from diffuse Galactic foregrounds; this discrepancy is
much smaller at 44 and 70\,GHz.

The red curve shows the sum of all systematic error corrections. In
$TT$, this exceeds white noise below
$\ell\lesssim100$--150, depending on channel, while for $EE$ it
dominates for $\ell\lesssim20$ at 30\,GHz, and for $\ell\lesssim7$--8
at 70\,GHz. The dominant systematic uncertainty in $EE$ at 70\,GHz on
the very largest scales is the orbital dipole (blue curves), which
implies gain fluctuations, as already noted through visual inspection
of the low-resolution covariance matrices in
Sect.~\ref{sec:error_propagation}; in comparison, the contribution
from correlated noise (orange curve) is about an order of magnitude
smaller at these $\ell$s.

This only holds true at 70\,GHz, however. Both at 30 and 44\,GHz, the
gain and correlated noise fluctuations are comparable in magnitude in $EE$,
and at 30\,GHz also the bandpass leakage variations (thin red curve)
are of the same order of magnitude. In contrast, this effect is
completely negligible at both 44 and 70\,GHz.

Next, we see that the sidelobe contribution (green curves) also
appears sub-dominant at all scales in all frequencies. However, this
picture is incomplete for at least two reasons. First of all, as
discussed by \citet{bp08}, uncertainties in the actual sidelobe
response function is not propagated in the current pipeline, and
support for this must clearly be added in a future extension of the
framework. Second, this plot does not account for the overall mean
sidelobe corrections, which, as seen in
Figs.~\ref{fig:corrmaps30}--\ref{fig:corrmaps70}, has a strong
coupling to the CMB Solar dipole, and thereby the overall calibration,
which is accounted for in the red (orbital dipole) curve. Therefore, 
the green curve in this particular figure should not be
interpreted as sidelobes being irrelevant. On the contrary, sidelobes
are important, and their coupling to the calibration serves as a
useful reminder that the entire analysis process is indeed global in
nature, and marginal distributions, like those shown in
Fig.~\ref{fig:corrmap_powspec_stddev}, offers a limited view of the
full joint distribution. Only the full sample set provides a complete
description of the full posterior.

\section{Conclusions}
\label{sec:summary}

In this paper, we have presented novel \Planck\ LFI frequency maps as
derived through the Bayesian \BP\ end-to-end analysis pipeline. These
maps have both lower absolute residual systematic uncertainties than
the corresponding \Planck\ 2018 products, and a substantially more complete
characterization of those residual uncertainties. One important
example of the former is the fact that the absolute calibration
uncertainty for \BP\ is, at least nominally, 25--40 times lower than
that of \Planck\ 2018, while an important example of the latter is
provided by the low-resolution covariance matrices of each pipeline;
the \BP\ covariances include all modelled sources of uncertainty,
while the \Planck\ 2018 covariance matrices only include correlated
noise and a few template-based corrections for gain and foregrounds.

These advances have been made possible through the development of the
first end-to-end CMB analysis pipeline, implemented in the form of a
statistically well-defined Gibbs sampler. The main advantage of this
framework is that all parameters, including relevant instrumental parameters, 
are explored \emph{jointly}, both in terms of individual frequency channels and model
parameters. Explicitly, rather than first calibrating and binning each
frequency channel separately, and then performing component separation 
frequency maps, the current method performs
calibration, mapmaking, and component separation in one tightly
integrated iterative loop in which all parameters communicate directly
with each other.

The importance of the end-to-end approach is clearly demonstrated 
by a careful analysis of the uncertainty budget of residual systematics 
for LFI. For the 70\,GHz $EE$ spectrum, for instance, we show that the 
dominant low-multipole systematic effect is actually detector- and time-dependent gain
variations, and not $1/f$-type correlated noise. For the 30\,GHz
channel, gain, bandpass, and correlated noise uncertainties are all of
comparable order of magnitude. These effects are difficult to
account for through traditional approaches, but easy to model in an a
tightly integrated pipeline.

We also argue that the novel Bayesian approach presented here has a
fundamentally different statistical interpretation than the
traditional frequentist simulation approach adopted by most
experiments to date. While both approaches are indeed
able to model and propagate systematic uncertainties using end-to-end
simulations, the fundamental difference lies in what parameters are
assumed when generating the simulated TOD in the first place. In the
traditional approach some astrophysical and instrumental parameters
are assumed at the start of the analysis with no formal propagation of
their errors; other parameters are drawn from random distributions and
propagated through the pipeline with no realistic reference to our
knowledge of their values. Conversely, in the novel Bayesian approach,
all parameters are derived as constrained realization from the real
dataset. On the other hand, we also note that even the traditional
approach typically adopts constrained realizations for some key
parameters, most importantly the CMB Solar dipole and the diffuse
Galactic model, and this makes its statistical interpretation somewhat
non-trivial. Focusing on the main CMB anisotropies, however, we argue
that the two approaches fundamentally address different questions. The
Bayesian approach is optimally tuned to answer questions like ``what
are the best-fit $\Lambda$CDM parameters of our universe?'', while the
frequentist approach is optimally tuned to answer questions like ``is
our dataset compatible with $\Lambda$CDM?'' For further details on this topic, see \citet{bp04}.

Clearly, after more than two decades of building powerful
community-wide analysis tools to analyze frequentist-style sky maps
and simulations, it will require a few code adjustments before the new
posterior-based products presented here will be used as widely as the
traditional products. The main difference is fortunately
straightforward: Rather than computing a given statistic from a single
best-fit sky map, one should now estimate the same statistic from an
ensemble of maps; for the current \BP\ LFI processing, this ensemble
consists of 3200 individual maps. As such, code parallelization will
be an important task going forward, but, fortunately, for many codes
this has already been done in order to analyze the end-to-end
simulations provided by \Planck. If so, generalization to the Bayesian
approach is indeed straightforward.

Both the current \BP\ code base and the associated LFI products are
made publicly available,\footnote{\citet{bp05}} and generalization to various
other datasets is currently on-going. The most advanced of these is
\WMAP, as reported by \citet{bp17}, for which significant improvements
are found with respect to the official
processing. \textsc{Cosmoglobe}\footnote{\url{http://cosmoglobe.uio.no}}
is an Open Source initiative aiming to coordinate such work for a wide
range of experiments and datasets, and we highly encourage everybody
interested to participate in this community-wide effort.

\begin{acknowledgements}
  We thank Prof.\ Pedro Ferreira and Dr.\ Charles Lawrence for useful suggestions, comments and 
  discussions. We also thank the entire \Planck\ and \WMAP\ teams for
  invaluable support and discussions, and for their dedicated efforts
  through several decades without which this work would not be
  possible. The current work has received funding from the European
  Union’s Horizon 2020 research and innovation programme under grant
  agreement numbers 776282 (COMPET-4; \BP), 772253 (ERC;
  \textsc{bits2cosmology}), and 819478 (ERC; \textsc{Cosmoglobe}). In
  addition, the collaboration acknowledges support from ESA; ASI and
  INAF (Italy); NASA and DoE (USA); Tekes, Academy of Finland (grant
   no.\ 295113), CSC, and Magnus Ehrnrooth foundation (Finland); RCN
  (Norway; grant nos.\ 263011, 274990); and PRACE (EU).
\end{acknowledgements}



\end{document}